\documentclass[12pt]{article}
   \usepackage{epsfig}
      \usepackage{amsmath}
      \usepackage{amssymb}
      \usepackage[font=footnotesize,labelfont=bf]{caption}
\usepackage{subcaption}
\usepackage{cite}

\usepackage{mylatex}

\usepackage{epstopdf}

\usepackage{amsmath}

\usepackage{setspace}
  \usepackage{graphicx}
  \usepackage{color}

 \newcommand{\be}{\begin{equation}}
\newcommand{\bea}{\begin{eqnarray}}
\newcommand{\eea}{\end{eqnarray}}
\newcommand{\beq}{\begin{equation}}
 \newcommand{\ee}{\end{equation}}

\newcommand{\sL}{\textsf{L}}

\def\tr{$\text{tr}$}

\begin{document}
  \renewcommand{\theequation}{\thesection.\arabic{equation}}

\begin{titlepage}

  \bigskip\bigskip\bigskip\bigskip

  \bigskip

\centerline{\Large \bf {Stringy $\mathrm{\mathbf{ER=EPR}}$}}

    \bigskip

  \begin{center}

 \bf { Daniel Louis Jafferis and Elliot Schneider}
  \bigskip \rm
\bigskip

{\it  Center for the Fundamental Laws of Nature, Harvard University, Cambridge, MA, USA}
\smallskip

\vspace{1cm}
  \end{center}

  \bigskip\bigskip

 \bigskip\bigskip
  \begin{abstract}

The $\mrm{ER=EPR}$ correspondence relates a superposition of entangled,
disconnected spacetimes to a connected spacetime with an
Einstein-Rosen bridge. We 
construct examples in which both sides may be 
described by weakly-coupled string theory. The
relation between 
them is given by a Lorentzian continuation of the
FZZ duality of the two-dimensional
Euclidean black hole CFT
in one example, and in another example
by continuation of a similar duality that we propose for  the
asymptotic Euclidean $\mrm{AdS}_3$ black hole. 
This gives a
microscopic understanding of $\mrm{ER=EPR}$: 
one has a worldsheet 
duality between string theory 
in a connected, eternal black hole, and in
a superposition of disconnected geometries in an entangled
state.
The disconnected description includes a condensate of
entangled folded strings emanating from a strong-coupling
region in place of a bifurcation point.
Our construction relies on a Lorentzian interpretation of
Euclidean time winding operators via angular quantization, as
well as some lesser known worldsheet string theories,
such as perturbation theory around a thermofield-double 
state, which we define using Schwinger-Keldysh contours in
target space.

 \medskip
  \noindent
  \end{abstract}

  \end{titlepage}

  \tableofcontents

\section{Introduction and Overview}

The beautiful relation between quantum entanglement and
geometric spatial connection, known colloquially as
$\mrm{ER=EPR}$, has given a 
new perspective on the 
gauge/gravity correspondence
\cite{VanRaamsdonk:2009ar,VanRaamsdonk:2010pw,Hubeny:2007xt,
  Maldacena:2013xja,Ryu_2006,Ryu2_2006}. 
The prototypical example is that a pair
of entangled black holes are joined by an
Einstein-Rosen bridge in their interiors.

\begin{figure}[t]
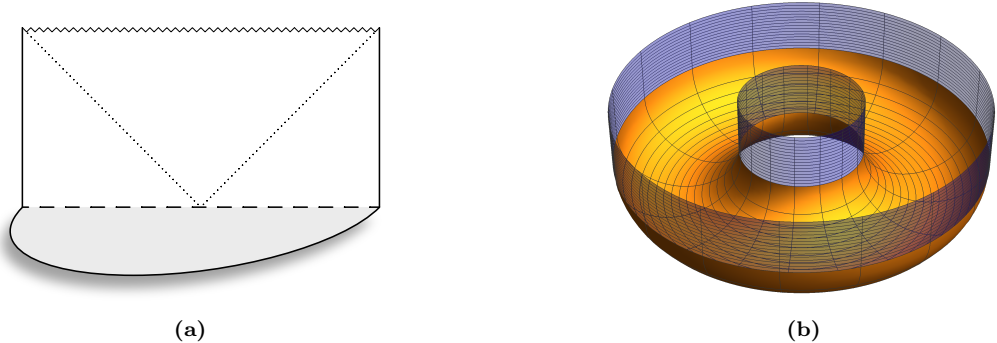

  \centering
  \begin{subfigure}[t]{.43\textwidth}
    \centering
    \ig{0cm}{width=.8\linewidth}{hh}
    \caption{}
    \label{fig:ads-hh}
  \end{subfigure}
\hspace{1cm}
\begin{subfigure}[t]{.43\textwidth}
  \centering
  \ig{0cm}{width=.8\linewidth}{hh-tfd}
  \caption{}
  \label{fig:ads-hh-tfd}
\end{subfigure}
\caption{\footnotesize
\bd{The $\mrm{AdS}_3$ HH State and the Dual TFD}. The conformal
diagram of the 
asymptotic $\mrm{AdS}_3$ two-sided black hole (or, rather, its
top half) is shown on the 
left. Over each point is an additional circle, which is
suppressed in the figure.
The left and right
asymptotic $\mrm{AdS}_3$ regions are causally separated by the
horizons, represented by the diagonal dotted lines. The future
singularity is the 
zigzag line at the top of the diagram. The Hartle-Hawking state
is prepared by halving the Euclidean continuation of the black
hole, shown by the half-disk, and gluing it to the
zero-time slice of the black hole on the horizontal dashed line.
The Euclidean time periodicity is $\beta$, 
the inverse Hawking temperature of the black hole.
The zero-time slice has the topology of an annulus and
the halved Euclidean black hole resembles a half-bagel, obtained by
revolving the dashed line and half-disk around the suppressed
circle. Above the Hawking-Page temperature, this bulk state is
dual to the thermofield-double state in two copies of the
boundary CFT on a circle (times time), shown on the right. The
yellow half-torus prepares the state on its two circle
boundaries, which then evolve forward in Lorentzian 
time along the two blue cylinders. The Hartle-Hawking cap that
prepared the bulk state corresponds to the solid half-torus obtained
by filling in the interior of the yellow surface.}
\label{fig:ads-dual}
\end{figure}

In asymptotic $\mrm{AdS}_{d+1}$, for
example, the two-sided (large) black hole in its Hartle-Hawking (HH)
state at inverse Hawking temperature $\beta$ is  
dual, above the Hawking-Page temperature,
to the thermofield-double (TFD) state 
in two copies of the boundary $\mrm{CFT}$ Hilbert space on
$\mrm{S}^{d-1}$ (Fig. \ref{fig:ads-dual})
\cite{Hartle:1976tp,Hartle:1983ai,Hawking:1982dh,Maldacena:2001kr}. 
The TFD is a maximally entangled state,
prepared by the Euclidean functional integral on
$[0,\beta/2]\times \mrm{S}^{d-1}$, and 
given by a
sum over product states of the two uncoupled CFTs,
$\ket{\mrm{TFD}}\propto \sum
e^{-\beta E_n/2} \ket{n^*}_\mrm{L}\otimes
\ket{n}_\mrm{R}$.\footnote{The star denotes the action of CPT.}
It is a purification of the thermal state at
inverse temperature $\beta$, the reduced density matrix in a
single copy of the CFT being $\tr_{\cH_\mrm{L}}
\ket{\mrm{TFD}}\bra{\mrm{TFD}} \propto e^{-\beta H}$ .

It is natural that
each side of the black hole is dual to a thermal
state of the CFT, the reduced density-matrix of the
HH state in the bulk likewise being
a thermal state of inverse Hawking temperature
$\beta$ \cite{Hawking:1974sw,Witten:1998zw}. 
Moreover,
that the two-sided black hole is dual to a state in two
independent copies of the CFT
is a boundary manifestation of the fact that the
left and right asymptotic $\mrm{AdS}$ regions in the bulk are
causally separated by the horizons.

On the other hand, it is surprising that a superposition of
product states in two
uncoupled CFTs admits a bulk description as a connected spacetime
\cite{VanRaamsdonk:2009ar,VanRaamsdonk:2010pw}.
Each term $\ket{n^*}_\mrm{L}\otimes \ket{n}_\mrm{R}$ in the
TFD sum
corresponds to a pair of disconnected spacetimes.
Thus, the linearity 
of quantum mechanics would seem to imply that their
superposition would in turn
be dual to a superposition of disconnected geometries.
And yet, the
superposition admits a connected description as
a two-sided black hole due to the entanglement\footnote{Below the
  Hawking-Page temperature, the boundary TFD state is instead dual to
  the bulk TFD in two disconnected copies of $\mrm{AdS}$
  (Fig. \ref{fig:ads-tfd}). Then the boundary TFD, and likewise
  the bulk disconnected superposition, is not
  sufficiently entangled to admit a dual semi-classical
  connected description.} 
between the two  
sides (Fig. \ref{fig:erepr}), 
so the existence of an Einstein-Rosen bridge is a
non-linear property of the state.   

There is no sharp contradiction because the Einstein-Rosen
bridge is cloaked behind a horizon, and its existence cannot be
measured at the boundary. Moreover,  scenarios in which it is
operationally observable, such as the traversable wormhole
protocol \cite{Gao:2016bin}, require that the two sides are
coupled together and only apply to a
small subspace of 
states, giving results consistent with linearity.  

\begin{figure}[t]
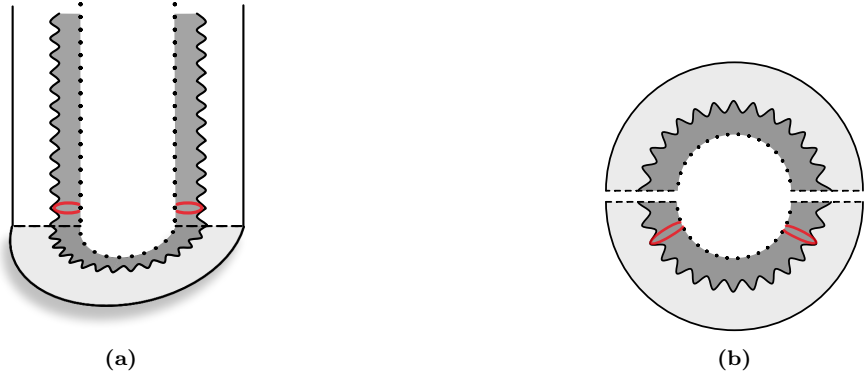

  \begin{subfigure}[t]{.43\textwidth}
    \centering
    \ig{0cm}{width=.5\linewidth}{tfd-string}
    \caption{}
    \label{fig:tfd}
  \end{subfigure}
  \hspace{1cm}
  \begin{subfigure}[t]{.43\textwidth}
    \centering
    \ig{0cm}{width=.5\linewidth}{annulus-string}
    \caption{}
    \label{fig:annulus}
  \end{subfigure}
  \caption{\footnotesize
    \bd{The Bulk TFD State}. The infinite
    cylinder geometry of the sine-Liouville background has the
    topology of an annulus (right). The linear dilaton implies
    that the string coupling vanishes at one asymptotic
    boundary and diverges at the other, represented by the solid
    and dotted circles. When the annulus is halved and glued to
    the Lorentzian continuation, it prepares the
    TFD state in two disconnected copies of flat
    linear-dilaton spacetime (left). The state prepared by
    halving the dual EBTZ background is similar but with an
    additional circle suppressed.
    Also pictured is the embedding of a
    string worldsheet with a pair of Euclidean time winding
    operator insertions
    from the sine-Liouville potential.
    The image of the worldsheet wraps the Euclidean time circle,
    extending out from the dotted strong-coupling boundary
    toward finite coupling, before folding back on itself and
    falling back to strong coupling. 
    When the worldsheet is sliced in angular quantization, 
    each spatial slice is then mapped to a folded string, shown in
    red, that comes 
    in and out of the strong-coupling region. The full
    sine-Liouville potential adds a condensate of such folded
    strings on top of the disconnected union of
    $\text{linear-dilaton} \times \text{time}$ backgrounds.
    Expanding the condensate, one obtains an EPR description
    of a string in a superposition of
    entangled disconnected spacetimes, dual to string theory in the
    connected black hole.} 
  \label{fig:spacetime-tfd}
\end{figure}

All this suggests that there should exist
examples of 
quantum gravity dualities,
relating a connected spacetime to an entangled superposition
of disconnected spacetimes. In this
work, we exhibit exact
string theory dualities of this type. The form of the relations is
between string 
theory in a two-sided black hole\footnote{Or 
  more simply in 
  a two-sided Rindler decomposition of $\mrm{AdS}_3$, with asymptotic
  $\mrm{AdS}_3$ regions separated by coordinate horizons.} made 
of fundamental strings in the HH state (Fig. \ref{fig:ads-hh}), and 
string theory in a pair of disconnected spacetimes in the
TFD state (in the bulk sense), with an entangled
condensate of folded strings (Fig. \ref{fig:tfd}).
Expanding the condensate, one indeed finds a dual EPR-like
description of 
the black hole given by a superposition of disconnected
geometries, each with a number of entangled strings.
On the disconnected side, the 
black hole bifurcation point is replaced by a strong-coupling region from
which the folded strings emanate.\footnote{The disconnected
  geometries may or may not be horizonless, however, once the
  backreaction of the folded strings is taken into account. It
  is reasonable to suspect that a typical microstate with many
  folded strings falling into the strong-coupling region will
  result in the formation of a black hole with a horizon, though we have not
  attempted to verify this.}
In the semi-classical limit,
the black hole description is weakly coupled in the $\alpha'$
sense, and the EPR description is strongly
coupled. Both sides may be taken at weak string coupling,
however.\footnote{In the sense that the boundary correlators or
  scattering amplitudes can be computed in string perturbation
  theory since they are governed by regions of the target space
  with tunably-weak string coupling. The target space in the EPR
  side of the 
  dualities includes a strong-coupling region, however.} 

\begin{figure}[t]
  \centering
  \ig{0cm}{width=.6\linewidth}{erepr}
    \caption{
      \footnotesize
      \bd{Schematic of} $\mrm{ER=EPR}$. The connected, two-sided,
      asymptotically 
      $\mrm{AdS}$ black hole admits a dual description as an
      entangled superposition of disconnected spacetimes,
      represented schematically by the wedges on the left, each
      of which is dual to an energy eigenstate $\ket{n}_\mrm{L},
      \ket{n}_\mrm{R}$ of the two copies of the boundary CFT.
      The $\mrm{ER=EPR}$ correspondence asserts that this
      equivalence of entangled quantum states to connected
      spacetimes holds more generally.
      We find examples of string dualities of this type,
      relating a string in a connected target space to a string
      in an
      entangled superposition of disconnected targets.
    }  
  \label{fig:erepr}
\end{figure}

These string dualities  
are defined by the Lorentzian continuation in the sense of the
target time coordinate
of CFT dualities for Euclidean black hole target spaces in two
and three dimensions. In two dimensions, the Euclidean duality
with which we begin is the well-known Fateev-Zamolodchikov-Zamolodchikov (FZZ)
duality of the asymptotic linear-dilaton Euclidean black hole CFT
\cite{FZZ,Kazakov:2000pm}. In three dimensions, one of the goals
of this paper is to propose that there exists an analogous duality of
the asymptotic Euclidean $\mrm{AdS}_3$ black hole CFT. We provide evidence for this duality by demonstrating that all symmetries match between the two sides, and by showing  that upon gauging  translations
symmetry along the ``third direction'', it reduces to the original FZZ duality. 

Each of these CFT dualities relates worldsheet string theory with a Euclidean black hole target
space characterized by a contractible Euclidean time circle, to a dual
geometry with a non-contractible time circle, together with a
potential built of winding operators. Continuing
the target time in the first case yields the ER description of a string in a
two-sided, Lorentzian black hole. The other yields the EPR
description of a string in a disconnected target. The challenge
in the latter case is to understand what it means to define a
Lorentzian
string theory by continuation from 
a Euclidean background containing insertions of Euclidean time winding
operators.

Developing a formalism for the Lorentzian treatment of
backgrounds built with Euclidean time winding operators is the
other main goal of our work.
We will argue that these string
theories are defined by choosing an appropriate contour of
integration on the string moduli space. In the neighborhood of
Euclidean time winding insertions, the contour involves a
deformation in the sense of angular quantization (as opposed to
the more familiar contour of radial quantization
\cite{Witten:2013pra}), that renders the moduli integral well-defined
when Lorentzian scattering operators collide with Euclidean winding
insertions. This results in a Hilbert space picture in which the euclidean time winding operators define particular states in the BRST quantization of extended strings with asymptotics determined by the angularly quantized operator.

We apply this formalism to the winding operators in sine-Liouville theory and its AdS$_3$ cousin described above, and show that they correspond to pairs of entangled folded strings
emanating from the strongly coupled interiors of the disconnected targets. The full theories have a condensate of such strings. In the ``infinitesimal'' versions of these dualities, which relate different descriptions of the same operator in the black hole target CFT, the winding side corresponds to a specific state in the extended string Hilbert space with both endpoints at the bifurcation point.

In the remainder of this introduction, we give an overview of 
these results. As just mentioned, the $\mathrm{ER = EPR}$ dualities
that we propose are defined by the Lorentzian continuation
of CFT dualities for Euclidean black hole target spaces. The essential 
feature of a Euclidean
duality required to realize $\mrm{ER=EPR}$ upon continuation is
that the 
Euclidean time circle should be contractible on one side of the
duality and non-contractible on the other. The contractible ER
description produces a state in a connected 
Lorentzian geometry  with a horizon upon continuation, while the
non-contractible EPR description gives a state in a 
disconnected geometry. By consistency, the non-contractible
description must feature
a mechanism that violates the winding
number conservation law around the 
Euclidean time circle, since the
string can unwind in the contractible dual. One could imagine
that this 
winding violation is accomplished by the inclusion of 
D-branes or by a potential that explicitly breaks the
symmetry.\footnote{Or by strong-coupling effects.}

The presence of D-branes would allow closed strings wrapping the
Euclidean time direction to break into pairs of open
strings ending on the brane. Equivalently, the boundary state of
the D-brane in the 
worldsheet CFT is a linear combination of states with
different Euclidean time winding numbers.
Such a duality would
necessarily involve the
cancellation of string loop diagrams on the EPR side with
tree-level effects on the ER side, however, which may be
complicated.

The EPR side of the duality can be interpreted as describing the
constituent objects that make up the black hole. In backgrounds with
Ramond-Ramond fluxes, one would therefore expect to find
D-branes. In our examples, however, the dualities only
involve closed strings, corresponding to black holes that
are made of fundamental strings.  
It will instead be a winding potential 
that is responsible for breaking the symmetry in these EPR
theories, 
corresponding to a condensate of strings that wrap
the non-contractible Euclidean time circle.

Our first example is found in
two-dimensional dilaton-gravity, where the
$\mrm{SL}(2,\reals)_k/\mrm{U}(1)$ CFT describes for large $k$ a
string in a 
cigar-shaped Euclidean black hole with an asymptotically linear
dilaton  (Fig.
\ref{fig:cigar}) \cite{Witten:1991yr}.
The dual description of
this CFT due to FZZ is given by a sigma-model into 
a cylinder, now with an infinite linear-dilaton direction, plus a
condensate of winding strings
(Fig. \ref{fig:sine-liouville}). It is known as the sine-Liouville
background, and the condensate is called the sine-Liouville
potential.

\begin{figure}[t]
  \centering
  \begin{subfigure}[t]{.43\textwidth}
    \centering
    \ig{0cm}{scale=.15}{cigar}
    \caption{
      \footnotesize
      \bd{The Cigar Background}. The cigar sigma-model is a
  weakly-coupled Lagrangian description of the
  $\mrm{SL}(2,\reals)_k/\mrm{U}(1)$ CFT when $k$ is large.
  For large $r$, the geometry is a cylinder of radius $\sqrt{\alpha'
    k}$, and as $r \to 0$ the cylinder smoothly caps off. The
  dilaton is a monotonically decreasing function of $r$. Its
  maximal value  $\Phi_0$ is attained at the tip, and at large
  $r$ it falls off linearly as $-r$ and the string coupling
  goes to zero. Although a string in the 
  weak-coupling region appears to be able to wind around the
  cylinder, there is no conserved topological charge because the
  string can unwind at the tip.}  
  \label{fig:cigar}
\end{subfigure}
\hspace{1cm}
\begin{subfigure}[t]{.43\textwidth}
  \centering
  \ig{0cm}{scale=.15}{sine-liouville}
  \caption{
    \footnotesize
    \bd{The sine-Liouville Background}. According to the
    FZZ duality, the sine-Liouville sigma-model is a dual
     description of the
    $\mrm{SL}(2,\reals)_k/\mrm{U}(1)$ CFT, better suited when $k$
    is small (compared to 2). The geometry is an infinite
    cylinder of radius 
    $\sqrt{\alpha' k}$. The dilaton is $\Phi = - Q\hat r$,
    so that the string coupling $e^{\Phi}$ diverges as
    $\hat r \to -\infty$ and vanishes as $\hat r \to \infty$. The
    sine-Liouville potential $e^{- \sqrt{(k-2)/\alpha'}
      \hat r} \mrm{Re}\, e^{i 
    \sqrt{k/\alpha'}(\hat \theta_\mrm{L} - \hat
    \theta_\mrm{R})}$ includes a pure-winding
  mode of $\hat \theta$ (represented by the circles wrapping
  the middle of the cylinder), times a linear-dilaton primary 
  (represented by the gradient).}
  \label{fig:sine-liouville}
\end{subfigure}
\caption{}
\end{figure}

The cigar description is weakly coupled for large
$k$, while the sine-Liouville description is strongly coupled.
Both backgrounds asymptote to 
identical cylinders in the region where the
linear dilaton goes to minus infinity and the string coupling
subsequently vanishes. Whereas the cigar geometry terminates at
finite string coupling at its tip, in sine-Liouville the
cylinder extends forever and the string coupling diverges.
In this description it is instead the sine-Liouville potential that is
responsible for  
suppressing string configurations that extend into the
strong-coupling region.
In both backgrounds
the apparent winding number conservation law of the common asymptotic
cylinder region is violated---in the cigar because a 
wound string may unwind at the tip, and in sine-Liouville
because the winding potential explicitly breaks the
symmetry.

\begin{figure}[t]
  \centering
  \begin{subfigure}[t]{.43\textwidth}
    \centering
    \ig{0cm}{width=.25\linewidth}{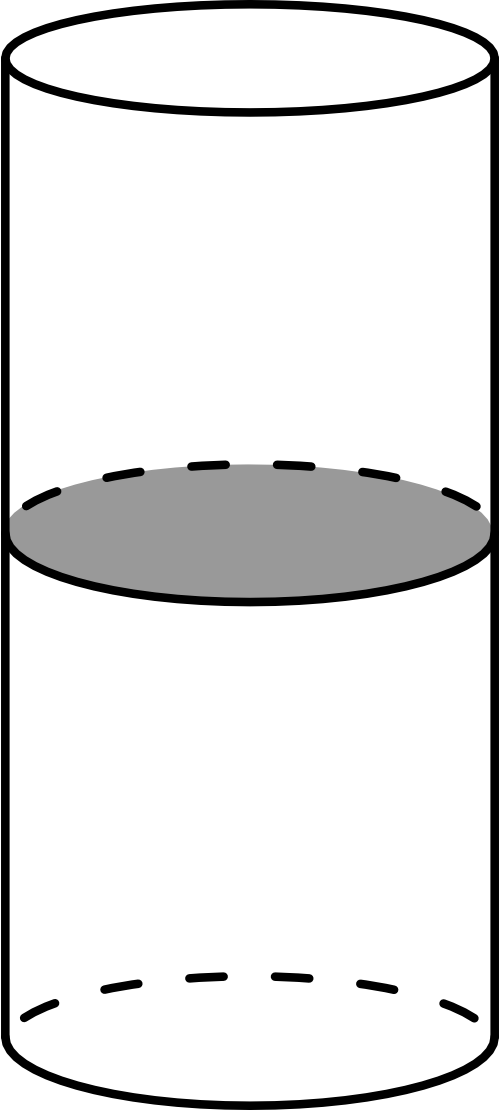}
    \caption{
      \footnotesize
      \bd{The Euclidean $\mrm{AdS}_3$ Black Hole}. 
      $\mrm{EAdS_3}$ may be described as a solid
      cylinder. Compactifying the length of the cylinder to
      form a solid torus yields the Euclidean continuation of
      the asymptotic $\mrm{AdS}_3$ black hole, where the
      continuation is performed with respect to the contractible
      cycle. The coset manifold
      $\mrm{SL}(2,\bbC)/\mrm{SU}(2)$ is equivalent to 
      $\mrm{EAdS}_3$, and a string in $\mrm{EAdS_3}$ may
      therefore be described by the
      $\mrm{SL}(2,\bbC)_k/\mrm{SU}(2)$ coset WZW model, where
      $\alpha' k = l_\mrm{AdS}^2$ sets the
      $\mrm{AdS}$ length. The quotient $\bbZ
      \bs \mrm{SL}(2,\bbC)_k/\mrm{SU}(2)$ thus describes a
      string in the asymptotic $\mrm{EAdS}_3$ black hole, known
      as Euclidean BTZ (EBTZ).}   
  \label{fig:ebtz}
\end{subfigure}
\hspace{1cm}
\begin{subfigure}[t]{.43\textwidth}
  \centering
  \ig{0cm}{width=.25\linewidth}{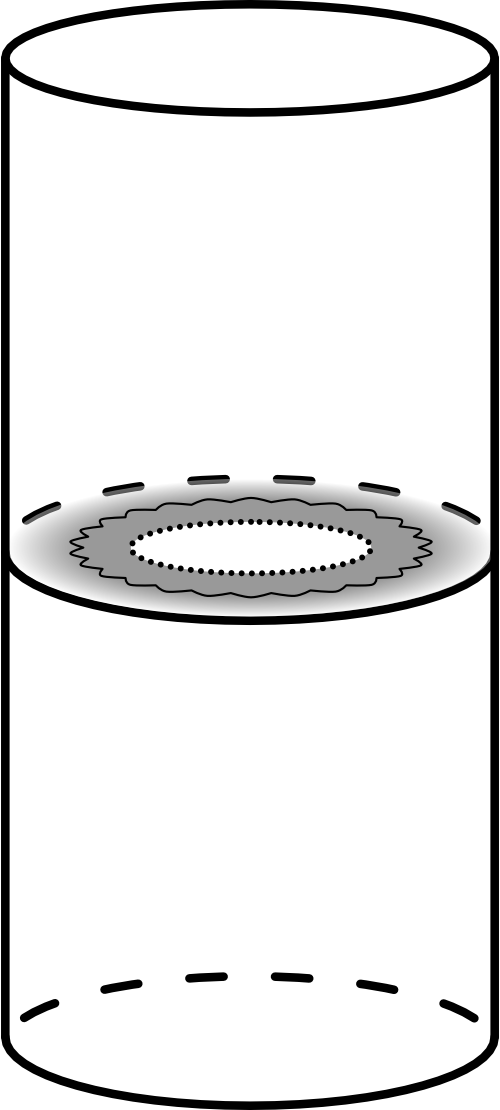}
  \caption{
    \footnotesize
    \bd{The 3D sine-Liouville Background}. In the dual
    description we propose for the
    $\mrm{SL}(2,\bbC)_k/\mrm{SU}(2)$ CFT and its black hole
    quotient $ \bbZ \bs
    \mrm{SL}(2,\bbC)_k/\mrm{SU}(2)$, the radial direction of the
    cylinder or solid torus is replaced by an infinite
    linear-dilaton direction, and a condensate of  strings
    winding the resulting non-contractible cycle is
    added. Gauging the translation symmetry along the length of
    the cylinder produces the
    two-dimensional 
    sine-Liouville background (Fig. \ref{fig:sine-liouville}), whereas
    gauging the same symmetry in the first description yields
    the two-dimensional cigar (Fig. \ref{fig:cigar}). Thus, this
    duality is an uplift 
    of the FZZ duality to a three-dimensional target space.}
  \label{fig:3d-sl}
\end{subfigure}
\caption{}
\end{figure}

In three-dimensional spacetime, we propose a similar duality of
the $\bbZ\bs \mrm{SL}(2,\bbC)_k/\mrm{SU}(2)$ CFT that describes
a string in the asymptotic Euclidean $\mrm{AdS}_3$ black
hole, known as Euclidean BTZ (EBTZ), which has the topology of
a solid torus (Fig. \ref{fig:ebtz}). The duality in fact applies
as well to  $\mrm{SL}(2,\bbC)_k/\mrm{SU}(2)$ itself, prior to
the $\bbZ$ quotient. The latter describes a string in Euclidean
$\mrm{AdS}_3$ ($\mrm{EAdS}_3$), which is equivalent to the
$\mrm{SL}(2,\bbC)/\mrm{SU}(2)$ coset manifold.
$\mrm{EAdS}_3$ may be described as a solid cylinder, and
compactifying its length yields the EBTZ solid torus
$\bbZ\bs \mrm{SL}(2,\bbC)/\mrm{SU}(2)$, where the circumference
$4\pi^2/\beta$ of the compactification is fixed by the inverse
Hawking temperature $\beta$ of the black hole. 
In the dual description, the radial direction
of the torus or cylinder is replaced by an infinite linear-dilaton
direction as in the sine-Liouville background, and a condensate
of winding strings wrapping the resulting
non-contractible cycle is again added (Fig.
\ref{fig:3d-sl}).\footnote{As we explain in
  Sec. \ref{sec:ads-duality}, the duality requires that we express
  the boundary cylinder or torus variables of
  $\mrm{SL}(2,\bbC)_k/\mrm{SU}(2)$ or $\bbZ \bs
  \mrm{SL}(2,\bbC)_k/\mrm{SU}(2)$ in the first-order formalism.}
The original description is weakly coupled for large
$k$ while the dual description is strongly coupled, and once
again $g_\mrm{s}$ diverges at the strong-coupling
boundary of the dual geometry.\footnote{\label{ft:small-k} For
  small $k$ (relative 
  to its minimal value $k = 2$), on the
  other hand, the original description of
  $\mrm{SL}(2,\bbC)_k/\mrm{SU}(2)$ is strongly coupled. 
  Then we expect that the dual
  sigma-model is the better description of the $\mrm{AdS}_3$
  vacuum for small $k$. It would be interesting to understand
  the connection between the sine-Liouville description and
  recent work on string theory in $\mathrm{AdS}_3$ at small $k$
  \cite{Eberhardt_2020,Gaberdiel_2018,Eberhardt_2019,Giribet_2018}.} 
Gauging the translation 
symmetry around the original non-contractible cycle of the torus
or the length of the cylinder in the
two descriptions reproduces the cigar and sine-Liouville
backgrounds. Thus, this duality may be thought of as a
three-dimensional uplift of the FZZ
duality.\footnote{$\mrm{SL}(2,\bbC)/\mrm{SU}(2)$ is a Euclidean
  continuation of 
$\mrm{SL}(2,\reals)=\mrm{AdS}_3$, and the two-dimensional black
hole may equivalently be thought of as a coset of the
former.}\footnote{Potentially related work on a
three-dimensional uplift of the FZZ duality was discussed in
\cite{Berkooz:2007fe}.}\footnote{We expect there is also a
supersymmetric version of this duality, as in the
supersymmetric FZZ duality of \cite{Hori:2001ax}. We
focus on the bosonic string in this paper for simplicity,
though we expect similar examples of $\mrm{ER = EPR}$ would hold
for the superstring in the cigar and the supersymmetric
$\mrm{AdS_3\times S^3}$ background.}

These dualities share the essential feature
that the two related
target space geometries are of different topologies.
In the cigar description of $\mrm{SL}(2,\reals)_k/\mrm{U}(1)$,
the geometry has the topology of a disk, with 
the asymptotic cylinder capping off at the origin.
By contrast, in the
sine-Liouville description the cylinder is infinite, and the
topology is an annulus. 
Thus, the
circle direction of the two geometries, which is defined as
the Euclidean time, is contractible on one side of the duality
and non-contractible on the other. Similarly, in the $\bbZ \bs
\mrm{SL}(2,\bbC)_k/\mrm{SU}(2)$ duality the contractible cycle
of the torus in the original description is replaced by a
non-contractible cycle in the dual, exchanging the $\mrm{disk}\times
\mrm{S}^1$ topology with an $\mrm{annulus}\times \mrm{S}^1$. It is again
this cycle that 
one defines as the Euclidean time in order to obtain the Lorentzian
black hole upon continuation.\footnote{In the three-dimensional
  case one may alternatively continue with respect to the other
  cycle, which prepares a
  thermal state in $\mrm{AdS}_3$, or simply the vacuum state
  prior to the $\bbZ$ quotient. See Ft. \ref{ft:small-k}.}

\begin{figure}[t]
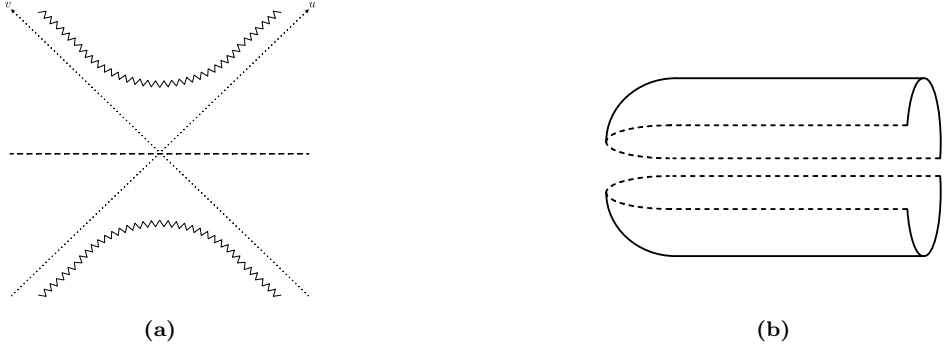

  \begin{subfigure}[t]{.43\textwidth}
    \centering
    \ig{0cm}{width=.6\linewidth}{kruskal}
    \caption{}
    \label{fig:black-hole}
  \end{subfigure}
  \hspace{1cm}
  \begin{subfigure}[t]{.43\textwidth}
    \centering
    \ig{.6cm}{width=.65\linewidth}{half-cigar}
    \caption{}
    \label{fig:half-cigar}
  \end{subfigure}
  \caption{\footnotesize
    \bd{The Two-Dimensional Black Hole}. The Lorentzian
    continuation of the Euclidean cigar 
    is a two-sided, eternal black hole. The horizons are
    the diagonal dotted lines, and the past and future
    singularities are the zigzag hyperbolas at the bottom and
    top. The geometry is invariant under time reflection about
    the dashed line, which enables the construction of the
    Hartle-Hawking state. The cigar, which has the
    topology of a disk, is cut in half and glued to the black
    hole along the fixed line of the reflection symmetry,
    similar to Fig. \ref{fig:ads-hh} but resembling the
    Schwarzschild causal diagram rather than BTZ. This
    Euclidean cap prepares the state on the dashed
    line, which is then evolved forward in Lorentzian time.}
\end{figure}

\begin{figure}[t]
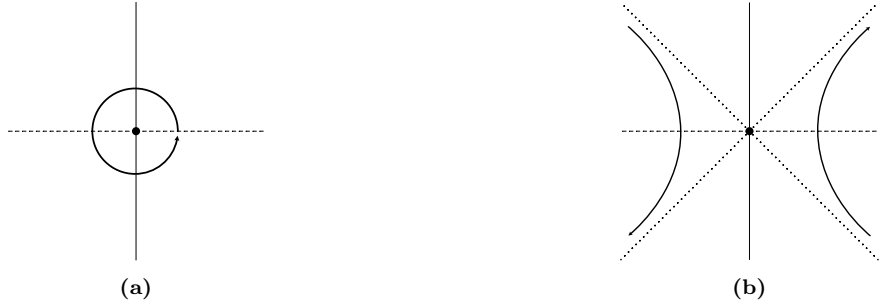

  \begin{subfigure}[t]{.43\textwidth}
    \centering
    \ig{0cm}{width=.5\linewidth}{angular-quantization}
    \caption{}
    \label{fig:angular-quantization}
  \end{subfigure}
  \hspace{1cm}
  \begin{subfigure}[t]{.43\textwidth}
    \centering
    \ig{0cm}{width=.5\linewidth}{rindler}
    \caption{}
    \label{fig:rindler}
  \end{subfigure}
  \caption{\footnotesize
    \bd{Angular Quantization}. When $\reals^2$ in polar
    coordinates (left) is continued with respect to the angular
    direction, the result is a wedge of the Rindler
    decomposition of Minkowski spacetime (right). With no
    operator insertions, the Euclidean functional integral over
    the half-plane prepares the Minkowski vacuum on the dashed
    line. In angular quantization, the Minkowski vacuum is
    identified with the TFD state in the two copies of the
    Hilbert space on the left and right Rindler wedges. The
    reduced density matrix in a single wedge is a thermal state
    at the Unruh temperature. To understand a Lorentzian
    string background with Euclidean time winding operator insertions,
    the neighborhood of the winding operator should be treated
    in angular quantization.
    The Hilbert space of angular quantization is a
    generalization of 
    the Rindler Hilbert space, now labeled by the asymptotic
    condition imposed by the boundary operator.}
    \label{fig:angular-continuation}
  \end{figure}

Indeed, continuing with respect to a contractible
Euclidean time circle yields a Lorentzian geometry with a
horizon; the vanishing coefficient of the
Euclidean time in the metric at the point where the circle
shrinks implies that the continuation is a
Lorentzian wedge bounded by a horizon at the same
point. The simplest example is the continuation of
$\reals^2$ with respect to angular Euclidean time $\theta = i
t$: $\diff s^2 = \diff r^2 + r^2 \diff \theta^2 \to \diff r^2 -
r^2 \diff t^2$. The result is the right wedge of the Rindler
decomposition of Minkowski spacetime, bounded by the Rindler
horizon at $r = 0$ where the coefficient of $\diff t^2$
vanishes (Fig. \ref{fig:rindler}).
The Lorentzian continuation of the cigar, whose metric is $\diff
s^2 \propto \diff r^2 + \tanh^2(r) \diff \theta^2$, is   
an eternal, two-sided black hole of two-dimensional
dilaton-gravity (Fig. \ref{fig:black-hole})
\cite{Witten:1991yr}. In fact, in the neighborhood of the tip  of
the cigar where $\tanh^2(r)$ approaches $r^2$, the geometry is simply
$\reals^2$, and thus the near-horizon geometry of the black hole
is again Rindler.\footnote{Of course, in Rindler the horizon is an
  artifact of the coordinates, whereas it is a genuine horizon 
  in the case of the black hole.}
Likewise, the continuation of the EBTZ solid
torus with respect to its contractible cycle is the asymptotic
$\mrm{AdS}_3$ black hole, known as BTZ
\cite{Banados:1992wn,Banados:1992gq}, whose near-horizon
geometry is $\text{Rindler}\times \mrm{S}^1$.

Continuing with respect to 
a non-contractible Euclidean time circle, on the other hand, naturally
produces  a disconnected Lorentzian geometry, as we 
recall momentarily. In both cases, the compactness of the
Euclidean time 
direction leads to a thermal state.

Although a black hole such as
Fig. \ref{fig:black-hole}  
is time dependent with respect to the global Kruskal time that
flows vertically, it is symmetric under
time reversal, which is a 
reflection about the horizontal dashed line in the figure. The
existence of this $\mrm{Z}_2$ symmetry ensures that the
Euclidean continuation of the geometry is real, and that the
fixed-point locus is common to both the Lorentzian and Euclidean
sections. Thus, the two may be cut in half and glued together
along this line as shown in Fig. \ref{fig:ads-hh} in the
$\mrm{AdS}$ case.
The functional integral over the Euclidean section prepares
the HH state on the asymptotic linear-dilaton
black hole background
\cite{Hartle:1976tp,Hartle:1983ai,Maldacena:2001kr}. This 
state is a generalized notion of a vacuum, and its existence is
due to the $\mrm{Z}_2$ symmetry of the black
hole. 

The infinite cylinder of the sine-Liouville background may
similarly be cut and continued to prepare a state in a
Lorentzian geometry (Fig. \ref{fig:spacetime-tfd}).
Now the half-annulus topology prepares a state
in the disconnected union of two copies of
$\reals^{1,1}$. This is the TFD state in the
bulk disconnected spacetime. On
each copy of $\reals^{1,1}$, one has a linear dilaton along the
spatial direction, with one asymptotic boundary at weak string
coupling and the other at strong
coupling, represented by the solid and dotted lines. 

On top of this free-field background, one has the condensate of
strings that wind the non-contractible Euclidean time
circle, which by the FZZ duality is equivalent to the connected
linear-dilaton black hole. A winding string worldsheet is also
shown in the  
figure. An important aspect of our construction is
to explain the interpretation of these 
winding strings in the Lorentzian continuation, which we will
argue produce pairs of entangled folded strings emanating from
the strong-coupling boundaries.

Note that although one could T-dualize the sine-Liouville
background, and thereby replace the winding potential with a
momentum potential whose Lorentzian interpretation is more straightforward,
its continuation would not be dual to the Lorentzian black
hole. To obtain a Lorentzian duality, one must continue with
respect to the same $\mrm{Z}_2$ symmetry on both sides of the
Euclidean duality. The continuation of the T-dualized
sine-Liouville background would instead be dual to the
continuation of the so-called trumpet geometry
\cite{Dijkgraaf:1991ba}, which has a naked singularity where the
$\theta$ circle of the cigar shrinks.

One may likewise prepare the HH state for the
$\mrm{AdS}_3$ black hole by slicing the EBTZ solid torus in half
across its contractible Euclidean time cycle, producing a
manifold in the shape of a halved bagel, and gluing its annulus
boundary to the identical
zero-time slice of the two-sided black hole
(Fig. \ref{fig:ads-hh}). The two circle 
boundaries of the annulus correspond to 
slices of the left and right asymptotic $\mrm{AdS}_3$
cylinder boundaries. Cutting the now non-contractible cycle of
the dual geometry, on the other hand, prepares a TFD state in
two disconnected copies of
$\text{linear-dilaton}\times\text{time}\times \mrm{S}^1$,
together with the condensate.\footnote{The
  $\text{linear-dilaton}\times\text{time}\times \mrm{S}^1$
  coincides with the asymptotics of $\mrm{AdS}_3$ written in
  first-order variables, as explained in
  Sec. \ref{sec:ads-duality}.}

Complexified spacetimes such as Fig. \ref{fig:ads-hh}
and \ref{fig:tfd}
are familiar in quantum field theory. The Euclidean cap
specifies the domain on which 
the fields in the functional integral are defined, and by gluing
two Euclidean caps together with a Lorentzian excursion in
between, one obtains a Schwinger-Keldysh contour on which the
functional integral computes expectation values in the
states specified by the caps \cite{Schwinger:1960qe,Keldysh:1964ud}. 
Such expectation values may be obtained
by computing the Euclidean correlation
function and then continuing the operator insertions to the
Lorentzian section.

From the point of view of string theory,
the Schwinger-Keldysh contour is now interpreted as
the integration cycle in a complexification of the target space over
which the worldsheet functional integral is evaluated.
Note that 
it is imprecise to merely ask for a string amplitude in
e.g. the black hole background---one must also specify a
state in order to define a string perturbation theory.
For example, one could ask for a string amplitude in
$\mrm{AdS}_3$ in the vacuum state or in a thermal state; the
Lorentzian section is the same in both cases, but the
string perturbation theories are different.
The state is fixed by the incoming and outgoing Euclidean
segments of the target space Schwinger-Keldysh contour.
Thus, the
target space contour obtained by gluing together two copies of
Fig. \ref{fig:ads-hh} produces string amplitudes for
the black hole background in the HH state. Similarly, the
contour obtained by gluing
together two copies of Fig. \ref{fig:tfd}---prior to adding the
condensate---computes amplitudes 
for string theory in 
$\reals^{1,1} \cup \reals^{1,1}$ in the TFD state, or $\reals^{1,1}\times \mrm{S}^1 \cup
\reals^{1,1} \times \mrm{S}^1$ in the three-dimensional case.

The two-dimensional linear-dilaton black hole is asymptotically
flat, and the simplest string amplitudes for the black hole in
the HH state compute conventional scattering amplitudes for
strings $\cO_{jE}$ scattering off the horizon
\cite{Dijkgraaf:1991ba}. They are labeled by their energy $E$
and linear-dilaton momentum $j \in \frac{1}{2} + i \reals$, subject
to the on-shell condition.\footnote{As well as a
  contribution $h$ from the internal CFT, in general, and a
  left/right label depending from which asymptotically flat
  region the particle originates.} The energy spectrum is
continuous at leading order in string perturbation theory, as 
is consistent with effective field theory in the black hole
background.
Such amplitudes may be obtained by
continuation from cigar amplitudes of operators $\cO_{jn}$ with
momentum $n \to i E$ 
around the compact Euclidean time circle.\footnote{More generally, the
  $\mrm{SL}(2,\reals)_k/\mrm{U}(1)$ CFT includes primaries 
  $\cO_{jnw}$ labeled by a third number $w$, related to the
  (non-conserved) 
  winding number of $\mrm{S}^1$ at infinity. To compute a
  scattering amplitude of a string in the Lorentzian black hole
  one continues $\cO_{j,n=iE,w=0}$.}  In
doing so, one must continue from the discrete set of Matsubara
frequencies $n$ of the Euclidean theory 
to the continuous set of Lorentzian energies $E$, as is
often necessary in thermal systems.

In three dimensions the continuation is in fact simpler. Because
$\mrm{AdS}$ is effectively a box, string amplitudes do not
compute an $S$-matrix, but rather correlation functions of the
dual CFT on the spacetime conformal boundary 
\cite{Giveon:1998ns,Kutasov:1999xu,deBoer:1998gyt,Maldacena:2001km}.
One may choose a basis of worldsheet
vertex operators labeled by 
the point on the conformal boundary where a dual operator is
inserted \cite{Teschner:1999ug}. Then Lorentzian string
amplitudes are obtained by 
continuation from Euclidean amplitudes in the usual sense of
continuation of the dual CFT. For example, a 
vertex operator $\Phi_j(z, \bar z; \xi, \theta)$ of
$\mrm{SL}(2,\bbC)_k/\mrm{SU}(2)$ describing string theory
in $\mrm{EAdS}_3$ in the spacetime vacuum state\footnote{Once
  combined with an appropriate internal CFT to cancel the
  conformal anomaly.} is labeled by a
point $(\xi,\theta) \in \reals\times \mrm{S}^1$ on the Euclidean
conformal boundary cylinder, where $z$ is a worldsheet coordinate. A
string amplitude of such operators computes a dual CFT
correlation function with insertions at those points, and by
continuing $\xi \to i t$ to
the Lorentzian section of the Schwinger-Keldysh contour, one
obtains an expectation value of the dual CFT in the vacuum
state.

Similarly, vertex operators of $\bbZ \bs
\mrm{SL}(2,\bbC)_k/\mrm{SU}(2)$ may be labeled by a point on the
spacetime conformal boundary $\mrm{T}^2$ of the EBTZ solid
torus. A Euclidean string amplitude computes a correlation
function of local operators inserted at those boundary points,
and by continuing the insertions one
obtains a string amplitude in the BTZ black hole in the HH state
\cite{Natsuume:1996ij,Hemming:2001we,Hemming:2002kd}. Above the
Hawking-Page temperature, these amplitudes compute the dominant
contribution to dual CFT expectation values in the TFD
state.

Alternatively, one may  Fourier transform the vertex operators
from the boundary position basis to obtain eigenstates of the
spacetime energy and momentum,
analogous to the aforementioned operators
in two dimensions. These correspond to the conventional modes of the
spacetime effective field theory.
Their energy spectrum is again
continuous in BTZ in the HH state---and discrete in
$\mrm{AdS}_3$ in the vacuum state. To directly compute the
Lorentzian string amplitudes of such vertex operators requires 
performing the continuation from the
Euclidean Matsubara frequencies, which may be more technically
challenging because one is continuing from a discrete set of data.
The amplitudes compute dual CFT expectation
values of the modes of its local operators. Of course, one may
alternatively compute a string amplitude in the boundary position basis
where the continuation to Lorentzian signature is easy, and
then Fourier transform the result to obtain an expectation value in
the momentum basis.

In this way, cutting and continuing the dual descriptions of the
$\mrm{SL}(2,\reals)_k/\mrm{U}(1)$ and 
$\bbZ \bs \mrm{SL}(2,\bbC)_k/\mrm{SU}(2)$ CFTs yield
Schwinger-Keldysh contours for the $\mrm{ER = EPR}$ dual string
theories that we propose in this paper. On the ER side, one has
string theory in the two or three-dimensional eternal black hole in the
HH state. On the EPR side, one has string theory in the
disconnected union of two copies of $\reals^{1,1}$ or
$\reals^{1,1}\times \mrm{S}^1$ in the TFD state, deformed by the
sine-Liouville condensate. The condensate is the most subtle
aspect of the continuation; it is built of strings that wind the
Euclidean time circle, and therefore its interpretation in the
Lorentzian continuation is not obvious. 

In two dimensions, the condensate takes the form
$V_\mrm{sL} \propto 
W_+ + W_-$, where $W_\pm = e^{-2b_\mrm{sL} \hat r}e^{\pm
  ik (\theta_\mrm{L} -  \theta_\mrm{R})}$ are
$\text{linear-dilaton}\times \mrm{S}^1$ 
vertex operators 
for a string with unit winding around the Euclidean time circle
$\theta$, times a Liouville-like 
factor in the linear-dilaton direction $\hat r$
(Fig. \ref{fig:sine-liouville}). The latter
serves 
to reflect strings away from the strong-coupling region $\hat r
\to - \infty$, where
$b_\mrm{sL}$ is a positive real number chosen such that these
vertex operators are marginal. 
Let us treat
the condensate as a large deformation of the
free $\text{linear-dilaton}\times \mrm{S}^1$
background; one obtains a series of
$W_+W_-$ insertions of the form $e^{-\int V_\mrm{sL}} \sim
\sum_{}\frac{1}{(N!)^2}\lp \int W_+ W_-\rp ^N$.\footnote{This
  expansion is formal and requires regularization in the
  strong-coupling region. Our aim, however, is to give an
  abstract picture of the EPR string background, not a
  practical scheme for computing amplitudes.}
Note that only paired operators $W_+W_-$ contribute
to the expansion due to the winding number conservation law of
the undeformed cylinder.
We emphasize that it is only after resumming the
series that one recovers the black hole
CFT.

Let us therefore consider the effect of a pair of $W_+$, $W_-$
insertions in the Lorentzian continuation of the
$\text{linear-dilaton} \times \mrm{S}^1$ string theory. Let
$W_+$ be inserted at the origin and $W_-$ at the
point-at-infinity on a Euclidean worldsheet $\mrm{CP}^1$.
One typically fixes the worldsheet diffeomorphism and
Weyl gauge redundancies in string theory by choosing a locally
flat metric $\diff 
s^2 = \diff z\, \diff \bar z$. Afterward, there remains a
residual gauge 
redundancy given by conformal transformations. 
In free theories, or more generally in 
backgrounds with a time-translation isometry,
one sometimes adopts the static gauge condition $\theta = \rho$
on-shell to fix the residual redundancy and obtain a target
space picture, wherein the target time $\theta$ 
is set equal to the worldsheet radial coordinate $\rho$, where $z =
e^{\rho + i \phi}$. Indeed,
in the usual radial quantization, 
one interprets $\rho$ as the worldsheet Euclidean time
coordinate, and the 
spatial slices are circles centered at the origin. The
functional integral over a disk with an operator
inserted at the origin prepares the corresponding state in the
CFT Hilbert space $\cH(\mrm{S}^1)$ on the boundary circle.

\begin{figure}[t]
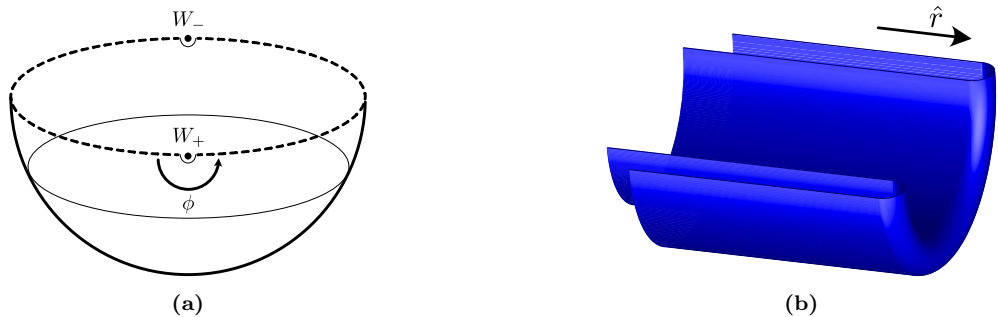

  \begin{subfigure}[t]{.43\textwidth}
      \centering
      \ig{0cm}{scale=1.1}{angular-tfd}
    \caption{}
    \label{fig:angular-tfd}
  \end{subfigure}
  \hspace{1cm}
  \begin{subfigure}[t]{.43\textwidth}
      \centering
      \ig{0cm}{scale=.8}{folded-string}
    \caption{}
    \label{fig:folded-string}
  \end{subfigure}
  \caption{\footnotesize
  \bd{The TFD State of a Pair of Folded Strings}. With insertions of
  Euclidean time winding operators $W_+$, $W_-$, the
  worldsheet should be treated in angular quantization
  to discuss the
  continuation to Lorentzian target time.
  One foliates the
  worldsheet in radial lines, such as the two dashed semi-circles
  shown on the left, with the angular direction
  interpreted as Euclidean time. The resulting Hilbert space
  $\cH_{+-}(\reals)$ lives on a line, labeled by asymptotic
  conditions associated to the operator insertions at either end.
  The functional integral over
  the halved worldsheet shown prepares the TFD state in
  two copies of $\cH_{+-}(\reals)$. The asymptotic conditions for
  $W_\pm$ send $\hat r \to -\infty$ with winding $\pm 1$. A
  schematic of the
  spacetime image of the halved worldsheet is shown on the
  right. In particular,  the two dashed spatial slices
  map to the folded strings bounding the blue figure, emanating
  from the strong-coupling region.  }
\end{figure}

However, the gauge choice $\theta = \rho$ is incompatible
with the operator insertion $W_+(0)$, which requires that
$\theta \to \phi$ has winding one as $\rho \to -\infty$.
Instead, one may adopt the 
angular gauge $\theta = \phi$, identifying the compact target
time with the angular coordinate on the worldsheet.
This suggests that in the neighborhood of a winding operator one
should treat the worldsheet in angular 
quantization
rather than the usual radial quantization. In this
formulation, the worldsheet is foliated in radial lines rather
than circles, and 
the angular direction $\phi$ running transverse to these spatial
slices is defined as the Euclidean time. The Hilbert space
$\cH_{+-}(\reals)$ 
in this quantization scheme lives on a line, labeled by
asymptotic conditions at either end associated to the insertions
of $W_+$ and $W_-$ there. When the worldsheet is cut in half
across this Euclidean time circle
as in Fig. \ref{fig:angular-tfd},
the functional integral on the
half-space prepares yet another TFD state, now valued in
$\cH_{+-}(\reals)\otimes \cH_{+-}(\reals)$.

Though less conventional than radial quantization, angular
quantization is likely familiar in the context of the Unruh effect
in Rindler spacetime. There, one considers a
field theory on $\reals^2$, again with the
angular direction interpreted as Euclidean time
(Fig. \ref{fig:angular-quantization}). After
the continuation one obtains the 
Rindler decomposition of Minkowski spacetime, with its
coordinate horizon separating the 
left and right Rindler wedges (Fig. \ref{fig:rindler}).
This coordinate horizon results in a quantization in a mixed
state on a space with an asymptotic endpoint.

With no insertions, the functional integral over the
Euclidean half-space prepares 
the Minkowski vacuum, which may equivalently be
understood as 
the TFD state in the product Hilbert space of the left and right
Rindler wedges in angular quantization.
The reduced density matrix in a single wedge is a
thermal state, resulting in the Unruh effect, and the TFD
state in the product Hilbert space is its purification.

\begin{figure}[t]
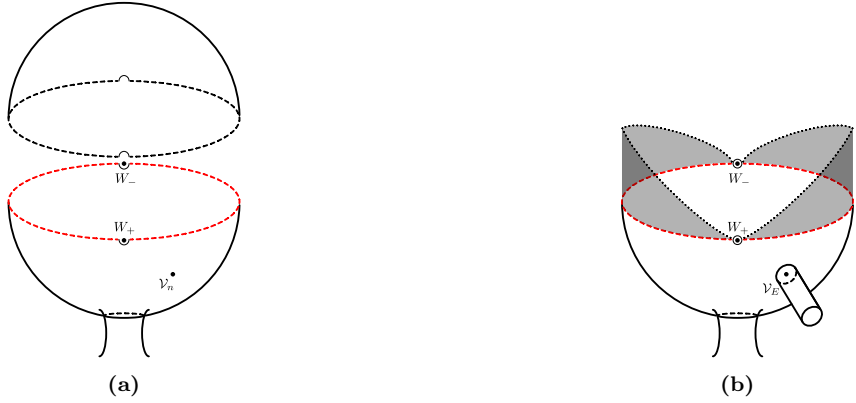

  \begin{subfigure}[t]{.43\textwidth}
    \centering
    \ig{0cm}{width=.45\linewidth}{local-tfd}
    \caption{}
    \label{fig:local-tfd}
  \end{subfigure}
  \hspace{1cm}
  \begin{subfigure}[t]{.43\textwidth}
    \centering
    \ig{0cm}{width=.45\linewidth}{contour-deformation}
    \caption{}
    \label{fig:contour-deformation}
  \end{subfigure}
  \caption{\footnotesize
    \bd{The Contour of Angular Quantization}. Near any pair of
    $W_+,W_-$ insertions on the worldsheet of a 
    $\text{linear-dilaton}\times \mrm{S}^1$ string diagram, one
    may slice their neighborhood as shown on the
    left. The upper cap prepares a pair of folded strings in the TFD
    state of angular quantization, which is glued to the rest of
    the worldsheet below. In computing a string amplitude
    of $\text{linear-dilaton} \times \mrm{time}$ including
    Euclidean time winding insertions, one must choose a
    deformed contour of integration in 
    evaluating the sum over moduli. On the right is sketched the
    contour used to evaluate the integral of a
    Lorentzian scattering operator. When the integrated
    insertion approaches another ordinary operator, the contour
    is deformed along the Lorentzian cylinder of radial
    quantization according to the prescription of
    \cite{Witten:2013pra}. When it approaches a winding
    operator, the  
    Euclidean cap that prepared the TFD state of folded strings
    is replaced by the pair of Rindler diamonds shown and glued
    along the red zero-time slice. The radial deformation prevents two scattering
    insertions from colliding, while the angular deformation
    prohibits a scattering operator from looping around a
    winding operator.}
    \label{fig:winding-worldsheet}
  \end{figure}

The angular quantization that we propose is a generalization of
the Rindler example, with operator insertions at one or both
ends of the spatial slice fixing asymptotic conditions for the
fields on the line. The asymptotic conditions for the marginal
winding operators $W_\pm$ of the $\text{linear-dilaton}\times
\mrm{S}^1$ background that make up the sine-Liouville potential
are\footnote{When inserted in the far past on the cylinder, the
  linear-dilaton asymptotic condition is modified to $\hat r \to
  2\alpha' (b_\mrm{sL} - Q/2) \rho$ due to the contribution from
  the background charge, where $Q = 1/\sqrt{\alpha'(k-2)}$ is
  the linear-dilaton slope. In the $k \to \infty$ limit,
  however, this contribution is sub-leading.}
\begin{subequations}
\begin{align}
  &\hat r \overright{\rho \to -\infty} 2\alpha' b_\mrm{sL} \rho\\
  &\theta \overright{\rho \to -\infty} \pm \phi.
\end{align}
\end{subequations}
Since $b_\mrm{sL}$ is positive, these asymptotic
conditions map the worldsheet neighborhood of each $W_\pm$
insertion to the strong-coupling region of the target cylinder,
with winding $\pm 1$, as shown in
Fig. \ref{fig:folded-string}. In 
particular, the spatial slices at $\phi = 0$ and $\pi$ that bound
the diagram in Fig. \ref{fig:angular-tfd}  map to folded
strings in spacetime at $\theta = 0$ and $\pi$ that emanate from the
strong-coupling region.\footnote{This picture is somewhat formal in that
  there  is no 
  saddle for the $\text{linear-dilaton}\times \mrm{S}^1$ theory
  plus the expanded condensate
  unless
  additional operators are inserted, due to the anomalous
  momentum conservation law. Moreover, to properly define
  string theory in this 
  background, one should introduce a regulator to suppress the
  strong-coupling region, which will completely break the
  translation symmetry.}
The Lorentzian interpretation of the pair of Euclidean time
winding insertions is thus that,
atop the disconnected union of $\text{linear-dilaton}\times
\mrm{time}$ backgrounds in the bulk TFD state, a pair of folded
strings  
is added, themselves entangled in the worldsheet TFD
state of angular quantization (Fig. \ref{fig:local-tfd}).
After
continuing\footnote{Or $\theta = \pi+ i t_\mrm{L}$ and $\phi 
  = \pi + i \Lt_\mrm{L}$ on the opposite side of the TFD.}
$\theta = i t$ and   
$\phi = i\Lt$, the folded string evolves forward in Lorentzian
time $t = \Lt$, as sketched in
Fig. \ref{fig:spacetime-tfd}.\footnote{For an ordinary
  insertion $e^{i n \theta}$, on the other hand, the 
  asymptotic condition $\theta \to -i
  \frac{\alpha'}{R^2} n
  \rho$ becomes $t \to E \Lt$ after continuing $\theta = it$,
  $\rho = i \Lt$, and $n = i \frac{R^2}{\alpha'} E$ in the
  usual sense of radial quantization.}

Ideas relating Euclidean time winding operators and
folded string solutions were previously explored in
\cite{Maldacena:2005hi}. In that context, the folded
strings emanated from the weak-coupling region. Here we extend
the connection between Euclidean time winding operators and
folded strings from the $c = 1$ analysis of
\cite{Maldacena:2005hi,Balthazar:2018qdv} to the true black hole
regime of $k > 3$.

There remains to address an important issue of mutual locality that
arises in computing Lorentzian string amplitudes in the EPR
microstates that describe strings scattering off the background
of entangled folded strings. 
In the ER description, 
scattering amplitudes in the HH state are obtained 
from cigar amplitudes with insertions $\cO_{jn}$ by continuing
from the discrete Matsubara frequencies $n$ to the continuous
Lorentzian energies $n \to i E$. One may perform the
continuation at the level of the CFT correlation function, and
then integrate over the moduli space to produce a Lorentzian string
amplitude.

In the asymptotic cylinder region of the cigar,
$\cO_{jn}$ 
approaches a superposition of $\text{linear-dilaton}\times
\mrm{S}^1$ primaries. To compute a Euclidean
amplitude in the sine-Liouville description, one would insert
these limiting free-field operators in the 
sine-Liouville functional integral and integrate over the moduli
space.

Upon attempting the same continuation $n \to i E$ in a
$\text{linear-dilaton}\times \mrm{S}^1$ correlator
after expanding the sine-Liouville condensate, however, one
will in general obtain a multi-valued function, the reason being
that the Euclidean time winding insertions and the Lorentzian
momentum operators are not mutually local. 
Namely, the OPE of winding $e^{\pm i k \til \theta(z,\bar z)}$ and
momentum $e^{i n \theta(z,\bar z)}$ operators of 
the compact boson is single-valued only for $n \in \bbZ$, which
is the expected momentum quantization law. When these insertions
approach one another, the correlation function behaves as
$(z/\bar z)^{\pm n/2} = e^{\pm i n \phi}$, yielding $e^{\mp E
  \phi}$ after continuation and violating the periodicity of
$\phi$. 
Thus, the result is single-valued only on an infinite-sheeted
cover of the 
worldsheet, over which the moduli integral would diverge in
one direction or the other.

On the other hand, the Lorentzian string amplitude for each
term in the expansion with a given number of winding insertions
may be obtained by 
integrating the Euclidean correlation function over the moduli
space, and only then continuing the Matsubara frequencies $n$ to
the Lorentzian energies $E$ in the final answer.
The question is how to evaluate
the integral of a Lorentzian correlation function over the
moduli, such that one obtains the same result.
To do so, one must define the moduli integral along a deformed
contour in a complexification of the moduli space on which the
integral converges and the
integrand is single-valued.

Even in the ER description---indeed, even in ordinary flat space
string theory---the naive integral over moduli 
diverges in general, and should instead be defined by
integrating  over a deformed complex contour \cite{Witten:2013pra}.
The prescription for ordinary operator insertions described in
\cite{Witten:2013pra} is to cut out the
neighborhood of each point in the original moduli space where
two insertions collide and glue in the Lorentzian cylinder of
radial quantization.

To evaluate a Lorentzian string amplitude
in the background of Euclidean time winding operators, we
similarly propose that their neighborhood 
should be replaced by a Rindler wedge of angular quantization
(Fig. \ref{fig:contour-deformation}).
In particular, continuing $\phi \to i \Lt$ in
the neighborhood where  Euclidean winding and Lorentzian
energy operators collide,  the problematic limiting behavior
$e^{\mp E \phi}$ is replaced by the single-valued and
oscillatory measure $e^{\mp 
  i E \Lt}$. 

In three dimensions,
the situation is again even simpler when working in
the basis of operators labeled by spacetime conformal boundary
points. Then there is no violation of mutual locality between
the sine-Liouville potential and vertex operators labeled by the
target Lorentzian section. Of course,
if one Fourier transforms to the momentum basis one must again
employ the deformed contour of angular quantization in
evaluating the sum over moduli, and the construction is similar
to the preceding discussion in two dimensions. Likewise, the
three-dimensional analog of the sine-Liouville potential again
produces a condensate of folded strings entangled between the
two disconnected copies of the  asymptotic $\mrm{AdS}_3$
spacetimes.

Having established the necessary ingredients, the Lorentzian
string dualities we propose now follow by continuation from the
$\mrm{SL}(2,\reals)_k/\mrm{U}(1)$ and $\bbZ \bs
\mrm{SL}(2,\bbC)_k/\mrm{SU}(2)$ CFT dualities. 
The weakly-coupled side (in the
$\alpha'$ sense) is string theory 
in the connected two-dimensional linear-dilaton
(or three-dimensional $\mrm{AdS}$) black hole
in the HH state. 
The strongly-coupled side is string theory in the disconnected
union of two copies of $\text{linear-dilaton}\times \text{time}$ (or
 $\text{linear-dilaton}\times \text{time} \times \mrm{S}^1$) in the 
bulk TFD state, with a condensate of pairs of entangled folded strings
emanating from the strong-coupling boundaries, themselves in the
worldsheet TFD state of angular quantization.\footnote{By
  continuing the duality of $\mrm{SL}(2,\bbC)_k/\mrm{SU}(2)$
  with respect to the same cycle, but
  prior to the $\bbZ$ quotient that makes the Euclidean black
  hole, one likewise obtains an even simpler duality between
  string theory in a connected Rindler decomposition of
  $\mrm{AdS}_3$ and the analogous EPR theory in
  $\text{linear-dilaton}\times \text{time} \times \reals$. The
  quotient replaces the ER side with the black hole and
  compactifies $\reals$ to $\mrm{S}^1$ on the EPR side.} 
By expanding the condensate, one obtains a superposition of
entangled disconnected microstates.
The dualities are therefore explicit examples of $\mrm{ER=EPR}$,
each relating a connected, two-sided spacetime 
to an entangled superposition of disconnected
geometries.

We point out that we will not construct string theories for the
individual disconnected microstates, but only for the thermal
gas of such states that constitutes the EPR description of the
black hole. It would however be very interesting to better
understand these individual  microstates in future work. Note that in contrast to microstates of extremal black holes such as those constructed in \cite{Bena:2016ypk}, the typical solutions in our context will be time dependent.  Moreover, one might expect the backreaction of the many folded
strings in a generic microstate to lead to the formation of a black hole, and it
would therefore be interesting if one could demonstrate that a
horizon forms around the folded strings in typical disconnected geometries. One open question is the extent to which perturbative string techniques can be used to characterize these solutions, given the presence of a strongly coupled region in the deep interior.

\begin{figure}[t]
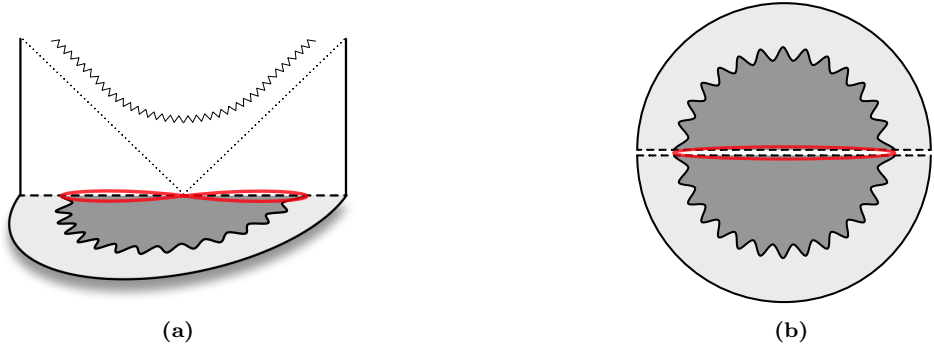

  \begin{subfigure}[t]{.43\textwidth}
    \centering
    \ig{0cm}{width=.7\linewidth}{horizon-string}
    \caption{}
    \label{fig:horizon-string}
  \end{subfigure}
  \hspace{1cm}
  \begin{subfigure}[t]{.43\textwidth}
    \centering
    \ig{0cm}{width=.6\linewidth}{tip-wrapping}
    \caption{}
    \label{fig:tip-wrapping}
  \end{subfigure}
  \caption{\footnotesize
    \bd{The Infinitesimal Duality}.
    When the $\mrm{SL}(2,\reals)_k/\mrm{U}(1)$ or $\bbZ \bs
    \mrm{SL}(2,\bbC)_k/\mrm{SU}(2)$ Euclidean black hole CFT
    is deformed by a marginal operator known as the
    sine-Liouville operator---so-called because its free-field
    limit is the sine-Liouville potential---a condensate of
    strings that wrap the Euclidean horizon is introduced
    (right). The red loop is a spatial slice of string,
    corresponding to the image of the dashed line in
    Fig. \ref{fig:angular-tfd} bounding the worldsheet TFD
    state with a pair of winding insertions. It may be thought
    of as a pair of folded strings 
    emanating from the horizon, contrasted with the disconnected
    folded strings emanating from the strong-coupling region in
    the EPR description 
    (Fig. \ref{fig:annulus} and \ref{fig:folded-string}). When
    continued to the Lorentzian ER string theory, one finds a
    pair of folded strings joined at the horizon bifurcation
    point in an entangled state. The deformation introduces a
    condensate of such strings, and the infinitesimal duality
    asserts that the same effect is described by shifting the
    constant mode of the dilaton. In both descriptions the mass
    of the black hole is in turn shifted.
  }
  \label{fig:infinitesimal-duality}
\end{figure}

Finally, we will also describe an ``infinitesimal'' version of
each of these dualities that relates two 
equivalent descriptions in the ER string theory of a marginal
deformation that shifts the mass of 
the black hole. The marginal operator is known as the sine-Liouville
operator, whose limiting form in the weak-coupling region is the
sine-Liouville potential. 
In one description of this operator, 
the deformation shifts the constant mode of the dilaton.
In the dual description,
the asymptotic condition for the sine-Liouville operator sends the string
to the horizon with unit winding \cite{cigar-note}, and so the
deformation introduces  
a condensate of horizon-wrapping strings.\footnote{Note that a 
  single abstract operator may admit multiple semi-classical
  descriptions via  asymptotic conditions.}
The identification of the
two descriptions is based on an isomorphism of the corresponding CFT states
discussed in \cite{Giveon:2016dxe}.
Whereas the condensate of Euclidean time winding strings as a
deformation of the free $\text{linear-dilaton}\times \mrm{S}^1
(\times \mrm{S}^1)$ produced in the Lorentzian continuation
pairs of disconnected folded 
strings emanating from the strong-coupling boundaries 
(Fig. \ref{fig:spacetime-tfd}), in the ER description of the
infinitesimal duality the perturbation introduces folded strings
joined at the horizon bifurcation point
(Fig. \ref{fig:infinitesimal-duality}). 
They may be thought of as the fundamental
strings that make up the black hole, whose
mass is in turn shifted under the deformation.

These horizon-bound strings, and their strong-coupling-bound EPR
counterparts, 
resemble the open strings conjectured in
\cite{Susskind:1993ws,Susskind_1994} to account for the black
hole entropy. It would be very interesting to understand how to
compute the entropy in these string backgrounds. Our work may
also be relevant to understanding entanglement entropy in string
theory, related to ideas discussed in
\cite{Witten:2018xfj,Dabholkar:1994ai}.

Recent potentially related work involving a condensate of folded
strings behind the 
black hole horizon and the Lorentzian continuation of the FZZ
duality has appeared in
\cite{Itzhaki_2018,Giveon_2019,itzhaki2019stringy,Giveon:2020xxh,Attali:2018goq,Itzhaki:2018rld,Yogendran:2015upa,Yogendran:2018ikf}.  
It would also be very interesting to establish the precise
connection between these results.

The remainder of the paper is organized as follows. In
Sec. \ref{sec:euclidean-dualities} we review the Euclidean FZZ duality of
$\mrm{SL}(2,\reals)_k/\mrm{U}(1)$, and we propose the new duality
of $\mrm{SL}(2,\bbC)_k/\mrm{SU}(2)$ and $\bbZ \bs
\mrm{SL}(2,\bbC)_k/\mrm{SU}(2)$. We also describe here the
infinitesimal interpretation of these Euclidean dualities.
In Sec. \ref{sec:string-states}
we discuss the formulation of string perturbation theory in
various states using Schwinger-Keldysh contours in a
complexified spacetime, which enables us to define the string
theories 
in the TFD and HH states that appear
in the Lorentzian dualities that we propose in Sec.
\ref{sec:erepr}. There, we formulate the Lorentzian string
dualities, and explain via angular quantization  the Lorentzian
interpretation of the Euclidean time winding operators, and the
corresponding deformation of the moduli space integration
contour. 

The fairly substantial length of the paper is in part due to an
effort to make it self-contained.
Expert readers may wish to skip our review of
the FZZ duality (Sec. \ref{sec:fzz-duality}) and the formulation
of string 
perturbation theory in $\mrm{AdS}_3$ in its vacuum state
(Secs. \ref{sec:ads} and 
\ref{sec:vacuum}). The discussion of Schwinger-Keldysh contours
for string 
theory in thermofield-double or Hartle-Hawking states
may also be more-or-less known to experts
(Sec. \ref{sec:thermal}-\ref{sec:black-hole}). The principal new 
ingredients that we discuss in order to establish our examples
of $\mrm{ER = EPR}$ include the uplifted duality of the
$\mrm{SL}(2,\bbC)_k/\mrm{SU}(2)$ and $\bbZ
\bs \mrm{SL}(2,\bbC)_k/\mrm{SU}(2)$ CFTs, the
Lorentzian interpretation of Euclidean time winding operators in
terms of  folded strings in the thermofield-double state of
angular quantization, the deformed contour of integration for
the sum over moduli in the background of Euclidean time winding
operators, and the infinitesimal interpretation of the various
dualities in terms of conformal perturbation theory.

\section{The Euclidean Dualities}
\label{sec:euclidean-dualities}

In this section, we review the FZZ
duality of the two-dimensional Euclidean
black hole CFT $\mrm{SL}(2,\reals)_k/\mrm{U}(1)$,
and we propose an uplifted three-dimensional duality of
the asymptotically Euclidean 
$\mrm{AdS_3}$ black hole $\bbZ\bs
\mrm{SL}(2,\bbC)_k/\mrm{SU}(2)$ and its Euclidean
$\mrm{AdS}_3$ parent $\mrm{SL}(2,\bbC)_k/\mrm{SU}(2)$. We also
discuss an   
infinitesimal interpretation of each duality that relates two
semi-classical descriptions of a conformal perturbation that
shifts the mass of the black hole.

\subsection{The FZZ Duality}
\label{sec:fzz-duality}

We begin\footnote{Readers familiar with the FZZ duality may wish to
  skip to Sec. \ref{sec:ads-duality}.} by reviewing the
two-dimensional FZZ duality 
\cite{FZZ,Kazakov:2000pm}, which relates two Lagrangian
descriptions of the $\mrm{SL(2,\reals)}_k/\mrm{U}(1)$
CFT.
This CFT is a coset of the $\mrm{SL}(2,\reals)_k$ WZW model.
The group  $\mrm{SL}(2,\reals)$ is equivalent to Lorentzian
$\mrm{AdS}_3$, and the WZW model describes a string propagating on
that manifold, with the WZW level $k$ corresponding to the
$\mrm{AdS}$ length, $l_\mrm{AdS}^2 = k l_\mrm{s}^2$
\cite{Maldacena:2000hw,Maldacena:2000kv,Maldacena:2001km,deBoer:1998gyt,Giveon:1998ns,Kutasov:1999xu,
  Gawedzki:1991yu,Teschner:1997ft,Teschner:1999ug}.
$\mrm{AdS}_3$ may be described as a solid cylinder with time
running along its length (Fig. \ref{fig:ebtz}), and the coset is defined by 
gauging this timelike isometry.
The result is the unitary $\mrm{SL}(2,\reals)_k/\mrm{U}(1)$ CFT,
of central charge
\begin{align}
  \label{eqn:central-charge}
  c = \frac{3k}{k-2} - 1,
\end{align}
the first term being the central charge of the
$\mrm{SL}(2,\reals)_k$ WZW model, less one for the quotient.
$k$ is a real number greater than two, which need not be an integer.

\subsubsection{The Cigar Background and the 2D Black Hole}
\label{sec:cigar}

The $\mrm{SL}(2,\reals)_k/\mrm{U}(1)$ coset is remarkable
because it describes, for large $k$, a string in a
Euclidean black hole of two-dimensional dilaton-gravity \cite{Witten:1991yr}.
In that limit, it admits a weakly-coupled Lagrangian description
in terms of the following sigma-model metric and
dilaton:
\begin{subequations}
  \label{eqn:cigar-bkgd}
  \begin{align}
    \label{eqn:metric}
    &\diff s^2 = \alpha' k \lp  \diff r^2 + \tanh^2(r)
      \diff\theta^2\rp \\
    \label{eqn:dilaton}
    &\Phi = -\log \cosh(r)+ \Phi_0.
  \end{align}
\end{subequations}
We will recall how this background follows by gauging the
$\mrm{SL}(2,\reals)_k$ 
WZW model in Sec. \ref{sec:ads}.
The action on a closed worldsheet $\Sigma$ of metric $h$ is
\begin{align}
  \label{eqn:cigar-S}
  S_\mrm{cigar}
  =& \frac{k}{4\pi} \int_\Sigma \diff^2 \sigma \,\sqrt{h}
     \lc
     (\del r)^2 + \tanh^2(r) (\del \theta)^2
     + \frac{1}{k} \cR[h] \lp - \log \cosh r +\Phi_0\rp\rc.
\end{align}
$r\in [0,\infty)$ and $\theta \sim \theta + 2\pi$ are
coordinates on a target space with the topology of a disk. In
the neighborhood of the origin, the geometry is simply
$\reals^2$ in polar coordinates,
\begin{subequations}
  \begin{align}
    \label{eqn:small-r}
    &\diff s^2 \to \alpha' k \lp \diff r^2 + r^2 \diff \theta^2
      \rp  + \cO(r^3)\\
    &\Phi \to \Phi_0 + \cO(r^2),
  \end{align}
\end{subequations}
while for large $r$ the geometry
approaches a cylinder of radius $\sqrt{\alpha' k}$:
\begin{subequations}
  \label{eqn:large-r}
\begin{align}
  &\diff s^2 \to \alpha' k (\diff r^2 + \diff
    \theta^2) + \cO\lp e^{-2r} \rp\\
  \label{eqn:large-r-dilaton}
  &\Phi \to -r  +\cO(1).
\end{align}
\end{subequations}
The target space therefore resembles a cigar (Fig. \ref{fig:cigar}),
with an asymptotic cylinder at large $r$  that caps off at the
tip $r = 0$, hence the background is
also known as the cigar sigma-model.
The leading $\cO(e^{-2r})$ correction to the metric is
$-4\alpha' k e^{-2r} \diff \theta^2$, and the corresponding operator
$e^{-2r}\partial \theta \bpartial \theta$ is the leading
correction to the cylinder background as one heads back toward
finite $r$. 
It is important to note that although the asymptotic cylinder
suggests the sigma-model has a topological winding
number, there is no conserved charge because a string that
appears to wind the cylinder can unwind at the tip.

The cigar is large and weakly curved for large $k$,
\begin{align}
  \label{eqn:curvature}
  \cR[g] = \frac{4}{\cosh^2(r)} \frac{1}{\alpha' k},
\end{align}
and the sigma-model is weakly coupled in the $\alpha'$ sense. 
The string coupling $g_\mrm{s} = e^\Phi$, meanwhile, is determined by
the dilaton Eqn. \ref{eqn:dilaton}. The unusual profile $\Phi(r)$
ensures conformal invariance of the curved background at leading
order,
\begin{align}
  \label{eqn:beta-function}
  \beta_{IJ} = \alpha'(R_{IJ}+2\del_I\del_J \Phi) + \cO(\alpha'^2)= \cO(\alpha'^2).
\end{align}

$\Phi(r)$ is a monotonically decreasing function,
with the constant $\Phi_0$ setting its maximal value 
at the tip of the cigar, $\Phi(0) = \Phi_0$. At large $r$
it falls off linearly. The string coupling therefore attains its
maximum $e^{\Phi_0}$ at the tip and decays to zero at large $r$,
which we refer to as the weak-coupling region.
The parameter $\Phi_0$ is a modulus of the
theory. It reflects the usual freedom to shift the dilaton by a
constant, the only effect being to shift the action by 
$\Phi_0 \chi$, with $\chi$ the Euler characteristic of $\Sigma.$

From the spacetime perspective, the cigar background $(g,\Phi)$ is a
solution of two-dimensional dilaton-gravity,
\begin{align}
  \label{eqn:dilaton-gravity}
  S_\mrm{spacetime} = - \frac{1}{2\kappa^2}\int \diff r\diff \theta \sqrt{g}\,
  e^{-2\Phi}
  \lp \cR[g] + 4 \lp \del \Phi \rp^2 +\frac{4}{\alpha' k} \rp,
\end{align}
whose equations of motion may be written
\begin{subequations}
  \label{eqn:2d-eom}
\begin{align}
  &R_{IJ}+2\del_I\del_J \Phi= 0\\
  &\lp \del \Phi \rp^2 - \frac{1}{2} \del^2
  \Phi-\frac{1}{\alpha' k} = 0. 
\end{align}
\end{subequations}
The first is again the leading order beta function equation from the
perspective of the worldsheet. The second computes the central
charge $c = 2 + 6\alpha' \lp (\del \Phi)^2-\frac{1}{2} \del^2
\Phi \rp$ at large $k$,
$\frac{3k}{k-2} - 1 = 2 + \frac{6}{k} + \cO(k^{-2})$.
At finite $k$, both the metric and
dilaton receive corrections \cite{Dijkgraaf:1991ba}.

With $\theta$ interpreted as the Euclidean time coordinate, this solution
describes a Euclidean black hole \cite{Witten:1991yr}. 
The horizon bifurcation point is found at the tip where the
$\theta$ circle 
shrinks. The geometry in that
neighborhood is $\reals^2$ (Eqn. \ref{eqn:small-r}), whose continuation with respect to angular
Euclidean time yields Rindler spacetime---the near-horizon geometry of the black hole.
Continuing $\theta = i t$ gives the Lorentzian metric in the
right wedge:
\begin{align}
  \label{eqn:lorentzian-metric}
  \diff s^2= \alpha'k \lp \diff r^2 - \tanh^2(r) \diff t^2 \rp.
\end{align}
The geometry approaches flat space at large $r$, $\diff s^2 \to \alpha' k (\diff r^2 -
\diff t^2)$, together with the asymptotically
linear dilaton. This coordinate patch ends at 
the horizon $r = 0$, where the coefficient of $\diff t^2$
vanishes. The complete, two-sided black hole may be described
in Kruskal coordinates
\begin{align}
  u = e^{-t} \sinh(r),\quad
  v = -e^{t} \sinh(r),
\end{align}
in terms of which the metric reads
\begin{align}
  \diff s^2 = \alpha' k \frac{\diff u\, \diff v}{uv-1},
\end{align}
and the dilaton is $\Phi = - \frac{1}{2} \log (1 - uv) + \Phi_0 $.
The extended black hole geometry is pictured in
Fig. \ref{fig:black-hole}. There is of course a second
asymptotic boundary in the left wedge. The singularities are
found at $uv = 1$. Note that the dilaton and string coupling
also diverge there. The mass of the black
hole is \cite{Witten:1991yr}
\begin{align}
  \label{eqn:cigar-mass}
  M = \frac{1}{\sqrt{\alpha' k}} e^{-2\Phi_0}.
\end{align}
Returning to the Euclidean theory, in the next sub-section we enumerate the spectrum
of the $\mrm{SL}(2,\reals)_k/\mrm{U}(1)$ CFT, after which we
discuss the FZZ dual description.

\subsubsection{$\mrm{SL}(2,\reals)_k/\mrm{U}(1)$  Spectrum}
\label{sec:operator-spectrum}

Above we have focused on the large $k$ limit where the cigar
sigma-model is weakly coupled, but thanks to
its relation to the $\mrm{SL}(2,\reals)_k$ WZW model
via the coset construction, which we review in Sec. \ref{sec:ads}, the exact spectrum of the
$\mrm{SL}(2,\reals)_k/\mrm{U}(1)$ CFT is known
\cite{Dijkgraaf:1991ba,Maldacena:2000hw,Hanany:2002ev}.
Its Virasoro primaries $\cO_{jnw}(z,\bar z)$ are labeled by
integers $n$ and $w$ and a complex number $j$, taking the
following values:
\begin{subequations}
\label{eqn:coset-spectrum}
  \begin{align}
  \label{eqn:complex-branch}
  &(\mrm{\bd{i}}) \quad j= \frac{1}{2}(1+i s),\quad s \in
    \reals_+\\
  \label{eqn:real-branch}
  &(\mrm{\bd{ii}})\quad  j_N = \frac{k|w|-|n|}{2}-N \in
    \lp \frac{1}{2}, \frac{k-1}{2}\rp,\quad N \in \bbN,
\end{align}
\end{subequations}
and carrying conformal weights
\begin{align}
  \label{eqn:coset-weights}
  h_{jnw}= -\frac{j(j-1)}{k-2} +
  \frac{(n-kw)^2}{4k},\quad\quad
  \bar h_{jnw} =-\frac{j(j-1)}{k-2} +
  \frac{(n+kw)^2}{4k}.
\end{align}
These two sets are referred to as the
complex and real branches of primaries based on the value of
$j$. The complex branch primaries correspond to
delta-function normalizable scattering states  on
the cigar with momentum  $s$, while the real branch 
primaries with $j=j_N$ correspond to bound states living at the
tip. The upper bound $j_N < \frac{k-1}{2}$ ensures
positivity of the weights $h_{j_N n w}$, while the 
lower bound $j_N >\frac{1}{2}$ ensures normalizability of the
wavefunctions. In all cases the spin
$h_{jnw}-\bar h_{jnw} = -nw$ is appropriately quantized.
One may also consider the continuation of $j$ to
general complex values, including 
real branch operators for
which $j$ is not valued in the discrete set $j_N$. The latter
map to non-normalizable states.

Since the cigar sigma-model approaches a free
$\text{linear-dilaton}\times \mrm{S}^1$ background at large $r$ (Eqn. \ref{eqn:large-r}), 
the abstract primaries $\cO_{jnw}$ may be expanded in free-field
primaries in that limit. To do so it is convenient to define 
canonically normalized coordinates, 
\begin{align}
  \label{eqn:canonical-coordinates}
  \hat r = \frac{1}{Q}r,\quad
  \hat \theta = \sqrt{\alpha' k} \theta,
\end{align}
where $Q$ is evidently given by $1/\sqrt{\alpha' k}$ in
the large $k$ limit. However, at finite $k$ it is corrected to
\cite{Dijkgraaf:1991ba} 
\begin{align}
  \label{eqn:Q}
  Q = \frac{1}{\sqrt{\alpha' (k-2)}},
\end{align}
and we use the exact value here so that we need not assume $k$
is large.\footnote{That is, the $k$-corrected background of
  \cite{Dijkgraaf:1991ba} still approaches the
  $\text{linear-dilaton}\times \mrm{S}^1$ background at large $r$, but
  with a modified scaling in the $r$ direction:
  $\diff s^2 \to \alpha' (k-2)\diff r^2 + \alpha' k \diff \theta^2$.} Note also that $\hat
\theta$ is periodic in $2\pi \sqrt{\alpha' k}$.
In the rescaled coordinates the asymptotic background is
\begin{subequations}
  \label{eqn:linear-dilaton}
\begin{align}
  &\diff s^2 \overright{\hat r \to \infty}
    \diff \hat r^2 + \diff \hat \theta^2 \\
  &\Phi \overright{\hat r \to \infty} -Q \hat r .
\end{align}
\end{subequations}

The Virasoro primaries of this free $\text{linear-dilaton}\times
\mrm{S}^1$ theory, considered in its own right with $\hat r$ permitted to
range over the entire real line, may be written
\begin{align}
  \label{eqn:primaries}
  \cV_{\alpha p_\mrm{L} p_\mrm{R}}(z, \bar z) = e^{-2\alpha \hat
  r(z, \bar z)} e^{i 
  p_\mrm{L} \hat \theta_\mrm{L}(z) + i p_\mrm{R} \hat
  \theta_\mrm{R}(\bar z)}.
\end{align}
$p_\mrm{L}$ and $p_\mrm{R}$ are valued in the lattice
\begin{align}
  \label{eqn:momentum-lattice}
  &p_\mrm{L} = \frac{n}{\sqrt{\alpha' k}} -
    \sqrt{\frac{k}{\alpha'}}w,
    \quad\quad
  p_\mrm{R} = \frac{n}{\sqrt{\alpha' k}} +
    \sqrt{\frac{k}{\alpha'}}w,\quad \quad
    n,w \in \bbZ,
\end{align}
where $n$ is the momentum number around the cylinder and $w$ is minus the
winding number.\footnote{We let $w$ denote the negative of the
  winding number so that it coincides with the spectral-flow
  number in $\mrm{SL}(2,\reals)_k$, as explained
  in Sec. \ref{sec:ads}.} The linear-dilaton momentum $\alpha$ may be
continued to a general complex number, and in a correlation
function of primaries it satisfies the anomalous conservation law
\begin{align}
  \label{eqn:anomaly}
  \sum_j \alpha_j = \frac{1}{2} Q \chi,
\end{align}
with $\chi$ the Euler characteristic of the worldsheet.
These primaries carry conformal weights
\begin{align}
  \label{eqn:free-weights}
  &h_{\alpha p_\mrm{L} p_\mrm{R}} = \alpha' \alpha(Q - \alpha) 
    +\alpha' \frac{p_\mrm{L}^2}{4},
    \quad\quad
  \bar h_{\alpha p_\mrm{L} p_\mrm{R}} = \alpha' \alpha(Q - \alpha) 
  +\alpha' \frac{p_\mrm{R}^2}{4},
\end{align}
with respect to the holomorphic stress-tensor
\begin{align}
  \label{eqn:ld-stress}
  T(z) = - \frac{1}{\alpha'} (\partial \hat r)^2 - Q \partial^2
  \hat r - \frac{1}{\alpha'} (\partial \hat \theta)^2
\end{align}
and its anti-holomorphic counterpart. The central charge of the
Virasoro algebra is $c_{\mrm{LD}\times \mrm{S}^1}=2 + 6 \alpha' Q^2,$ 
which reproduces the exact central charge of the coset Eqn. 
\ref{eqn:central-charge} using Eqn. \ref{eqn:Q}.

In the asymptotic region the CFT primary $\cO_{jnw}$ may then be
expanded in the free-field primaries $\cV_{\alpha p_\mrm{L}
  p_\mrm{R}}$ \cite{Dijkgraaf:1991ba}:
\begin{align}
  \label{eqn:asymptotic}
  \cO_{jnw}\overright{\hat r \to \infty}
  \lp e^{-2Q(1-j)\hat r} + R(j,n,w) e^{-2Qj\hat r} \rp
  e^{ip_\mrm{L}  \hat\theta_\mrm{L} + i p_\mrm{R} \hat \theta_\mrm{R}}.
\end{align}
$p_\mrm{L}$ and $p_\mrm{R}$ are as in
Eqn. \ref{eqn:momentum-lattice}; namely, the operator labels $n$ and
$-w$ correspond to the momentum and winding numbers
around the asymptotic cylinder. $j$ is meanwhile related to the 
asymptotic linear-dilaton momentum.
$R(j,n,w)$ is the reflection coefficient
\cite{Teschner:1999ug,Giveon:1999tq,Maldacena:2001km}:
\begin{align}
  \label{eqn:r}
  R(j,n,w)
  =& 
     \lp  \nu(k)\rp^{2j-1}
     \frac{\Gamma \lp 1 - \frac{2j-1}{k-2} \rp}
     {\Gamma \lp 1 + \frac{2j-1}{k-2} \rp}\\
   &\times
     4^{2j-1}
     \frac{\Gamma\lp 1 - 2j \rp}{\Gamma\lp 2j - 1\rp}
     \frac{\Gamma\lp j+\frac{|n|-kw}{2}\rp
     \Gamma\lp j + \frac{|n|+kw}{2}\rp}
     {\Gamma\lp 1-j+\frac{|n|-kw}{2}\rp
     \Gamma\lp 1-j+\frac{|n|+kw}{2}\rp}.\nt
\end{align}
Semi-classically, $R$ is the amplitude for a string sent from the
weak-coupling region to reflect and return to infinity. 
As an abstract CFT quantity, it
characterizes a redundancy in the space of CFT operators
$\cO_{jnw}$ when analytically continued to the complex
$j$-plane: operators labeled by $j$ and $1-j$ are identical, up
to rescaling by the reflection coefficient.
To avoid double-counting operators, one restricts
the domain to $\mrm{Re}(j) > \frac{1}{2}$ or $j \in \frac{1}{2}
+ i \reals_+$ as in Eqn. \ref{eqn:coset-spectrum}. 
$R$ satisfies $R(1-j,n,w) R(j,n,w) = 1$. $\nu(k)$ is a
convention-dependent function of $k$, but not of $j$. 

The zero-mode wavefunction of the state prepared by
$\cO_{jnw}$ may likewise be expanded at large $\hat r$.
The radial wavefunction differs from the operator by a factor of
$e^{-\Phi} \to e^{Q \hat r}$ due to the dilaton's coupling to
curvature:\footnote{\label{footnote:background-charge} With the cylinder metric on $\mrm{S}^2$, $\diff
  s^2 = 
\frac{\diff z\diff \bar z}{z \bar z}$, the scalar curvature
$\cR[h] = \frac{4\pi}{\sqrt{h}}\lp \delta(z, \bar z) + \delta(z-z_\infty, \bar
z-\bar z_\infty) \rp$ is singular at the two poles. Then the
effect of the dilaton coupling to curvature
$e^{-\frac{1}{4\pi} \int \diff^2\sigma\,
\sqrt{h} \cR[h] \Phi}$
is to insert so called background-charge operators
$e^{-\Phi}$ at the ends of the cylinder which contribute to the
wavefunction. See e.g. \cite{cigar-note} for a review.}
\begin{align}
  \label{eqn:wavefunction}
  \Psi_{jnw}(\hat r) \underset{\propto}{\overright{\hat r \to \infty}} 
  e^{2Q\lp j- \frac{1}{2}\rp \hat r} + R(j,n,w) e^{-2Q \lp j-\frac{1}{2}
  \rp \hat r}.
\end{align}
With $j \in \frac{1}{2} + i \reals_+$, both exponentials are
oscillatory, and one obtains a delta-function normalizable
state. These are the scattering states of
Eqn. \ref{eqn:complex-branch}, the two terms in the asymptotic
wavefunction describing the incoming and reflected waves at
infinity. The asymptotic operator is identified with the
linear-dilaton primary $e^{-2\alpha \hat r}$ with $\alpha =
Q(1-j)$ plus its reflection $e^{-2(Q-\alpha) \hat r}$, together with the
compact boson primary of momentum $n$ and winding $-w$. The corresponding
free-field weights (Eqn. \ref{eqn:free-weights}), which
are invariant under $\alpha \to Q - \alpha$,
reproduce the exact weights (Eqn. \ref{eqn:coset-weights}).

Away from the complex branch, the first exponential in
Eqns. \ref{eqn:asymptotic} and \ref{eqn:wavefunction} dominates
the second for $\mrm{Re}(j) > \frac{1}{2}$. Then, 
generically, the operator approaches $\cV_{\alpha p_\mrm{L}
  p_\mrm{R}}$ with $\alpha = Q(1-j)$ at weak coupling,
the wavefunction diverges exponentially, and the associated state is
non-normalizable.
However, at the discrete values $j = j_N$ of
Eqn. \ref{eqn:real-branch}, 
$R(j_N,n,w)$ is singular due to one of the two Gamma functions
$\Gamma \lp j + \frac{|n| \pm 
  kw}{2} \rp$. 
Then it is the reflected
term $e^{-2Qj_N \hat r}$ of Eqn. \ref{eqn:asymptotic}
that dominates in the asymptotic region,
and the radial wavefunction decays:
\begin{align}
  \label{eqn:bound-wavefunction}
  \frac{1}{R}\Psi_{j_N n w}(\hat r)
  \underset{\propto}{\overright{\hat r \to \infty}} 
  e^{-2Q \lp j_N-\frac{1}{2}\rp \hat  r}.
\end{align}
In this way, one finds a discrete spectrum of normalizable bound
states on Eqn. \ref{eqn:real-branch}, the lower bound
$j_N>\frac{1}{2}$ ensuring the wavefunction decays at infinity.

The simplest bound states have $n=0$, $w = \mp 1,$
and $j = \frac{k}{2} -1$.\footnote{Note that the set
  Eqn. \ref{eqn:real-branch} is empty for $w = 0$, consistent
  with the fact that there are no normalizable solutions of the
  Laplace equation on the cigar.}
The asymptotic form of these operators, which we will denote by $\cW_\pm$, is
\begin{align}
  \label{eqn:wpm-asymptotic}
\cW_\pm \equiv \frac{1}{R}\cO_{j = \frac{k}{2}-1,n=0,w=\mp 1}
  \overright{\hat r \to \infty}
  e^{-\sqrt{\frac{k-2}{\alpha'}}\hat r} e^{\pm i
  \sqrt{\frac{k}{\alpha'}}(\hat \theta_\mrm{L}-\hat
  \theta_\mrm{R})},
\end{align}
with radial wavefunction $  \Psi_\pm(\hat r) \to  e^{-Q(k-3)\hat r}$.
Their sum,
\begin{align}
  \label{eqn:sl-op}
  \cO_{\mrm{sL}} \equiv \cW_+ + \cW_-,
\end{align}
is called the sine-Liouville operator \cite{Giveon:2016dxe}, and
will be important in what follows.
It is of conformal weight $(1,1)$, and one may therefore
consider the effect of deforming the CFT by this operator, which
we will take up in Sec. \ref{sec:inf-fzz}.
The wavefunction is normalizable for $k > 3$, consistent 
with the lower bound $j > \frac{1}{2}$. 

Finally, we note for later use the asymptotic conditions
associated to these operators
\cite{Zamolodchikov:1995aa,Harlow:2011ny,cigar-note}.
Beginning with the free theory,
when $\cV_{\alpha p_\mrm{L}  p_\mrm{R}}$ is inserted in the functional
integral it adds a source term to free the equations of
motion. The solutions in the neighborhood of the source inserted
in the far past on the cylinder are the Green functions\footnote{The shift $\alpha \to
\alpha - \frac{Q}{2}$ in Eqn. \ref{eqn:r-asymp} is due to the
background charge at the end of the cylinder. See Ft. \ref{footnote:background-charge}.}
\begin{subequations}
  \label{eqn:ld-circle-green}
  \begin{align}
    \label{eqn:r-asymp}
  &\hat r(\rho,\phi) \overright{\rho\to -\infty}
    2\alpha' \lp \alpha - \frac{Q}{2}\rp \rho+ \cO(1)\\
  &\hat \theta(\rho,\phi)\overright{\rho\to-\infty}
    -i \sqrt{\frac{\alpha'}{k}} n \rho - \sqrt{\alpha' k}w \phi+ \cO(1),
\end{align}
\end{subequations}
where $z = e^{\rho + i \phi}$. These asymptotic conditions
specify how the fields behave as one approaches the operator
insertion point on the worldsheet.
For $\mrm{Re}(\alpha) <
\frac{Q}{2}$ the asymptotic condition maps the insertion point
to the weak-coupling region $\mrm{Re}(\hat r) \to \infty$, while 
for $\mrm{Re}(\alpha) > \frac{Q}{2}$ it is mapped to the
strong-coupling region $\mrm{Re}(\hat r) \to -\infty$ where $g_\mrm{s} =
e^{-Q\hat r}$ diverges.
The zero-mode wavefunction $\Psi(\hat r) = e^{2 \lp \frac{Q}{2}-\alpha\rp \hat r}$
correspondingly diverges at weak or strong coupling depending on
the sign of $\mrm{Re}(\alpha) - \frac{Q}{2}$.

In the cigar, meanwhile, for generic $\mrm{Re}(j) >
\frac{1}{2}$ the coset primary $\cO_{jnw}$ approaches the
free-field primary $\cV_{\alpha p_\mrm{L} p_\mrm{R}}$ in the
weak-coupling region, where $\alpha = Q(1-j)$ satisfies
$\mrm{Re}(\alpha) < \frac{Q}{2}$.  Then the
neighborhood of the insertion is mapped to the asymptotic cylinder region
where the curvature corrections are sub-leading, and the
free-field solution is self-consistent. 
The asymptotic condition describing such an
operator is thus as in the free theory,
Eqn. \ref{eqn:ld-circle-green} with $\alpha = Q(1-j).$ Complex
branch operators may similarly be understood by deforming $j \to
j + \vep$.

On the bound state spectrum $j=j_N$, however, we have seen that
the asymptotic linear-dilaton momentum of $\cO_{jnw}$ is instead
$\alpha = Q j$. Now $\alpha > \frac{Q}{2}$, and the free-field
asymptotic condition would send $\hat r \to - \infty$. In this case the
string is mapped out of the free-field region, where the
curvature corrections from the cigar are important, and
Eqn. \ref{eqn:ld-circle-green} is no longer a self-consistent
solution of the cigar equations of motion.

Since the neighborhood of a bound state insertion is not mapped
to the weak-coupling region,
the limiting form Eqn. \ref{eqn:asymptotic} of the operator is insufficient
to determine the behavior of the string.
The radial dependence of the operator on the full
cigar was determined in \cite{Dijkgraaf:1991ba}. For example,
for the bound states $\cW_\pm$,
\begin{align}
  \cW_\pm \propto \lp\sinh^2(r) - \frac{1}{k-2} \rp\sech^k(r).
\end{align}
In \cite{cigar-note} we found the associated asymptotic conditions:
\begin{subequations}
  \label{eqn:tip-asymptotic}
\begin{align}
  &r\overright{\rho\to-\infty} e^\rho\\
  &\theta \overright{\rho \to -\infty} \pm \phi.
\end{align}
\end{subequations}
Thus, in the neighborhood of the bound state insertions
$\cW_{\pm}$, the string asymptotically wraps the tip of the
cigar with winding $\pm 1$.

\subsubsection{The Dual sine-Liouville Background}
\label{sec:sl}

Having reviewed the spectrum of the
$\mrm{SL}(2,\reals)_k/\mrm{U}(1)$ CFT and its description in terms
of the cigar sigma-model at large $k$, we now come to the dual
description of Fateev, Zamolodchikov, and Zamolodchikov \cite{FZZ,Kazakov:2000pm}.

We have seen that the cigar asymptotes to the free
$\text{linear-dilaton}\times \mrm{S}^1$ background
at large $\hat r$ (Eqn. \ref{eqn:linear-dilaton}).
The full linear-dilaton background, with $\hat r$ permitted to
range over the whole real line, is not a unitary CFT or a
well-defined string background. As a CFT, the operator product
does not close on the space of delta-function normalizable
states (for which $\alpha \in \frac{Q}{2} + i \reals$), and as a
string background the string coupling $g_\mrm{s} = e^{-Q\hat r}$
diverges as $\hat r \to - \infty.$

In the cigar background, this strong-coupling region
is eliminated by ending the geometry at $r = 0$.
Another familiar way of
regulating the free linear dilaton is to turn on a potential 
$V_\mrm{L} \propto e^{-2b_\mrm{L}  \hat r}$ that serves as a barrier,
suppressing string configurations which extend too deeply into
the strong-coupling region.
The linear-dilaton momentum $b_\mrm{L}$ is chosen such that the
potential is of weight $(1,1)$: $\alpha' b_\mrm{L}(Q-b_\mrm{L}) =1$.
The result is the Liouville CFT, together with the free compact boson.

The cigar and $\mrm{Liouville}\times \mrm{S}^1$ backgrounds
are identical at large $\hat
r$ and reproduce the same central charge (Eqn. \ref{eqn:central-charge}). 
One might ask if they are dual descriptions of
the same CFT. The 
answer is no, as is clear, for example, from the fact that the
$\text{Liouville} \times \mrm{S}^1$ theory conserves 
the string winding number around the cylinder, which is broken
in the cigar.\footnote{However, for $2<k<3$, the
  $\text{Liouville}\times \mrm{S}^1$ and 
  $\mrm{SL}(2,\reals)_k/\mrm{U}(1)$ CFTs are conjectured to be
  connected by a conformal manifold \cite{Kazakov:2000pm}.}
There is, however, a close relative of the $\text{Liouville} \times
\mrm{S}^1$ background, called sine-Liouville, that is
dual to the cigar, and this is the content of the FZZ
duality.

In sine-Liouville, the $\text{linear-dilaton}\times \mrm{S}^1$ is
instead deformed by $V_\mrm{sL} \propto e^{-2b_\mrm{sL} \hat
r} \mrm{Re}\, e^{i \sqrt{\frac{k}{\alpha'}} (\hat \theta_\mrm{L} -
\hat \theta_\mrm{R})}$. The potential consists of a
Liouville-like radial factor  $e^{-2b_\mrm{sL}\hat r}$,
together with the unit-winding operator around the $\mrm{S}^1$
direction. The presence of the winding operator 
explicitly breaks the winding number conservation law of the free
theory, consistent with winding non-conservation in the cigar. 
The momentum of the linear-dilaton factor,
\begin{align}
  \label{eqn:bsl}
  b_\mrm{sL} = \frac{1}{2}\sqrt{\frac{k-2}{\alpha'}},
\end{align}
is again chosen such that the potential is weight $(1,1)$,
\begin{align}
  \label{eqn:sl-weight}
  \alpha' b_\mrm{sL} (Q - b_\mrm{sL}) + \frac{k}{4} = 1,
\end{align}
$k/4$ being the contribution of the unit-winding operator.
At large $\hat r$, the potential decays and one recovers the
same asymptotic $\text{linear-dilaton}\times \mrm{S}^1$ theory as for
the cigar.
One thinks of the sine-Liouville background as being built up of
a condensate of winding strings on top of the cylinder,
as pictured in Fig. \ref{fig:sine-liouville}.
The dual description of the coset is
best\footnote{Though not strictly-speaking weakly coupled, since
the circle radius is not large.} near $k
= 2$, where $b_\mrm{sL}$ is small. At large $k$ it is a
strongly-coupled description of the CFT.

The sine-Liouville action on a closed worldsheet $\Sigma$ is thus
\begin{align}
  \label{eqn:sl}
  S_\mrm{sL}
  =& \frac{1}{4\pi \alpha'} \int_\Sigma \diff^2\sigma \sqrt{h}
     \lc
     (\del \hat r)^2 + (\del \hat \theta)^2
     + 4\pi \lambda (W_+ + W_-) 
  -\alpha' Q \cR[h] \hat r
     \rc, 
\end{align}
where
\begin{align}
  \label{eqn:wpm}
  W_\pm = e^{-2b_\mrm{sL} \hat r} e^{\pm i
  \sqrt{\frac{k}{\alpha'}}\lp \hat \theta_\mrm{L} - \hat
  \theta_\mrm{R} \rp}
\end{align}
are the winding $\pm 1$ components of the sine-Liouville potential.
$\lambda$ is a positive constant.

The duality relates two target spaces of different
topologies. In the cigar description the target space is a
disk, whose contractible Euclidean time circle indicates the
Lorentzian continuation is a connected geometry with a
horizon (Eqn. \ref{eqn:lorentzian-metric}). In the
sine-Liouville description the target space is 
an annulus, and continuation with respect to the
non-contractible Euclidean time circle produces a disconnected
Lorentzian geometry. Both descriptions share the same asymptotic
$\text{linear-dilaton}\times \mrm{S}^1$ region. The leading departure
from the free theory
at finite $r$ in the cigar is 
the metric deformation $e^{-2r} \partial \theta \bpartial
\theta$, while
sine-Liouville results from the potential deformation
$e^{-2b_\mrm{sL}\hat r} \cos \big(\sqrt{k/\alpha'}(\hat
\theta_\mrm{L} - \hat \theta_\mrm{R}) \big)$. 

One is again free to add
a constant mode $\Phi_0$ to the dilaton, but it may be
eliminated by the field redefinition $\hat r \to \hat r +
\frac{\Phi_0}{Q}$, up to a rescaling of $\lambda$
by $e^{-(k-2)\Phi_0}$. In particular, only the combination
$e^{-2\Phi_0}\lambda^{\frac{2}{k-2}}$ is a meaningful parameter of the
string theory, which sets the mass of the black hole; we can
therefore choose $\Phi_0 = 0$ in the sine-Liouville description.

Relatedly, although the action Eqn. \ref{eqn:sl} takes the
form of the free $\text{linear-dilaton}\times \mrm{S}^1$ action
deformed by the sine-Liouville potential with coefficient
$\lambda$, the sine-Liouville background is not a small
perturbation of the free theory. The freedom to rescale
$\lambda$ by a field redefinition of $\hat r$ implies that
correlation functions are not analytic functions of $\lambda$,
and one cannot in general write a sine-Liouville correlator as a Taylor
expansion in $\lambda$ with coefficients computed from the free
theory. As in Liouville \cite{Knizhnik:1988ak}, 
the $\lambda$ dependence of a correlator of 
primaries in the sine-Liouville frame may be evaluated by
performing the zero-mode integral over the linear-dilaton
coordinate \cite{Kazakov:2000pm}. The functional integral with
asymptotic primary insertions $\prod_N
e^{-2Q(1-j_N)\hat r}\cS_N(\hat\theta)$, where $\cS_N$ are
$\mrm{S}^1$ primaries,  
may be written
\begin{align}
  \int \mrm{D}\hat r \mrm{D}\hat \theta~e^{-S_\mrm{sL}[\hat
  r,\hat \theta]}
  &\prod_N
  e^{-2Q(1-j_N) \hat r}\cS_N\\
  =&
     \int \mrm{D}\hat r' \mrm{D}\hat \theta~e^{-S_\mrm{LD}[\hat
     r'] - S_\mrm{S^1}[\hat \theta]}
     \prod_N  e^{-2Q(1-j_N) \hat r'}\cS_N\nt\\
  &\hspace{1cm}\times\int \diff \hat r_0~
    e^{Q\lp \chi -2\sum_N(1-j_N) \rp  \hat r_0
    -\lp \frac{\lambda}{\alpha'}
     V_\mrm{sL}[\hat r', \hat \theta] \rp
     e^{-2b_\mrm{sL} \hat r_0} }\nt,
\end{align}
where $\hat r(z, \bar z) = \hat r_0 + \hat r'(z, \bar z)$ and
$V_\mrm{sL}[\hat r',\hat \theta] = 2\int \diff^2 \sigma \sqrt{h}
e^{-2 b_\mrm{sL} \hat r'} \cos \lp \sqrt{k/\alpha'} (\hat
\theta_\mrm{L} - \hat \theta_\mrm{R}) \rp$. Using the identity
\begin{align}
  \int\limits_{-\infty}^\infty \diff\xi~e^{\kappa \xi - \beta
  e^{-\xi}} = \beta^{\kappa}\Gamma(-\kappa),
  \quad\quad
  \mrm{Re}(\kappa) < 0,~~ \mrm{Re}(\beta)> 0,
\end{align}
the zero-mode integral may be evaluated 
for $ \sum_N(\mrm{Re}(j_N)-1) < -\frac{1}{2}\chi$ as
\begin{align}
  \label{eqn:zero-mode-integral}
  \int\limits_{-\infty}^\infty \diff \hat r_0~
    e^{2b_\mrm{sL} \kappa  \hat r_0
    -\lp \frac{\lambda}{\alpha'}
     V_\mrm{sL}[\hat r', \hat \theta] \rp
  e^{-2b_\mrm{sL} \hat r_0} }
  =&\frac{1}{2b_\mrm{sL}}
     \lp \frac{\lambda}{\alpha'} V_\mrm{sL}[\hat r', \hat
     \theta] \rp^\kappa
     \Gamma \lp -\kappa \rp,
\end{align}
where
\begin{align}
  \label{eqn:kappa}
  \kappa = \frac{Q}{2b_\mrm{sL}}\lp \chi -2\sum_N(1-j_N)
  \rp.
\end{align}
When $\mrm{Re}(\kappa)>0$, the zero-mode integral
over the real line diverges, and must be deformed to a complex
contour that preserves convergence and analyticity.\footnote{Of
  course, since one is in 
general interested in complex values of $j$, it is inevitable
that the functional integral is complexified.}

Thus, the sine-Liouville correlation function is reduced to a
$\text{linear-dilaton}\times \mrm{S}^1$ correlation function,
with the linear-dilaton zero-mode measure omitted, plus $\kappa$
powers of the integrated sine-Liouville potential:
\begin{align}
  \label{eqn:sl-zero}
  &\vev{\prod_N  e^{-2Q(1-j_N) \hat r}\cS_N}_\mrm{sL}\\
  &\hspace{1cm}=
     \frac{1}{2b_\mrm{sL}}
     \lp \frac{\lambda}{\alpha'} \rp^\kappa
     \Gamma \lp -\kappa \rp
     \vev{V_\mrm{sL}[\hat r, \hat
     \theta]^\kappa \prod_N  e^{-2Q(1-j_N) \hat r}\cS_N}_\mrm{LD\times
     S^1,\fs{0}}.\nt
\end{align}
Such an expression with the integrated 
potential inserted $\kappa$ times, which is in general a complex
number, does not admit an
obvious interpretation as a correlation function of local
operators, and is defined by the functional integral.

On a genus $g$ Riemann surface, one obtains the scaling
$\lambda^{\frac{2}{k-2} \lp 1-g + \sum_N(j_N-1)
  \rp}$. The answer is analytic 
in $\lambda$ only when $\kappa$ is a natural number, in which
the case the $\kappa$ insertions of the potential
$V_\mrm{sL}{}^\kappa$ is most straightforward. The subsequent
divergence of the pre-factor 
$\Gamma(-\kappa)$ is attributable to the
volume of the non-compact target space. 

Observe that the sine-Liouville potential $W_+ + W_-$ coincides
with the weak-coupling limit of the previously defined coset
operator $\cO_\mrm{sL} = \cW_+ + \cW_-$
(Eqns. \ref{eqn:wpm-asymptotic}-\ref{eqn:sl-op}), hence the
common terminology.
A conformal
perturbation of $\mrm{SL}(2,\reals)_k/\mrm{U}(1)$ by the
marginal operator $\cO_\mrm{sL}$ is trivial at the level of the
CFT---the deformation merely shifts the coefficient $\lambda$
of the sine-Liouville potential, which may be undone by a field
redefinition of $\hat r$ at the cost of introducing a dilaton
zero-mode. The latter is a trivial improvement term from the
perspective of the CFT. As a string background, on the other
hand, the deformation by $\cO_\mrm{sL}$ is important because it
shifts the mass of the black hole. We explore this point further
in Sec. \ref{sec:inf-fzz}.

\subsection{An $\mrm{AdS}_3$ Duality}
\label{sec:ads-duality}

In this section we propose a new duality that may be considered
the uplift of FZZ  to the Euclidean $\mrm{AdS}_3$ CFT
$\mrm{SL}(2,\bbC)_k/\mrm{SU}(2)$, or its quotient $\bbZ \bs
\mrm{SL}(2,\bbC)_k/\mrm{SU}(2)$ that describes a string in
the asymptotic Euclidean $\mrm{AdS}_3$
black hole.\footnote{Or, equivalently, thermal
  $\mrm{AdS}_3$.} The familiar description of 
these CFTs is of a string
propagating in a solid cylinder or solid
torus geometry, with respect to which the disk topology of the cigar may be thought of
as a cross-sectional slice (Fig. \ref{fig:ebtz}).
As in
FZZ, one may attempt to define a dual description by
extending the semi-infinite radial direction $r \in [0,\infty)$
to an infinite line $\hat r \in
(-\infty,\infty)$, with the dual action given by the same limiting form
at infinity plus the marginal, unit-winding operator
around the resulting non-contractible cycle (Fig. \ref{fig:3d-sl}). To do so requires
adopting a first-order description near the conformal boundary,
such that the limiting theories are free. Gauging the 
translation symmetry along the length of the cylinder  in the two
descriptions yields the 
same cigar and sine-Liouville backgrounds of the two-dimensional
duality, providing strong evidence that the uplifted duality
holds in the parent CFTs.

\subsubsection{  $\mrm{SL}(2,\reals)_k$ and $\mrm{SL(2,\bbC)}_k/\mrm{SU}(2)$}
\label{sec:ads}

We begin by reviewing relevant details of $\mrm{AdS}_3$, the
$\mrm{SL}(2,\reals)_k$ and $\mrm{SL}(2,\bbC)_k/\mrm{SU}(2)$ WZW
models, and the coset construction of
$\mrm{SL}(2,\reals)_k/\mrm{U}(1)$.

$\mrm{AdS}_3$ may be described as a solid cylinder
(Fig. \ref{fig:ebtz}) with 
radial coordinate $r \in[0,\infty)$, angular coordinate
$\theta \sim \theta + 2\pi$, and Lorentzian time coordinate $t\in
(-\infty,\infty)$ running along its length. The metric in these
cylinder coordinates is
\begin{align}
  \label{eqn:ads-metric}
  \diff s^2_{\mrm{AdS}} =
  l_\mrm{AdS}^2\lp -\cosh^2(r) \diff t^2 + \diff r^2 +
  \sinh^2(r) \diff \theta^2\rp.
\end{align}
As a (pseudo) Riemannian manifold, $\mrm{AdS}_3$ is
equivalent to the Lie group
$\mrm{SL}(2,\reals)$, as may be seen from the parameterization\footnote{$e^{\pm
    2\pi i \mrm{T}_3} =- \bd{1}$, and therefore $g$ is invariant under
both $t \to t + 2\pi$ and $\theta \to \theta + 2\pi$. By $\mrm{AdS}_3
= \mrm{SL}(2,\reals)$, we mean the covering space where $t$ is
decompactified.} 
\begin{align}
  \label{eqn:global-matrix}
  g = e^{i(t+\theta)\mrm{T}_3} e^{2ir
  \mrm{T}_2}e^{i(t-\theta)\mrm{T}_3} \in 
  \mrm{SL}(2,\reals),
\end{align}
where $\mrm{T}_1=-\frac{i}{2} \sigma_1$,
$\mrm{T}_2=-\frac{i}{2} \sigma_3,$ and $\mrm{T}_3 = \frac{1}{2} \sigma_2$
are a basis of $\Lsl(2,\reals)$, satisfying $[\mrm{T}_i,\mrm{T}_j]
=i\epsilon_{ijk}\eta^{kl} \mrm{T}_l$, with metric $\eta_{ij} = 2 \mrm{tr}(\mrm{T}_i\mrm{T}_j)=
\mrm{Diag}(-1,-1,1)_{ij}.$
The $\mrm{AdS}_3$ metric Eqn. \ref{eqn:ads-metric}
is then reproduced by the
usual group metric $\frac{1}{2}l_\mrm{AdS}^2 \mrm{tr} \lp g^{-1} \diff g
g^{-1} \diff g \rp.$

One may therefore describe a string propagating in
$\mrm{AdS_3}$ by the $\mrm{SL}(2,\reals)_k$
WZW model 
\cite{Maldacena:2000hw,Maldacena:2000kv,Maldacena:2001km,deBoer:1998gyt,Giveon:1998ns,Kutasov:1999xu}. 
The WZW level $k$, which is a real number greater than 2, sets
the $\mrm{AdS}$ length scale, $l_\mrm{AdS}=\sqrt{\alpha' k}$.
$\mrm{AdS_3}$ has constant
negative curvature $\cR = -6/l_\mrm{AdS}^2$, and conformal symmetry on the worldsheet is
supported by a $B$-field that is contributed by the Wess-Zumino
term, $\mrm{tr} \big( \lp g^{-1} \diff g\rp^{\wg 3} \big) \propto \diff
B$, with $B \propto  \cosh(2r) \diff t \wg \diff \theta$. The
dilaton, meanwhile, is a constant, and the central charge is $c
= 3k/(k-2).$

However, one must be careful to define what one means by the
$\mrm{SL}(2,\reals)_k$ WZW model. Since the metric Eqn. \ref{eqn:ads-metric} is
Lorentzian, the WZW action includes wrong-sign kinetic terms,
and the functional integral is ill-defined. 
The ambiguity is meaningful and corresponds to the need to choose a state in
spacetime around which one wishes to build a string
perturbation theory.
The Lorentzian string
dualities that we describe in this paper will refer to string
perturbation theories in various states. The appropriate
worldsheet theories are
defined by analytic continuation from unitary CFTs, which
encode the choice of state. We elaborate on this point
in Sec. \ref{sec:string-states}. For now, by the
$\mrm{SL}(2,\reals)_k$ WZW model we mean
string perturbation theory in $\mrm{AdS}_3$ in the spacetime
vacuum state, which is defined by continuation from the
theory with target Euclidean $\mrm{AdS}_3$ $(\mrm{EAdS_3})$.

$\mrm{EAdS_3}$ is obtained from Eqn. \ref{eqn:ads-metric} by continuing $\xi  = i t$,
\begin{align}
  \label{eqn:eads-metric}
  \diff s^2_{\mrm{EAdS}} =
  l_\mrm{AdS}^2 \lp 
  \cosh^2(r) \diff \xi ^2 + \diff r^2 +
  \sinh^2(r) \diff \theta^2\rp,
\end{align}
with $\xi  \in (-\infty,\infty)$. The same continuation
in Eqn. \ref{eqn:global-matrix} yields not a group, but a parameterization of the 
vector space of $2\times 2$ Hermitian matrices with unit
determinant and positive eigenvalues. The group $\mrm{SL}(2,\bbC)$ acts
transitively on this space by conjugation, $g \to \Lambda g
\Lambda^\dg$, with stabilizer $\mrm{SU}(2)$. Thus, 
$\mrm{EAdS_3}$ is equivalent to the coset
manifold $\mrm{SL}(2,\bbC)/\mrm{SU}(2)$, and one may describe a string in $\mrm{EAdS}_3$ by the
$\mrm{SL}(2,\bbC)_k/\mrm{SU}(2)$ coset CFT
\cite{deBoer:1998gyt,Giveon:1998ns,Kutasov:1999xu,
  Gawedzki:1991yu,Teschner:1997ft,Teschner:1999ug}.

The $\mrm{SL}(2,\bbC)_k/\mrm{SU}(2)$ action is most conveniently
written in Poincar\`e coordinates,
\begin{align}
  \label{eqn:poincare-metric}
  \diff s^2_\mrm{EAdS} = l^2_\mrm{AdS}
  \lp \diff \sigma^2 + e^{2\sigma}   \diff \gamma \diff   \bar \gamma \rp,
\end{align}
where $\gamma$ is a complex coordinate, $\bar \gamma$ is its
complex conjugate, and 
$\sigma \in (-\infty,\infty)$. They are related to the cylinder coordinates by
\begin{subequations}
  \label{eqn:poincare-cylinder}
\begin{align}
  &\sigma = -\xi  + \log \cosh r\\
  &\gamma = \tanh(r) e^{\xi  +i \theta}\\
  &\bar \gamma = \tanh(r)e^{\xi  - i \theta}.
\end{align}
\end{subequations}
At the conformal boundary $r \to \infty$, note that
$\gamma \to e^{\xi +i\theta}$ is the usual map between the
boundary sphere with complex coordinate $\gamma$ and the
boundary cylinder with complex coordinate $W \equiv \xi  + i
\theta$. 

The action, including the contribution from the $B$-field, is
then \cite{Giveon:1998ns}
\begin{align}
  \label{eqn:poincare-action}
  S = \frac{k}{2\pi} \int \diff^2z \lp \partial \sigma
  \bar\partial \sigma + e^{2\sigma}\partial \bar \gamma
  \bar\partial \gamma\rp.
\end{align}
By the change of variables Eqn. \ref{eqn:poincare-cylinder} we
obtain the action in cylinder coordinates\footnote{We drop an
  additional total derivative term $i \tanh(r) (\partial r
  \bpartial \theta - \bpartial r \partial \theta)$ that
  corresponds to an exact $B$-field $\diff (\log \cosh(r) \diff \theta)$.}
\begin{align}
  \label{eqn:cylinder-action}
  S= \frac{k}{2\pi} \int \diff^2z\big\{
  \partial r \bpartial r + \partial \xi  \bpartial \xi 
  + (\partial \xi  - i \partial \theta)(\bpartial \xi  + i \bpartial
  \theta) \sinh^2(r)
  \big\}.
\end{align}

The cigar is obtained by gauging the translation isometry along
the length of the $\mrm{AdS}_3$ or $\mrm{EAdS}_3$ cylinder.
Promoting $\partial \xi  \to \partial \xi  + A$, $\bpartial 
\xi  \to \bpartial \xi  + \bar A$, one obtains the gauged action
\begin{align}
  S \to S + \frac{k}{2\pi} \int \diff^2z \lp 
  J^3 \bar A + \bar J^3 A + \cosh^2(r) A \bar A
  \rp,
\end{align}
where 
\begin{align}
\label{eqn:j3}
J^3(z) = \partial \xi + \sinh^2(r) (\partial \xi - i \partial
\theta)    
\end{align}
is the holomorphic component of the current for
translations in $\xi$, and $\bar J^3$ is the anti-holomorphic
component.
Solving the auxiliary equations of motion for $A, \bar A$ and evaluating the action on
the solution one obtains
\begin{align}
  \label{eqn:cigar-action}
  S_\mrm{cigar} = \frac{k}{2\pi} \int \diff^2z\lp \partial r
  \bpartial r + \tanh^2(r) \partial \theta \bpartial \theta \rp.
\end{align}
Thus, classically gauging the length of the cylinder produces a disk
topology with the cigar metric $\diff s^2 = \alpha' k (\diff r^2
+ \tanh^2(r) \diff \theta^2)$, as claimed in
Eqn. \ref{eqn:cigar-bkgd}. Quantum mechanically, integrating
out the gauge fields produces the dilaton, as necessitated
by conformal invariance (Eqn. \ref{eqn:beta-function}), as well
as additional sub-leading corrections to the
background Eqn. \ref{eqn:cigar-bkgd} \cite{Dijkgraaf:1991ba}.

One may then obtain the spectrum Eqn. \ref{eqn:coset-spectrum}
of the gauged WZW model
by starting from the $\mrm{SL}(2,\reals)_k$ spectrum and
applying the coset construction.
To do so we first review the Hilbert space of  
$\mrm{SL}(2,\reals)_k$ as obtained in \cite{Maldacena:2000hw}.

Observe from Eqn. \ref{eqn:global-matrix}
that the obvious isometries of the $\mrm{AdS}_3$ metric
in cylinder coordinates (Eqn. \ref{eqn:ads-metric}), namely
time translations $t \to t + \delta t$ and rotations
$\theta \to \theta + \delta \theta$, 
are implemented in $\mrm{SL}(2,\reals)$ by
$g \to e^{i \delta t \mrm{T}_3}g e^{i \delta t \mrm{T}_3}$ and 
$g \to e^{i \delta \theta \mrm{T}_3} g e^{-i\delta \theta \mrm{T}_3}.$ The
time translation and rotation generators are therefore the WZW charges
$J^3_0 +\bar J^3_0$ and $J^3_0 - \bar J^3_0$, whose eigenvalues
are the spacetime energy and angular momentum.
The complete isometry algebra of
$\mrm{AdS}_3=\mrm{SL}(2,\reals)$ is $\Lsl(2,\reals)_\mrm{L}
\oplus \Lsl(2,\reals)_\mrm{R} \simeq \Lso(2,2)$, whose
generators $\{J^3_0,J^\pm_0\}\cup \{\bar J^3_0, \bar J^\pm_0 \}$
in the raising/lowering basis satisfy
\begin{align}
  \label{eqn:sl2-global}
  [J^3_0,J^\pm_0]=\pm J^\pm_0,\quad
  [J^+_0,J^-_0]=-2J^3_0,  
\end{align}
and similarly for $\bar
J^a_0$, with $[J^a_0,\bar J^b_0] = 0$.

As is usual in WZW models, the target isometry algebra
$\Lsl(2,\reals)_\mrm{L}
\oplus \Lsl(2,\reals)_\mrm{R}$
is extended to a current algebra 
$\what\Lsl_k(2,\reals)_\mrm{L}\oplus \what\Lsl_k(2,\reals)_\mrm{R}$:
\begin{subequations}
\begin{alignat}{3}
  &[J_A^3,J_B^3]& &=&\, &-\frac{k}{2} A \delta_{A+B}\\
  &[J_A^3,J^\pm_B]& &=& &\pm J^\pm_{A+B}\\
  &[J_A^+,J_B^-]& &=& &-2J^3_{A+B}+kA\delta_{A+B},
\end{alignat}
\end{subequations}
$J_A^a$ being the modes of the currents
$\{J^a(z)\}_{a=3,\pm}$,
$J^a(z) = \sum_{A\in \bbZ} J^a_A /z^{A+1}$. Likewise one has
anti-holomorphic modes $\bar J_A^a$ of the currents $\bar
J^a(\bar z)$,
which satisfy the same algebra and commute with the holomorphic
modes. The isometry algebra corresponds to the global
sub-algebra generated by the zero-modes $J^a_0, \bar J^a_0$.

The $\mrm{SL}(2,\reals)_k$ Hilbert space is organized in
representations of this current algebra \cite{Maldacena:2000hw}:
\begin{align}
  \label{eqn:sl2r-spectrum}
\cH_{\mrm{SL}(2,\reals)_k} = \bigoplus_{j,\alpha,w}
  \begin{dcases}
  \what D^{+,w}_j\otimes \what D^{+,w}_j & j \in \lp \frac{1}{2}, 
  \frac{k-1}{2}\rp,\, w \in \bbZ\\
  \what C^w_{j,\alpha}\otimes \what C^w_{j,\alpha} & j \in
  \frac{1}{2} + i \reals_+,\, \alpha \in [0,1),\, w \in
  \bbZ. 
\end{dcases}
\end{align}
The notation here is as follows. $D_j^+$ 
denotes the spin $j\in \reals_+$ 
lowest-weight discrete-series representation of the global 
$\Lsl(2,\reals)$ sub-algebra, which consists of states $\lc 
\ket{j,m}\cn m \in j + \bbN \rc$, with $m$ being
the eigenvalue of $J^3_0$ and $-j(j-1)/(k-2)$ being the
conformal weight with respect to the $\mrm{SL}(2,\reals)_k$
stress tensor. The lowest-weight state $\ket{j,j}$ is
annihilated by $J^-_0$, and the remainder of the representation
is obtained by acting with $J^+_0$.
Similarly, one may consider 
highest-weight discrete-series representations $D_j^-$, consisting of
states $\lc \ket{j,m} \cn m \in -j - \bbN \rc$. $C_{j,\alpha}$ denotes
the spin $j\in \frac{1}{2} + i \reals$ continuous-series representation of
$\Lsl(2,\reals)$, consisting of states $\lc \ket{j,m} \cn m \in
\alpha + \bbZ \rc$, with $\alpha \in [0,1)$ an additional
parameter that fixes the $J^3_0$ eigenvalue modulo $\bbZ$. Note
that continuous-series representations are 
infinite towers with no relation between $j$ and $m$, while the
discrete series are semi-infinite towers in which $j$ and $m$
are related by integer shifts. 

On any such global $\Lsl(2,\reals)$ representation, one
may in the usual way build an $\what \Lsl_k(2,\reals)$ current-algebra
representation by demanding that the states $\ket{j,m}$ are
primary with respect to the positive current modes, $J^a_{A>0}
\ket{j,m} = 0$. Then the 
current-algebra representations, denoted $\what D_j^\pm$ and
$\what C_{j,\alpha}$, are built by the action of the
negative modes $J^a_{A<0}$ on the primaries. 

In addition to these standard varieties of current-algebra
representations built atop primaries, the $\mrm{SL}(2,\reals)_k$
spectrum includes less familiar
representations $\what D_j^{\pm,w}$ and $\what C_{j,\alpha}^w$,
labeled by an integer $w$, which arise  
from an automorphism $J^a_A \to J^a_A[w]$ of the current algebra
known as the spectral-flow automorphism: 
\begin{subequations}
\label{eqn:spectral-flow-auto}
  \begin{align}
  &J^3_A = J^3_A[w] + \frac{k}{2} w \delta_A\\
  &J^\pm_A = J^\pm_{A\mp w}[w].
\end{align}
\end{subequations}
$w$ may roughly be thought of as a winding number around the
$\theta$ circle of $\mrm{AdS}_3$ \cite{Maldacena:2000hw}. Of
course, since this cycle is contractible, $w$ is not conserved.

States in spectral-flowed representations transform as ordinary
representations under $J^a_A[w]$, with the action of $J^a_A$
then determined by Eqn. \ref{eqn:spectral-flow-auto}.
The Virasoro modes, being obtained from
the currents by the Sugawara construction, in turn transform
under spectral flow:
\begin{align}
  \label{eqn:spectral-virasoro}
  L_A = L_A[w]-w J^3_A[w]-\frac{k}{4} w^2 \delta_A. 
\end{align}
For example, the $J^3$ and conformal weights of 
a spectral-flowed primary state $\ket{j,m;w}$ are
\begin{subequations}
\begin{alignat}{4}
  \label{eqn:spectral-weight}
  &J^3_0& &\ket{j,m ; w}& &=&\, &\lp m + \frac{k}{2}w\rp \ket{j,m 
    ;w}\\
  \label{eqn:spectral-dimension}
  &L_0& &\ket{j,m ; w}& &=&\, &\lp -\frac{j(j-1)}{k-2} - w m -
  \frac{1}{4} k w^2\rp \ket{j,m ;w}.  
\end{alignat}
\end{subequations}
In general, we denote by $M = m + \frac{k}{2}w$ the eigenvalue of $J^3_0$ to
distinguish it from the eigenvalue $m$ of $J^3_0[w]$ prior to
spectral flow. Of course, in unflowed representations the two coincide.

A current-algebra primary $\ket{j,m}$ is also Virasoro
primary, and from
Eqns. \ref{eqn:spectral-flow-auto}-\ref{eqn:spectral-virasoro} it
follows that the spectral-flowed state $\ket{j,m;w}$ remains
Virasoro primary.
We also point out for later use that, for $w>0$, $\ket{j,m;w}$
transforms as a lowest-weight state $\ket{J,M=J}$ with respect to
the global sub-algebra, $J^-_0\ket{j,m;w} = J^-_w[w]\ket{j,m;w} 
= 0$. Thus, $\ket{j,m;w}$ sits at the bottom of a lowest-weight
representation $D_J^+$ of spin $J=m+\frac{k}{2}w$.
Likewise, for $w< 0$,
$\ket{j,m;w} \in D^-_{J'}$
is a highest-weight state $\ket{J',M'=-J'}$ of spin $J' = - \lp m + \frac{k}{2} w\rp$.
For further details of these representations, see
\cite{Maldacena:2000hw}.

The complete spectrum Eqn. \ref{eqn:sl2r-spectrum} of the $\mrm{SL}(2,\reals)_k$ WZW model
consists of the current-algebra representations
built on the lowest-weight discrete-series representations
$D_j^+\otimes D_j^+$ for $\frac{1}{2} < j < \frac{k-1}{2}$ and
on the continuous-series representations $C_{j,\alpha}\otimes
C_{j,\alpha}$ for $j \in \frac{1}{2} + i \reals_+$, as well as
their associated spectral-flowed representations 
for all $w \in \bbZ$.
Note that the spectrum does not explicitly list both the lowest-weight
and highest-weight discrete-series representations $\what
D_j^{\pm,w}\otimes \what D_{j}^{\pm,w}$.
These representations are not independent. Rather, one has
the following
isomorphism, which exchanges lowest and highest weights, shifts
the spectral flow by one unit, and reflects $j \to
\frac{k}{2}-j$ \cite{Maldacena:2000hw}:
\begin{align}
  \label{eqn:spectral-iso}
  \what D_j^{-,w}\simeq \what D_{\frac{k}{2}-j}^{+,w-1}. 
\end{align}
Note that the 
interval $j \in \lp \frac{1}{2}, 
\frac{k-1}{2} \rp$ is mapped to itself under the reflection $j
\to  \frac{k}{2} - j,$ so that if $j$ lies in the interval
then so does $\frac{k}{2} - j$ and vice-versa.

The $\mrm{SL}(2,\reals)_k/\mrm{U}(1)$ spectrum
Eqn. \ref{eqn:coset-spectrum} then follows from
Eqn. \ref{eqn:sl2r-spectrum} by the coset construction
\cite{Maldacena:2000hw}. The time translation symmetry being
generated by $J^3_0 + \bar J^3_0$,
the Virasoro primaries of $\mrm{SL}(2,\reals)_k/\mrm{U}(1)$
descend from those $\mrm{SL}(2,\reals)_k$ states which are
(i) Virasoro primary, (ii) $J^3$ and $\bar J^3$ primary, and
(iii) satisfy the
projection $J^3_0+\bar J^3_0 = 0$. These are of the form
\begin{align}
  \label{eqn:discrete-parent}
  \ket{j_N, m = \frac{-kw+n}{2}, \bar m = \frac{-kw-n}{2}; w}
  \in
  \begin{dcases}
    \what D_{j_N}^{+,w}\otimes \what D_{j_N}^{+,w}
    & w < 0\\
    \what D_{j_N}^{-,w}\otimes \what D_{j_N}^{-,w}
    & w > 0,
  \end{dcases}
\end{align}
and
\begin{align}
  \label{eqn:continuous-parent}
  \ket{j = \frac{1}{2}(1 + i s), m = \frac{-kw+n}{2}, \bar m =
  \frac{-kw-n}{2}; w} 
  \in \what C_{j,\alpha}^w\otimes \what C_{j,\alpha}^w.
\end{align}
Here, $m + \bar m = -kw$ is the projection condition $J^3_0 +
\bar J^3_0 = 0$, the respective eigenvalues being $m + kw/2$
and $\bar m + kw/2$. $m - \bar m = n$ is the eigenvalue of
$J^3_0 - \bar J^3_0$, i.e. the angular momentum around the
$\mrm{AdS}_3$ cylinder, which is therefore integer
quantized. 
In Eqn. \ref{eqn:discrete-parent},
$j_N = \frac{1}{2} (k|w|-|n|) - N$
is as in Eqn. \ref{eqn:real-branch}, with $N \in \bbN,$
determined in this context by the requirement
that the states with the necessary values of $m$, $\bar
m$ indeed fit into discrete-series representations.
Namely,
for $w<0$ one finds $j_N - m, j_N - \bar m \in -
\bbN$ corresponding to a lowest-weight representation, while for $w>0$,
$j_N+m, j_N + \bar m \in - \bbN$ giving a
highest-weight representation.
In
Eqn. \ref{eqn:continuous-parent}, there is no relation between $j$ and $m, \bar
m$, and one simply chooses the appropriate value of $\alpha$ to
identify the relevant state in the continuous series.

As mentioned above, spectral-flowed current-algebra primaries
$\ket{j,m,\bar m; w}\equiv\ket{j,m;w}\otimes 
\ket{j,\bar m;w}$  are Virasoro primary, and from
Eqn. \ref{eqn:spectral-flow-auto} they are primary with respect
to $J^3,\bar J^3$ as well. Thus, the states
Eqn. \ref{eqn:discrete-parent}-\ref{eqn:continuous-parent} indeed
satisfy the necessary conditions, and descend to Virasoro
primaries of $\mrm{SL}(2,\reals)_k/\mrm{U}(1)$.
The quantum number $j$, related to the linear-dilaton momentum
in the asymptotic cylinder of the cigar or sine-Liouville
backgrounds, descends from the $\Lsl(2,\reals)$ spin. The
quantized momentum $n$ around the asymptotic cylinder
corresponds to the angular momentum around $\mrm{AdS}_3$. And
the winding number $-w$ follows from the spectral-flow parameter
of the $\what\Lsl_k(2,\reals)$ representations.

The conformal
weights of these states with respect to the coset stress tensor are as in
Eqn. \ref{eqn:coset-weights}, obtained from the
$\mrm{SL}(2,\reals)_k$ weights Eqn. \ref{eqn:spectral-dimension}
less the $\mrm{U}(1)$ contribution.
Note that in writing Eqn. \ref{eqn:discrete-parent} we have applied the
isomorphism Eqn. \ref{eqn:spectral-iso} to convert
current-algebra descendent states from lowest-weight
representations in Eqn. \ref{eqn:sl2r-spectrum} to
spectral-flowed primaries from highest-weight representations.
This identification is important in the infinitesimal
interpretation of the FZZ duality discussed in
Sec. \ref{sec:inf-fzz}.

Finally, we point out that the components $\cW_\pm =
\cO_{j=\frac{k}{2}-1,n=0,w=\mp 1}$ of the sine-Liouville
operator $\cO_\mrm{sL} = \cW_+ + \cW_-$ (Eqn. \ref{eqn:sl-op})
descend from the states \cite{Giveon:2016dxe}
\begin{align}
  \label{eqn:wpm-components}
  \cW_\pm = \ket{\frac{k}{2}-1,\pm \frac{k}{2}, \pm \frac{k}{2};
  \mp 1} \in \what D_{\frac{k}{2}-1}^{\pm,\mp 1}\otimes \what
  D_{\frac{k}{2}-1}^{\pm,\mp 1}.
\end{align}
In fact, since the $J^3_0, \bar J^3_0$ eigenvalues of these
states independently vanish (as opposed to merely their sum),
the coset states and $\mrm{SL}(2,\reals)_k$ states are
identical, there being no $\mrm{U}(1)$ factors to strip
away. Thus the weights Eqns. \ref{eqn:coset-weights} and
\ref{eqn:spectral-dimension} are identical---namely,
$(1,1)$---and $\cO_\mrm{sL}$ defines a marginal operator of both
$\mrm{SL}(2,\reals)_k$ and
$\mrm{SL}(2,\reals)_k/\mrm{U}(1)$. This fact will be important
in the next sub-section when we uplift the FZZ duality to
$\mrm{AdS}_3$, and again in Sec. \ref{sec:inf-fzz} when we
discuss the infinitesimal versions of these dualities.

\subsubsection{Uplifting the Duality}
\label{sec:uplifting-duality}

Now we would like to lift the FZZ duality from the
$\mrm{SL}(2,\reals)_k/\mrm{U}(1)$ coset CFT to its Euclidean
$\mrm{AdS}_3$ 
parent $\mrm{SL}(2,\bbC)_k/\mrm{SU}(2)$, and the associated
asymptotic $\mrm{EAdS}_3$ black hole $\bbZ \bs
\mrm{SL}(2,\bbC)_k/\mrm{SU}(2)$.   

The cigar topology, being obtained from $\mrm{EAdS_3}$ by gauging the
translation isometry along the length of the cylinder, may be
thought of as a disk sliced from the cylinder
(Fig. \ref{fig:ebtz}). In the FZZ dual 
description, the disk topology of the cigar is
replaced by the annulus topology of the sine-Liouville cylinder,
plus the condensate of winding strings. The two descriptions share 
the same free $\text{linear-dilaton}\times \mrm{S}^1$ limit in the
weak-coupling region. Whereas the cigar geometry terminates at
the origin of the disk $r = 0$, the sine-Liouville cylinder continues
into the strong-coupling region $\hat r \to -\infty$, the potential wall instead
taking responsibility for reflecting strings away.

It is natural to wonder if there exists a similar duality of the
$\mrm{SL}(2,\bbC)_k/\mrm{SU}(2)$ CFT in which the solid
cylinder target space of Eqn. \ref{eqn:eads-metric} with its semi-infinite radial coordinate $r
\in [0,\infty)$ is replaced by a fully infinite radial direction $\hat r
\in (-\infty,\infty)$ and a condensate of winding strings that
wrap the resulting non-contractible cycle
(Fig. \ref{fig:3d-sl}). If such a description 
existed, such that gauging the translation symmetry reproduced the sine-Liouville
sigma-model, one would obtain a three-dimensional uplift of the FZZ
duality.

The interpretation of the cigar as a Euclidean black hole followed
upon identifying its angular coordinate with Euclidean
time. One is likewise free to continue $\mrm{EAdS_3}$ with
respect to its contractible angular direction $\theta=iT$,
as opposed to ordinary $\mrm{AdS}_3$ which was obtained by
continuing  $\xi = i t$.
The result is a Lorentzian wedge with a coordinate horizon, 
which we will refer to as $\mrm{AdS}_3$-Rindler:
\begin{align}
  \label{eqn:ads-rindler}
  \diff s^2_{\mrm{AdS}\text{-Rindler}} =
  l_\mrm{AdS}^2\lp -
  \sinh^2(r) \diff T^2 + \diff r^2+ \cosh^2(r) \diff \xi^2\rp.
\end{align}
$\mrm{AdS}_3$-Rindler is an analogue of ordinary Rindler
spacetime, the latter obtained from 
$\reals^2$ by continuing with 
respect to angular Euclidean time, yielding the right wedge
of a two-sided decomposition of Minkowski spacetime separated by
a coordinate horizon  
(Fig. \ref{fig:angular-continuation}). The coordinates of
Eqn. \ref{eqn:ads-rindler} likewise cover a patch of 
$\mrm{AdS}_3$ up to a coordinate horizon at $r = 0$,
where the coefficient $\sinh^2(r)$ of $\diff T^2$ vanishes.

One may furthermore obtain the asymptotic $\mrm{AdS}_3$ black
hole by compactifying $\xi\sim \xi +
4\pi^2/\beta_\mrm{BH}$, where $\beta_\mrm{BH}$ is the inverse
Hawking temperature of the black hole, as we review in Sec. \ref{sec:black-hole} 
\cite{Banados:1992wn,Banados:1992gq}. Thus, the continuation of
the would-be uplifted duality likewise relates a connected black
hole to a disconnected spacetime.
The Euclidean duality will apply both to
$\mrm{SL}(2,\bbC)_k/\mrm{SU}(2)$ and the quotient $\bbZ \bs \mrm{SL}(2,\bbC)_k/\mrm{SU}(2)$ that
compactifies $\xi$, and so for
the purposes of the present discussion it makes little
difference whether $\xi$ is compact or not.

One immediately encounters a problem with the above proposal,
however, on examining the $\mrm{SL}(2,\bbC)_k/\mrm{SU}(2)$
action (Eqn. \ref{eqn:cylinder-action}). Whereas the cigar and
sine-Liouville Lagrangians approach the same free theory at
infinity, which may then be deformed by either the cigar-capping
or sine-Liouville operators, the $\mrm{EAdS}_3$ Lagrangian is
singular at $r \to \infty$.

This singular asymptotic behavior may be remedied by applying
the first-order formalism.\footnote{See
  e.g. \cite{Losev_2006,Nekrasov:2005wg} for reviews.}
Classically, one has the identity of Lagrangians
\begin{align}
  f \partial \bar W \bpartial W = \chi \bpartial W + \bar\chi
  \partial \bar W - \frac{1}{f} \chi \bar \chi,
\end{align}
the auxiliary equations of motion for $\chi,\bar \chi$ setting
$\chi = f \partial \bar W$ and $\bar \chi = f \bpartial W$,
which recover the left-hand-side upon substitution. Quantum
mechanically, the change of variables introduces a dilaton
$\log \sqrt{f}$ from the transformation of the functional
integral measure. Thus, the action
Eqn. \ref{eqn:cylinder-action} may be replaced by
\begin{align}
  \label{eqn:first-order}
  S=\frac{k}{2\pi} \int \diff^2 z\lc
  \partial r \bpartial r + \partial \xi \bpartial \xi
  + \chi (\bpartial \xi + i \bpartial \theta)
  + \bar\chi (\partial \xi - i \partial \theta)
  - \frac{1}{\sinh^2(r)} \chi \bar \chi
  \rc,
\end{align}
together with a dilaton $\Phi = -\log \sinh r + \Phi_0$. This is the
cylinder version of the standard Wakimoto form of 
the action Eqn. \ref{eqn:poincare-action} in
Poincar\`e coordinates \cite{Giveon:1998ns}.\footnote{We use
  $\chi$ and $W$ to denote the cylinder-valued first-order coordinates rather
  than the typical $\beta$ and $\gamma$ to avoid confusion with
  the first-order formalism in Poincar\`e coordinates.}
$\chi$ is a $(1,0)$
form, set to $\chi = \sinh^2(r) \partial \bar W$ by the
equations of motion, where $W = \xi + i \theta$ is the complex
coordinate on the asymptotic cylinder, $W\sim W + 2\pi
i$.

Gauging the $\xi$ translation symmetry in
Eqn. \ref{eqn:cylinder-action} produced the cigar action
at leading order (Eqn. \ref{eqn:cigar-action}).
In the form Eqn. \ref{eqn:first-order}, the current for $\xi$ translations is
$J^3(z) = \partial \xi + \chi$, which coincides with
Eqn. \ref{eqn:j3} when evaluated on the auxiliary solution. The gauged
first-order action becomes
\begin{align}
  S \to S + \frac{k}{2\pi} \int \diff^2z\lp
  J^3 \bar A + \bar J^3 A + A \bar A
  \rp.
\end{align}
Classically integrating out $A,\bar A$ yields a first-order description of
the cigar,\footnote{In this description one has contributions to
the dilaton both from the first-order formalism and from
integrating out $A,\bar A$. Further integrating out $\chi, \bar
\chi$ eliminates the former contribution, leaving the 
dilaton profile Eqn. \ref{eqn:dilaton} of the
cigar.} 
\begin{align}
  \label{eqn:first-order-cigar}
  S_\mrm{cigar} = \frac{k}{2\pi} \int \diff^2z\lp
  \partial r \bpartial r
  +i \chi \bpartial \theta -i \bar\chi \partial \theta -
  \coth^2(r) \chi \bar \chi
  \rp,
\end{align}
and integrating out $\chi,\bar \chi$ again reproduces the cigar
action. 

The presentation Eqn. \ref{eqn:first-order} is advantageous because
it is non-singular at $r \to \infty$, where the potential $\chi
\bar \chi /\sinh^2(r)$ goes to zero. In that limit, the gauged action after
classically integrating out the gauge fields is then as in
Eqn. \ref{eqn:first-order-cigar} but with $\coth^2(r) \to 1$,
and further integrating out the auxiliaries yields the expected
asymptotic cylinder background,
\begin{align}
  S_\mrm{cigar}\overright{r\to\infty}
  \frac{k}{2\pi} \int \diff^2z \lp \partial r \bpartial r+
  \partial \theta \bpartial \theta \rp.
\end{align}
Thus, Eqn. \ref{eqn:first-order} is a preferable description of
$\mrm{SL}(2,\bbC)_k/\mrm{SU}(2)$ for the purposes of uplifting
the FZZ duality because the free theory it approaches at $r\to
\infty$ is transparently the uplift of the same limit of the
cigar background.

This large $r$ limit of Eqn. \ref{eqn:first-order} is a
linear-dilaton plus first-order cylinder system. One may define
canonically normalized fields
\begin{align}
  \label{eqn:hatted-vars}
  \hat r = \frac{1}{Q} r,\quad\quad
  \hat W = \sqrt{\alpha' k} W,\quad\quad
  \hat \chi = \sqrt{\alpha' k} \lp \chi + \frac{1}{2}\partial \xi\rp,
\end{align}
such that the asymptotic action appears
\begin{align}
  \label{eqn:first-order-asymp}
  S_{\mrm{LD}\times \cF(\bbC/\bbZ)} =
  \frac{1}{2\pi\alpha'}\int \diff^2 z \lp
  \partial \hat r\bpartial \hat r + \hat\chi \bpartial \hat W + \hat{\bar\chi} \partial
  \hat{\bar W}
  \rp,\quad  \Phi(\hat r) = - Q \hat r.
\end{align}
We denote by $\cF(X)$ the first-order system
valued in $X$; in this case $X = \bbC/\bbZ$ is the cylinder
$\hat W \sim \hat W + 2\pi i \sqrt{\alpha' k }$. 
As before, $Q = 1/\sqrt{\alpha'(k-2)}$ after accounting
for quantum corrections. 
The asymptotic potential $e^{-2r}\chi \bar \chi$ in
Eqn. \ref{eqn:first-order}  is the leading correction to the
free theory at finite $r$, analogous to $e^{-2r}\partial \theta
\bpartial \theta$ in the cigar. 

It is important to understand once more how the gauging of the
translation symmetry along the length of the $\mrm{EAdS}_3$
boundary cylinder is implemented in the description
Eqn. \ref{eqn:first-order-asymp}.
The holomorphic current $\partial \xi +\chi$ in the unhatted
variables becomes $J^3(z) = \hat \chi + \frac{1}{4} (\partial \hat
W+\partial \hat{\bar W})$ 
(after rescaling by $\sqrt{\alpha' k}$ for convenience).\footnote{
Note that the appropriate current is not simply $\hat \chi$, as might be
suggested from the coefficient of $\bpartial \hat W$ in
Eqn. \ref{eqn:first-order-asymp}. 
In writing Eqn. \ref{eqn:first-order-asymp},
we have dropped a term $-\frac{i}{2}
(\partial \hat\xi 
\bpartial \hat\theta 
- \bpartial \hat\xi \partial \hat\theta)$, which is total
derivative in $\mrm{EAdS}_3$, though one should  include it in
EBTZ where it contributes a 
non-trivial $B$-field. It does not
contribute to the equations 
of motion, but it does contribute $\frac{i}{2} \partial \hat
\theta = \frac{1}{4} (\partial \hat W - \partial \hat{\bar W})$
to the holomorphic current for translations in $\hat\xi=\mrm{Re}(\hat
W)$.
The correction is important in order to identify the appropriate
current whose gauging reproduces the
$\text{linear-dilaton}\times \mrm{S}^1$ background. 

The same
  subtlety arises in the ordinary complex boson, $\partial X
  \bpartial X + \partial Y \bpartial Y = \partial \bar Z
  \bpartial Z -i (\partial X \bpartial Y - \bpartial X \partial
  Y),$ where $Z=X+i Y$. The usual holomorphic current for translations
  in $X$ is $\partial X$. In the complex description, the
  corresponding current including the contribution from the
  exact $B$-field is $\partial \bar Z+i \partial Y =
  \frac{1}{2} (\partial Z + \partial \bar Z) = \partial X$, as
  desired.

Note that the current in the
hatted variables, $\frac{1}{\sqrt{\alpha' k}} \lp \hat \chi +
\frac{i}{2} \partial \hat \theta\rp =  \chi +
\frac{1}{2} \partial W $, differs from the current
$\chi + \partial \xi  = \chi + \frac{1}{2} (\partial W +
\partial \bar W)$ in the unhatted variables by $\frac{1}{2} \partial \bar W$, which
vanishes by the equations of motion. The discrepancy arises
because the change of variables Eqn. \ref{eqn:hatted-vars}
shifts $\chi$ by $\partial \xi$, preserving the translational
symmetry of $\xi$ under which both $\chi$ and $\hat \chi$ are
invariant, while producing currents that differ by irrelevant
terms proportional to the equations of motion.}
Gauging this symmetry of the free $\text{linear-dilaton}\times
\cF(\bbC/\bbZ)$ system is then
implemented by 
\begin{align}
  \label{eqn:first-order-gauged}
  S_{\mrm{LD}\times \cF(\bbC/\bbZ)} \to S_{\mrm{LD}\times 
  \cF(\bbC/\bbZ)} +\frac{1}{2\pi\alpha'} \int \diff^2z\lp
  J^3 \bar A + \bar J^3 A + A \bar A
  \rp,
\end{align}
which is invariant under $\hat W \to \hat W + \vep$, $\hat {\bar
  W} \to \hat{\bar W} + \vep$, $A \to A - \partial \vep$, $\bar
A \to \bar A - \bpartial \vep$, $\hat \chi \to \hat
\chi+\frac{1}{2} \partial \vep$, and $\hat{\bar\chi} \to \hat{\bar
  \chi} + \frac{1}{2} \bpartial \vep.$ Integrating out $A, \bar
A$ and $\chi, \bar \chi$, we recover the
$\text{linear-dilaton}\times \mrm{S}^1$ background
Eqn. \ref{eqn:linear-dilaton} as expected.

The $\hat \chi \hat W$ system is the familiar ($c=2$) bosonic ghost
``$\beta \gamma$'' system, except that $\hat W \sim \hat W + 2\pi i
\sqrt{\alpha' k}$ is cylinder-valued. The holomorphic stress
tensor is
\begin{align}
  \label{eqn:first-order-stress}
  T(z) = -\frac{1}{\alpha'} (\partial \hat r)^2 - Q \partial^2
  \hat r -\frac{1}{\alpha'} \hat \chi \partial \hat W,
\end{align}
whose central charge $3 + 6\alpha' Q^2$ reproduces the exact
central charge $3k/(k-2)$ of $\mrm{SL}(2,\bbC)_k/\mrm{SU}(2)$.
$\hat \chi(z)$ carries conformal weight
$(1,0)$ and $\hat W(z)$ is dimensionless, where the equations of motion imply $\bpartial
\hat W = \bpartial \hat \chi = 0$. Their OPE is
\begin{align}
  \label{eqn:w-chi-ope}
  \hat W(z) \hat \chi(0) \sim \frac{\alpha'}{z},
\end{align}
with $\hat W(z)  \hat W(0)$ and $\hat \chi(z) \hat \chi(0)$
non-singular. The anti-holomorphic sector is analogous.

Having obtained the free-field limit Eqn. \ref{eqn:first-order-asymp}
at the conformal boundary of $\mrm{EAdS}_3$, we now follow FZZ
and attempt to define a dual description by taking the free
theory over the infinite $\hat r$ line and deforming it by an
appropriate uplift of the sine-Liouville potential.
As explained following Eqn. \ref{eqn:wpm-components},
$\mrm{SL}(2,\reals)_k$ and $\mrm{SL}(2,\reals)_k/\mrm{U}(1)$
contain an identical marginal operator $\cO_\mrm{sL}$, whose
weak-coupling limit in the coset is the sine-Liouville potential
(Eqn. \ref{eqn:wpm-asymptotic}). One may identify the same operator
in the continued space of $\mrm{SL}(2,\bbC)_k/\mrm{SU}(2)$
operators. 
We wish to find the
asymptotic form of this operator in the
$\text{linear-dilaton}\times \cF(\bbC/\bbZ)$ description of the
$\mrm{EAdS}_3$ boundary.
We will therefore search for a marginal operator of the
free theory that carries unit winding around $\mrm{Im}(\hat W)$ and
no momentum, such that it reduces to the familiar sine-Liouville
potential upon gauging the symmetry.

The winding operators of the first-order cylinder system are
less familiar than those of the ordinary compact boson \cite{Frenkel_2006}. From
Eqn. \ref{eqn:w-chi-ope} one obtains 
\begin{align}
  \label{eqn:winding-ope}
  &\hat W(z) e^{\mp \sqrt{\frac{k}{\alpha'}} \int^{0} \diff z'
  \,\hat \chi + \diff \bar z' \hat {\bar \chi}}
  \sim
  \pm \sqrt{\alpha' k} \log(z)
  e^{\mp \sqrt{\frac{k}{\alpha'}} \int^{0} \diff z'
    \,\hat \chi + \diff \bar z' \hat {\bar \chi}},
\end{align}
and likewise for $\hat {\bar W}(\bar z)$.
Thus, 
$e^{\mp \sqrt{\frac{k}{\alpha'}} \int^{0} \diff z'
  \,\hat \chi + \diff \bar z' \hat {\bar \chi}}$ carries winding
$\pm 1$ with respect to 
$\hat \theta = \frac{1}{2i}(\hat W - \hat
{\bar W})$, while
$\hat \xi = \frac{1}{2} (\hat W + \hat{\bar
  W})$ is single-valued. The integral $\int^{0} \diff z'
\,\hat \chi + \diff \bar z' \hat {\bar \chi}$ is evaluated
along a contour ending at the origin, where the winding operator is
inserted. Demanding that observables be independent of the
choice of contour constrains the spectrum of the CFT, such that
the integrated expression defines 
a local operator \cite{Frenkel_2006}. 

This winding operator alone is not
annihilated by the current $J^3 = \hat \chi +
\frac{1}{4}(\partial \hat W+ \partial \hat{\bar W})$,
however:\footnote{The last term $\partial \hat{\bar W}$ in the
  current is trivial by the equations of motion, and does not
  contribute to any OPEs.}
\begin{align}
  \label{eqn:non-zero-ope}
  \lp\hat \chi(z) + \frac{1}{4} \partial \hat W(z)\rp
  e^{\mp \sqrt{\frac{k}{\alpha'}} \int^{0} \diff z'
  \,\hat \chi + \diff \bar z' \hat {\bar \chi}}
  \sim
  \pm \frac{\sqrt{\alpha' k}}{4z}
  e^{\mp \sqrt{\frac{k}{\alpha'}} \int^{0} \diff z'
  \,\hat \chi + \diff \bar z' \hat {\bar \chi}}.
\end{align}
To obtain a winding operator with no momentum along the
cylinder one must append the factor $e^{\pm \frac{1}{4}\sqrt{\frac{k}{\alpha'}}
  (\hat W+ \hat{\bar W})}$, which cancels against the OPE Eqn. \ref{eqn:non-zero-ope}.
The OPE of $\hat W$ with itself being non-singular, the inclusion
of this factor preserves the winding OPE Eqn. \ref{eqn:winding-ope}.

Thus, the combination
\begin{align}
  \label{eqn:winding-op}
e^{\pm \frac{1}{4}\sqrt{\frac{k}{\alpha'}}
  (\hat W+ \hat{\bar W})}e^{\mp \sqrt{\frac{k}{\alpha'}} \int \diff z'
  \,\hat \chi + \diff \bar z' \hat {\bar \chi}},
\end{align}
or more simply $ e^{\mp k \int
  \diff z' \chi + \diff \bar z' \bar \chi}$ in the unhatted
variables, carries zero
momentum and winding $\pm 1$, as desired. Its conformal
weight is $k/4$ on the left and right
due to the double-contraction with the stress tensor
Eqn. \ref{eqn:first-order-stress}, just as for the
ordinary winding operators $e^{\pm i\sqrt{\frac{k}{\alpha'}}
  (\hat\theta_\mrm{L}-\hat \theta_\mrm{R})}$ of the
two-dimensional cylinder (Eqn. \ref{eqn:sl-weight}).

In fact,  Eqn. \ref{eqn:winding-op} is precisely
the  dimensional uplift of these ordinary winding operators. Gauging the
translation symmetry as in Eqn. \ref{eqn:first-order-gauged} and
solving the auxiliary equations of motion, one finds $\hat\chi =
\frac{1}{2} \partial \hat \xi - i \partial \hat \theta.$
Evaluating Eqn. \ref{eqn:winding-op} on this solution yields 
\begin{align}
  e^{\pm \frac{1}{2} \sqrt{\frac{k}{\alpha'}}\hat \xi}
  e^{\mp \frac{1}{2}\sqrt{\frac{k}{\alpha'}} \int (\diff z'
     \partial \hat  \xi + \diff \bar z' \bpartial \hat \xi)}
  e^{\pm i \sqrt{\frac{k}{\alpha'}} \int (\diff z' \partial \hat
  \theta - \diff \bar z' \bpartial \hat \theta)}
  = e^{\pm i \sqrt{\frac{k}{\alpha'}} (\hat \theta_\mrm{L} -
     \hat \theta_\mrm{R})},
\end{align}
the equations of motion of the gauged action implying $\partial
\bpartial (\hat W - \hat{\bar W}) = 0$ and therefore $\hat
\theta(z,\bar z) = \hat \theta_\mrm{L}(z) + \hat
\theta_\mrm{R}(\bar z)$. 

Finally, to this unit winding operator we append the same
linear-dilaton primary 
$e^{-\sqrt{\frac{k-2}{\alpha'}} \hat r}$ as in
Eqn. \ref{eqn:wpm} to obtain a marginal operator.
The three-dimensional sine-Liouville potential is then 
$W_+ + W_-$, where
\begin{align}
  \label{eqn:3d-sl-comps}
  W_\pm(z,\bar z) = e^{-\sqrt{\frac{k-2}{\alpha'}} \hat r(z,\bar
  z)}e^{\pm \frac{1}{4}\sqrt{\frac{k}{\alpha'}}
  (\hat W(z)+ \hat{\bar W}(\bar z))}e^{\mp
  \sqrt{\frac{k}{\alpha'}} \int^{z,\bar z} \diff z'
  \,\hat \chi + \diff \bar z' \hat {\bar \chi}},
\end{align}
and which reduces to the original two-dimensional sine-Liouville
potential upon gauging.

The proposal is that the deformation of
the $\text{linear-dilaton}\times \cF(\bbC/\bbZ)$ background
(Eqn. \ref{eqn:first-order-asymp}) by $V_\mrm{sL}$ yields a dual
description of $\mrm{SL}(2,\bbC)_k/\mrm{SU}(2)$, which is a
better description of the CFT when $k -2$, and therefore the
linear-dilaton momentum, is small. If one further identifies
$\hat W \sim \hat W + 
4\pi^2 \sqrt{\alpha' k}/ \beta_\mrm{BH}$ (corresponding to compactifying
$\xi \sim \xi + 4\pi^2/\beta_\mrm{BH}$), the same deformation of
the $\text{linear-dilaton}\times \cF(\bbC/(\bbZ\times \bbZ))$ yields
a dual description of 
the Euclidean black hole CFT
$\bbZ \bs \mrm{SL}(2,\bbC)_k/\mrm{SU}(2)$.

For string theory in $\mrm{SL}(2,\bbC)_k/\mrm{SU}(2)$ and $\bbZ
\bs \mrm{SL}(2,\bbC)_k/\mrm{SU}(2)$, the natural vertex
operators are labeled by points on the spacetime conformal
boundary cylinder or torus, where they correspond to an
insertion of a local operator of the boundary CFT. In the
dual description, such an
operator behaves asymptotically as a linear-dilaton primary
times a delta-function primary of the first-order CFT  at that
point, plus  
a sub-leading contribution due to reflection off the
sine-Liouville potential in the interior. 

Eqn. \ref{eqn:3d-sl-comps} gives the
limiting form of the components 
$\cW_\pm$ of the $\mrm{SL}(2,\bbC)_k/\mrm{SU}(2)$ sine-Liouville operator near the
conformal boundary of $\mrm{EAdS}_3$. Recall from
Eqn. \ref{eqn:wpm-components} that $\cW_\pm$ is obtained from
the primary $\ket{k/2 - 1,\pm k/2, \pm k/2}$ by applying $w=\mp 1$
units of spectral flow. The asymptotic limit of the former is
$e^{-(k-2)r}e^{\pm k \xi}$. Spectral flow is meanwhile
implemented by the operator $e^{kw \int^z \diff z' J^3 + \diff
  \bar z' \bar J^3} = e^{kw \xi} e^{kw \int^z \diff z' \chi + \diff
  \bar z' \bar \chi}$, where $J^3(z) = \partial \xi +
\chi$ \cite{Argurio_2000}. Together, one obtains $\cW_\pm \to
e^{-(k-2)r}e^{\mp k \int \diff z' \chi + \diff
  \bar z' \bar \chi}$, which is identical to
Eqn. \ref{eqn:3d-sl-comps} with the coordinate transformation
Eqn. \ref{eqn:hatted-vars}.

In summary, the proposed duality of
$\mrm{SL}(2,\bbC)_k/\mrm{SU}(2)$, or its asymptotic
$\mrm{EAdS_3}$ black hole quotient 
$\bbZ \bs \mrm{SL}(2,\bbC)_k/\mrm{SU}(2)$, is as follows. On the
one hand one has the familiar description, weakly coupled for
large $k$, of a string propagating in a solid cylinder or torus
supported by a $B$-field:
\begin{align}
  \label{eqn:eads-action-cov}
  S_\mrm{EAdS_3} = 
  \frac{k}{4\pi} \int_\Sigma \diff^2\sigma \sqrt{h}\,
  \bigg\{
  &(\del r)^2 + \cosh^2(r) (\del \xi)^2 + \sinh^2(r) (\del
    \theta)^2\\
  &- 2\epsilon^{ab} \sinh^2(r) \del_a \xi \del_b \theta
  + \frac{\Phi_0}{k} \cR[h] \bigg\}\nt,
\end{align}
where $\sqrt{h}\epsilon^{z\bar z} = - \sqrt{h} \epsilon^{\bar z
  z} = -i$. 
In first-order variables (Eqn. \ref{eqn:first-order}), 
this background approaches a free
linear-dilaton times first-order cylinder system in the $r \to
\infty$ limit (Eqn. \ref{eqn:first-order-asymp}). The dual
sine-Liouville  
description is given by the same free theory, defined now with
an infinite linear-dilaton direction extending into
the strong-coupling region, deformed by the marginal potential
with unit winding and zero momentum,
\begin{align}
\label{eqn:3d-sl-cov}
  S_{\mrm{sL}} = 
  \frac{1}{4\pi\alpha'} \int_\Sigma \diff^2\sigma \sqrt{h}\,
  \bigg\{
  (\del \hat r)^2 &+ 2 h^{ab} (\hat \chi_a \del_b \hat W +
  \hat{\bar \chi}_a \del_b \hat{\bar W})\\
  &+ 4\pi\lambda (W_+ + W_-)
  -\alpha' Q \cR[h] \hat r
  \bigg\}.\nt
\end{align}

Gauging the translation symmetry in the two descriptions
recovers the cigar and 
two-dimensional sine-Liouville backgrounds, reproducing the
original FZZ duality and strongly suggesting the validity of this
three-dimensional proposal. We expect there is an analogous
supersymmetric duality given by the uplift of the supersymmetric
FZZ duality of Hori and Kapustin \cite{Hori:2001ax}.

This three-dimensional sine-Liouville background shares many
similarities with its two-dimensional counterpart. For example,
by integrating over the zero-mode one obtains a relation
analogous to Eqn. \ref{eqn:sl-zero}, where now $\cS_N$ denote
operators of the first-order system rather than the compact
boson. In particular, the correlation functions obey the
scaling relation $\lambda^\kappa$, where $\kappa$ is given in
Eqn. \ref{eqn:kappa}.
The same type of scaling is predicted by the dual Wakimoto
description of the 
$\mrm{SL}(2,\bbC)_k/\mrm{SU}(2)$ CFT
\cite{Giveon_2002,Giribet_2001,Kim_2015}. 
Once again the lack of analyticity is due
to the freedom to rescale $\lambda$ by field redefinitions up to
shifts of the dilaton zero-mode. In an action with a given coefficient
$\lambda$ and zero-mode $\Phi_0$, the latter may always be
eliminated with the former rescaled to $e^{-2b_\mrm{sL}
  \Phi_0/Q}\lambda$. 

The $\mrm{SL}(2,\bbC)_k/\mrm{SU}(2)$ CFT may alternatively be described
in Poincar\`e coordinates by the
$\text{linear-dilaton} \times \cF(\bbC)$ system deformed by the
Wakimoto potential $\beta \bar \beta e^{-2\sigma}$
\cite{Giveon:1998ns}. We 
relabel here the linear-dilaton field as $\sigma$ and the
first-order fields as $\beta$ and $\gamma$ to be consistent with
the notation in Eqn. \ref{eqn:poincare-action}, which is
reproduced upon integrating out $\beta,\bar \beta$.
If one were to quotient
$\gamma \sim \gamma + 2\pi i$ in this description,
i.e. replacing $\cF(\bbC)$ by $\cF(\bbC/\bbZ)$, one would obtain
a singular background with a cusp at $\sigma \to - \infty$ where
the transverse Poincar\`e metric $e^{2\sigma} \diff \gamma \diff
\bar \gamma$ shrinks. Continuing with respect to the compact
Euclidean time $\mrm{Im}(\gamma)$ would then yield a thermal
state in the Poincar\`e patch at inverse temperature $2\pi$.

If one deforms this cusp theory by the spectral flow of the
Wakimoto operator around the compactified cycle, it is natural
to conjecture that an RG flow is initiated back to the
$\mrm{SL}(2,\bbC)_k/\mrm{SU}(2)$ CFT. The spectral flow in this
case preserves the marginal conformal weight of the
operator, and introduces winding one around
$\mrm{Im}(\gamma)$. One may further deform the 
cusp theory by any number 
of such spectral-flowed operators with any values of 
spectral flow without changing the endpoint of the RG flow.
These flows are reminiscent of discussions of closed string tachyon
condensation in string theory \cite{Barbon:2002nw,Barbon:2001di,Horowitz_2006,Harvey:2001wm,Horowitz_2005,Silverstein:2006tm,Adams_2005,Adams_2001}. By
adding all the spectral-flowed operators with particular
coefficients, one expects to obtain another
description of the $\mrm{SL}(2,\bbC)_k/\mrm{SU}(2)$ CFT, as a
condensate of winding strings on the thermal Poincar\`e
orbifold. In principle, smoothness of the interior, as encoded in
conformal invariance of the worldsheet theory, would determine
the coefficients of all of the spectral-flowed operators.

\subsection{Infinitesimal Dualities}
\label{sec:inf-fzz}

Finally, we describe an infinitesimal interpretation of the FZZ
duality and its three-dimensional uplift described in the previous
sub-sections. This interpretation is based on the isomorphism
Eqn. \ref{eqn:spectral-iso} 
that relates the spectral-flowed components
$\cW_\pm$ of the sine-Liouville operator to unflowed
current-algebra descendents, as identified by the authors of
\cite{Giveon:2016dxe}. We apply their identification of these
vertex operators to discuss
an infinitesimal version of each duality in the sense of
conformal perturbation theory 
around the $\mrm{SL}(2,\reals)_k/\mrm{U}(1)$ or
$\bbZ \bs \mrm{SL}(2,\bbC)_k/\mrm{SU}(2)$ CFTs:  there are two
semi-classical descriptions of 
perturbations by the sine-Liouville operator 
which shift the mass of the black hole. In the unflowed
description of the perturbation, the value of the dilaton
zero-mode is shifted. In the flowed description, the
perturbation introduces a
condensate of strings that wrap the horizon. 

As reviewed in Secs. \ref{sec:operator-spectrum} and
\ref{sec:ads}, the $\mrm{SL}(2,\reals)_k$ and
$\mrm{SL}(2,\reals)_k/\mrm{U}(1)$ CFT spectra share a marginal operator
$\cO_\mrm{sL} = \cW_+ + \cW_-$,
normalizable\footnote{Delta-function normalizable in the
  $\mrm{SL}(2,\reals)_k$ case.} for $k>3$
and non-normalizable for $2 < k < 3$. One may likewise identify
$\cO_\mrm{sL}$ in the continued spectrum of
$\mrm{SL}(2,\bbC)_k/\mrm{SU}(2)$ operators, again
(delta-function) normalizable for $k>3$.
The sine-Liouville potential is
the limiting form of $\cO_\mrm{sL}$ near the weak-coupling
boundary in the various sigma-model descriptions of these CFTs.

The sine-Liouville Lagrangians for these CFTs make evident that
a conformal perturbation by $\cO_\mrm{sL}$ is trivial at the CFT
level. The deformation merely shifts the coefficient $\lambda$
of the sine-Liouville potential, which may be undone by a field
redefinition that shifts $\hat r$ by a constant. Then the
perturbation leaves the same sine-Liouville background, except
that the dilaton has been shifted by a constant, resulting from the
linear-dilaton in $\hat r$. The latter is a trivial improvement
term of the CFT, the only effect being to multiply the
functional integral by $e^{-\delta\Phi_0 \chi}$, where $\chi$
is the worldsheet Euler characteristic.\footnote{Not to be
  confused with the first-order auxiliary field.}

In the string theory, however,
$e^{-2\Phi_0}\lambda^{\frac{2}{k-2}}$ controls the mass of the
black hole, which is deformed under the
perturbation. That the perturbation is normalizable for $k>3$
implies that the black hole mass may fluctuate
\cite{Seiberg_1992,Karczmarek_2006}.\footnote{For $2<k<3$ the CFTs
likely no longer admit a black hole
interpretation \cite{Giveon:2005mi}.}  

Consider the $\mrm{SL}(2,\reals)_k/\mrm{U}(1)$ case.
The infinitesimal duality will equate two descriptions
of the perturbation by $\cO_\mrm{sL}$ in the cigar sigma-model
at large $k$. 
In one description, the deformation shifts the
value $\Phi_0$ of the dilaton at the tip of the cigar, and in
the other it introduces a condensate of strings that wrap the tip. 

\begin{figure}[t]
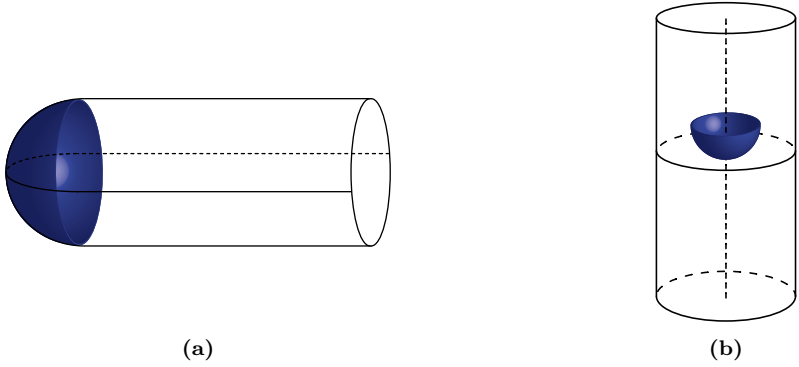

  \centering
  \begin{subfigure}[t]{.43\textwidth}
    \centering
    \ig{1cm}{scale=.5}{cigar-wrapping}
    \caption{}
    \label{fig:cigar-wrapping}
  \end{subfigure}
  \begin{subfigure}[t]{.43\textwidth}
    \centering
    \ig{0cm}{scale=.6}{cylinder-wrapping}
    \caption{}
    \label{fig:cylinder-wrapping}
  \end{subfigure}
  \caption{\footnotesize
  \bd{sine-Liouville Asymptotic Conditions}. The spacetime image of the
  string worldsheet in the neighborhood of a $\cW_+$ insertion
  is shown in the cigar description of
  $\mrm{SL}(2,\reals)_k/\mrm{U}(1)$ (left) and the cylinder
  description of $\mrm{SL}(2,\bbC)_k/\mrm{SU}(2)$ (right).}
\label{fig:wrapping-saddles}
\end{figure}

The latter description follows from the asymptotic conditions
Eqn. \ref{eqn:tip-asymptotic} that characterize insertions of
$\cW_\pm$ in the cigar. 
In conformal perturbation theory, a
deformation by $\vep \int \diff^2z\, \cO_\mrm{sL}$
is expanded in powers of $\vep$:
\begin{align}
  \label{eqn:sl-perturbation}
  e^{-\vep \int(\cW_+ + \cW_-)}
  =\sum_{N_1,N_2=0}^\infty \frac{(-\vep)^{N_1+N_2}}{N_1!N_2!}
  \lp \int \cW_+\rp^{N_1} \lp \int \cW_-\rp^{N_2}. 
\end{align}
At large $k$, $\cW_\pm\propto \sech^k(r)$ is a heavy operator
that changes the 
behavior of the saddles of the functional integral.
In the neighborhood of each $\cW_\pm$ insertion,
the asymptotic condition implies that the image of the string
worldsheet asymptotically wraps the tip 
of the cigar with winding $\pm 1$
(Fig. \ref{fig:cigar-wrapping}). In this way, the conformal
perturbation by $\cO_\mrm{sL}$ 
introduces a condensate of strings that wrap the horizon.

To understand the dual semi-classical description of the
perturbation, recall from Eqn. \ref{eqn:wpm-components} that
$\cW_\pm$ correspond to the states $\ket{j = k/2 -
  1,m=\bar m = \pm k/2; w = \mp 1}$ in the
$\what\Lsl_k(2,\reals)_\mrm{L} \oplus \what
\Lsl_k(2,\reals)_\mrm{R}$ representations $\what
D_{\frac{k}{2}-1}^{\pm,\mp 1} \otimes \what D_{\frac{k}{2} -
  1}^{\pm,\mp 1}$. By Eqn. \ref{eqn:spectral-iso}, $\what
D_{\frac{k}{2}-1}^{\pm, \mp 1}$ is isomorphic to 
$\what D_{1}^{\mp}$, under which the spectral-flowed primaries
$\cW_\pm$ are identified with unflowed descendents at
level one \cite{Giveon:2016dxe}:
\begin{align}
  \label{eqn:sl-iso}
  &\ket{\frac{k}{2}-1,\pm \frac{k}{2},\pm \frac{k}{2}; \mp 1}
  \simeq
    J_{-1}^\pm \bar J_{-1}^\pm \ket{1,\mp 1,\mp 1;0}.
\end{align}
For example, $J_{-1}^- \ket{j=1,m=1;w=0}$ carries zero $J^3_0$
charge, is of holomorphic conformal weight one, and is a
lowest-weight state of $\Lsl(2,\reals)_\mrm{L}$, consistent with
the properties of $\ket{k/2-1,-k/2;1}$. The latter belongs to the normalizable
spectrum for $k > 3$, such that $j = k/2 - 1$ satisfies the lower
bound
$j> 1/2$. The former likewise belongs to the
spectrum for $k > 3$, such that $j = 1$ satisfies the
upper bound $j < \frac{k-1}{2},$ the spectral-flow isomorphism  $j \to k/2 - j$
exchanging the upper and lower bounds.

The sine-Liouville operator $\cO_\mrm{sL}$ is the sum of the $w
= \mp 1$ operators on the left of Eqn. \ref{eqn:sl-iso}. Let us
denote by $\cO_\Phi$ 
the sum on the right.\footnote{One could also consider 
  the marginal operator 
defined by the differences of these operators, but this deformation breaks
worldsheet parity \cite{Hori:2001ax}.} The isomorphism
identifies the two.
On a flat worldsheet, the vertex operator for $\cO_\Phi $
in the cigar description is, at large $k$ \cite{Hori:2001ax},
\begin{align}
  \cO_\Phi (z, \bar z)\big|_\mrm{flat} = \frac{k}{2\pi}\sech^2(r) \lp
  \partial r \bpartial r+ 
  \tanh^2(r) \partial \theta \bpartial \theta \rp. 
\end{align}
Recalling the cigar metric Eqn. \ref{eqn:metric}, this operator
is evidently a metric deformation. Namely, adding to the cigar action $\vep
\int \diff^2z\, \cO_\Phi $ results in a sigma-model with
shifted metric
\begin{align}
  \diff s^2[\vep] = \alpha' k \lp \diff r^2 + \tanh^2(r) \diff
  \theta^2 \rp \lp 1 + \vep \sech^2(r) \rp.
\end{align}
The deformed metric is in fact related to the original metric by a reparameterization 
$\til r = r + \frac{1}{2} \vep \tanh(r)$:
\begin{align}
  \diff s^2[\vep]=
  \alpha' k \lp \diff \til r^2 + \tanh^2(\til r) \diff
  \theta^2 \rp + \cO(\vep^2).
\end{align}

On a curved worldsheet, the deformation by $\cO_\Phi$ must
simultaneously transform the dilaton $\Phi \to \Phi[\vep] =
-\log \cosh (\til r) + \til \Phi_0 + \cO(\vep^2)$ so as to
preserve the conformal symmetry of the background:
\begin{align}
  \Phi[\vep]
  =& -\log \cosh(r)+\Phi_0 - \frac{1}{2} \vep \tanh^2(r) + \vep \delta
     \Phi_0 + \cO(\vep^2). 
\end{align}
We have also allowed for the deformation to shift the zero-mode $\Phi_0\to
\til \Phi_0 = \Phi_0 + \vep \delta \Phi_0$, Eqn. \ref{eqn:beta-function}
placing no constraint on the constant mode of
the dilaton.
Since the operator is normalizable at large $k$, one expects the deformation
to vanish at large $r$, from which  
we infer that $\delta \Phi_0 = \frac{1}{2}$.
Then the vertex operator for $\cO_\Phi $ is given by 
\begin{align}
  \cO_\Phi (z, \bar z) = \sech^2(r) \lp \frac{k}{2\pi}\lp
  \partial r \bpartial r+ 
  \tanh^2(r) \partial \theta \bpartial \theta \rp 
  + \frac{1}{16\pi} \cR[h] \rp.
\end{align}

The deformation of the cigar sigma-model by 
$\vep \int \diff^2 z\, \cO_\Phi$ is thus again almost trivial. By the
field redefinition $r 
\to \til r$ one may undo the deformation, up to
a shift $\Phi_0 \to \Phi_0 + \frac{\vep}{2}$ of the value of the
dilaton at the tip. This is a
trivial improvement term from the perspective of the
CFT. 
In the string theory,
however, $\Phi_0$ sets the mass of the black hole
(Eqn. \ref{eqn:cigar-mass}), which is 
shifted under the
perturbation,  $\delta M = -\vep M$.

Thus, the identification Eqn. \ref{eqn:sl-iso} leads to 
an infinitesimal form of the FZZ
duality, relating superficially different marginal deformations
of the cigar that shift the mass of the black hole. One
description simply shifts the value of the dilaton at the tip
of the cigar, while the 
dual description introduces a condensate of winding 
strings that wrap the tip. 

For a black hole of a given mass $M$, one may formally apportion
the mass between the value of the dilaton at the tip and the
strength of the condensate---as for $\Phi_0$ and $\lambda$ in
the sine-Liouville 
description, neither of the two parameters is
independently meaningful.
One may always trade away the contribution
from the condensate in favor of a shifted value of the dilaton by
applying the duality. The
black hole may therefore be described by the pure cigar background
Eqn. \ref{eqn:cigar-bkgd} of mass Eqn. \ref{eqn:cigar-mass} 
with the condensate turned off, just as one may set the constant
mode of the dilaton
to zero in the sine-Liouville background
Eqn. \ref{eqn:sl}. Alternatively, one may trade $\Phi_0$ in
favor of the condensate. Then as in the sine-Liouville
description one may think of the black hole as
being made up of winding strings. The conformal perturbation adds
additional strings and so increases the black hole mass.

The same isomorphism Eqn. \ref{eqn:sl-iso} between
$\cO_\mrm{sL}$ and $\cO_\Phi$ holds in three dimensions, and therefore
the duality of $\mrm{SL}(2,\bbC)_k/\mrm{SU}(2)$ and its black
hole quotient $\bbZ\bs\mrm{SL}(2,\bbC)_k/\mrm{SU}(2)$ admits a similar
infinitesimal interpretation. The black hole mass is given by
\cite{Banados:1992wn,Banados:1992gq} 
\begin{align}
  \label{eqn:btz-mass}
  M = \frac{1}{8 G_\mrm{N}} \frac{R_\mrm{s}^2}{l_\mrm{AdS}^2},
\end{align}
where the horizon radius is related to the inverse Hawking
temperature by $\beta_\mrm{BH} = 2\pi l_\mrm{AdS}/R_\mrm{s}$. In
three-dimensional 
gravity, $G_\mrm{N} \propto l_\mrm{p} \propto g_\mrm{s}^2 l_\mrm{s}$. Thus, $M
\propto e^{-2\Phi_0}$ as in the two-dimensional black hole.  
A conformal perturbation by
$\cO_\Phi$ again shifts the constant mode of the dilaton, and in
turn the mass. To
understand the dual 
interpretation, one needs the asymptotic conditions
for $\cW_\pm$ in $\mrm{EAdS_3}$. The asymptotic conditions
Eqn. \ref{eqn:tip-asymptotic} in the cigar followed from the
$\rho\to -\infty$ limit of the cigar-wrapping saddle $r =
\sinh^{-1}(e^\rho),$ $\theta = \pm \phi$ \cite{cigar-note}. This
configuration may be uplifted to a solution of the
$\mrm{SL}(2,\bbC)_k/\mrm{SU}(2)$ equations of motion with
\begin{align}
  \label{eqn:uplifted-solution}
\xi =\pm \frac{1}{2} \log (1+e^{2\rho}).  
\end{align}
We conjecture that the
asymptotic condition for $\cW_\pm$ in $\mrm{EAdS_3}$ and the
Euclidean black hole is then as in
Eqn. \ref{eqn:tip-asymptotic}, together with
$\xi \to \pm \frac{1}{2} e^{2\rho}$, up to shifts by a constant.
Then the perturbation by $\cO_\mrm{sL}$
again introduces a condensate of horizon-crossing
strings, the horizon now corresponding to the one-dimensional
locus $r = 0$ (Fig. \ref{fig:cylinder-wrapping}).

This completes our discussion of the Euclidean dualities of the
$\mrm{SL}(2,\reals)_k/\mrm{U}(1)$,
$\mrm{SL}(2,\bbC)_k\ab/\mrm{SU}(2)$, and
$\bbZ\bs\mrm{SL}(2,\bbC)_k/\mrm{SU}(2)$ CFTs. Our goal in the
remainder of the paper is to understand the Lorentzian string
dualities that follow by analytic continuation in the sense of
the target time coordinate. In the next section we discuss how
such continuations define Lorentzian string theories in various
spacetime states, before finally applying this machinery to
describe the $\mrm{ER=EPR}$ string dualities in
Sec. \ref{sec:erepr}.

\section{State Dependence of String Perturbation Theory }
\label{sec:string-states}

String perturbation theory is often phrased as an
expansion around a Lorentzian spacetime solution. However, this
is not entirely precise. Even in the non-linear sigma-model
approximation at leading order in $\alpha'$, the target space
time direction has wrong-sign kinetic terms, and so the definition of
the worldsheet functional integral requires the specification of an
appropriate contour in a complexification of the target
space. Such contours 
are string theory analogues of the Schwinger-Keldysh contours
familiar from field theory. They
consist of  
Euclidean caps that specify the spacetime state, with the 
Lorentzian background glued between them. In this section, we
describe this formulation of string perturbation theory,
emphasizing in particular string perturbation theory around the
thermofield-double and Hartle-Hawking states that appear in the
$\mrm{ER=EPR}$ dualities we propose in Sec. \ref{sec:erepr}.

\subsection{Schwinger-Keldysh Contours for Lorentzian String
  Theory}
\label{sec:sk}

Suppose one wishes to construct a string perturbation theory
around a Lorentzian geometry $M$.
Roughly speaking, one often
thinks of the worldsheet CFT as a sigma-model\footnote{Combined
  with background fields to support conformal symmetry, and
  a unitary 
  internal CFT to cancel the conformal anomaly.} into $M$, but this
is imprecise for two related reasons. First, the Lorentzian
signature of $M$ implies that the kinetic term for the target
time coordinate has the wrong sign. Then the sigma-model action on a
Euclidean worldsheet is unbounded below, and the
functional integral over the real target fields
diverges.\footnote{In this section, we always choose a Euclidean
metric on the worldsheet. When we speak of continuing between
Lorentzian and Euclidean signatures, we mean with respect
to the time coordinate of the target space $M$. We will describe
the continuation of the worldsheet metric that is appropriate
for computing Lorentzian string amplitudes in the
next section. We argue, there, that in the vicinity of a Euclidean
time winding operator insertion, the Euclidean worldsheet (or,
more precisely, the string moduli space) should
be continued to Lorentzian signature in the sense of angular
quantization. This prescription generalizes that of
\cite{Witten:2013pra} for 
ordinary momentum operators,
whose neighborhood is continued in the sense of radial
quantization when integrating over the moduli.}\footnote{The
functional integral over real spatial 
directions can also diverge, such as along the asymptotic
linear-dilaton directions present in several of the backgrounds
we consider in this paper
\cite{Seiberg:1990eb,Harlow:2011ny,cigar-note}. Along such
directions the 
functional integral should again be defined over a complex cycle
in the target space. But that issue is separate from the
divergence of the functional integral over the time coordinate
in Lorentzian backgrounds.}
Second,
a string perturbation theory is not in general defined by
choosing a target geometry $M$ alone; one must also choose a
state of semi-classical quantum gravity in spacetime around
which one wishes to construct 
the perturbation theory.\footnote{When computing the string
  $S$-matrix in Minkowski spacetime, it is usually implicit that
  one has chosen the vacuum state. But one could also consider,
  for example, string perturbation theory in Minkowski spacetime in
  a thermal state \cite{Atick:1988si,Horowitz_1998}, or the
  HH state of a black hole in asymptotic Minkowski spacetime, and so on.}

For example, suppose $M =
\mrm{AdS_3}$, on which we will focus in this section because it
is one of 
the simplest and best understood examples.
In constructing string
perturbation theory in $\mrm{AdS}_3$, one can choose from
a variety of states, such as
the global $\mrm{AdS}$ vacuum, an excited state above the
vacuum, a thermal state, the thermofield-double (TFD) state in two
copies of $\mrm{AdS}_3$, and so on.

The divergence of the functional integral over Lorentzian target
fields and the ambiguity in the choice of state are closely
related. The functional integral should instead be defined by
continuation from a sigma-model with Euclidean target. Such a 
continuation is not unique, and the choice one makes encodes the
state around which the perturbation theory is defined. Thus, the
divergence of the functional integral and the necessity of
choosing a state are reconciled by defining the functional
integral along a contour in a complexification of the target
space. The contour consists of Euclidean caps at either end,
which ensure convergence of the functional integral, joined along a
Lorentzian excursion in the
middle
\cite{horava2020string,Horava:2020apz,Horava:2020val,Skenderis_2008,Skenderis:2008dg}. 

\begin{figure}[t]
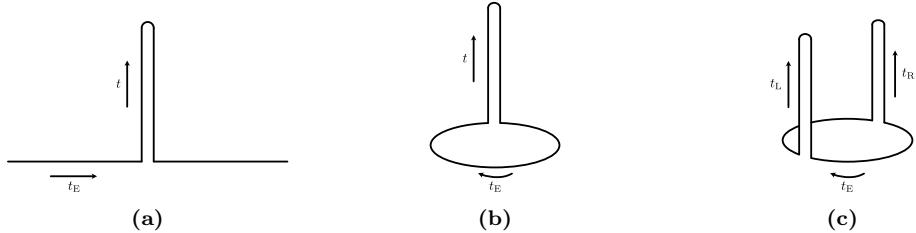

  \centering
  \begin{subfigure}[t]{.28\textwidth}
    \centering
    \ig{0cm}{scale=.7}{time-contour}
    \caption{}
    \label{fig:time-contour-vacuum}
  \end{subfigure}
  \begin{subfigure}[t]{.28\textwidth}
    \centering
    \ig{0cm}{scale=.7}{time-contour-thermal}
    \caption{}
    \label{fig:time-contour-thermal}
  \end{subfigure}
  \begin{subfigure}[t]{.28\textwidth}
    \centering
    \ig{0cm}{scale=.7}{time-contour-tfd}
    \caption{}
    \label{fig:time-contour-tfd}
  \end{subfigure}
  \caption{\footnotesize
  \bd{Schwinger-Keldysh Contours}. A Schwinger-Keldysh contour
  is a contour in the complex time plane that consists of
  Euclidean caps joined to Lorentzian excursions. The functional
  integral defined along such a contour computes the expectation
  value of operators inserted on the Lorentzian section
  between the initial and final states specified by the
  Euclidean caps. In field theory, the contour defines the
  domain on which the fields are defined. In string theory, it is
  the contour in
  target space over which the worldsheet
  functional integral is to be performed. The first contour
  pictured computes expectation values in a pure state such as
  the vacuum. The second computes expectation values in a
  thermal state, and the third in a TFD state. 
}
\label{fig:time-contours}
\end{figure}

Several such contours in the complex target time plane are pictured in
Fig. \ref{fig:time-contours}. The first computes expectation values
in a pure state such as the vacuum, the second in a thermal
state, and the third in 
a TFD state. They share a
common Lorentzian section (with two copies thereof in the
last example), but correspond to different
complexifications of the target time $t_\mrm{E} + i t$,
$t_\mrm{E}$ being non-compact in the first and compact in the
second and third. Moreover, the periodicity $t_\mrm{E} \sim
t_\mrm{E} + \beta$ in the latter cases is an additional choice,
specifying the inverse temperature of the thermal
state.\footnote{For a black hole geometry, the periodicity of
  the Euclidean continuation is fixed to obtain a smooth
  geometry. Then the black hole in the HH state has a fixed temperature.
  However, a thermal state in e.g. $\mrm{AdS}$ can
  have any temperature (at least, until the Hagedorn limit).}

The continuation is real when the Lorentzian geometry has a
$\mrm{Z}_2$ time-reflection isometry. Moreover
the Euclidean and Lorentzian
sections share a common zero-time slice---the fixed-point locus
of the symmetry---along which they may be glued together to
define the contour.

Complex time contours of this form are familiar from ordinary field
theory, where they are known as Schwinger-Keldysh contours. In
that context, they are contours in the base spacetime, rather than in the target space. Then the contour
specifies the domain on which the fields
are defined, and the functional integral computes expectation
values between the states specified by the caps. The functional
integral over the incoming
Euclidean cap prepares a state in the Hilbert space of the
quantum field theory on the spatial slice to which it is
glued. Then the state is 
evolved forward in Lorentzian time and any desired local operators are
inserted. Finally, the Lorentzian segment is reversed and 
glued to the outgoing Euclidean cap which prepares the outgoing
state. 

Such complexified functional integrals may be computed
by starting from a Euclidean correlation function and 
moving the operators to the Lorentzian section by
continuing the insertion times in the Euclidean answer.
One thinks of cutting open the Euclidean manifold on one or more
constant Euclidean time slices and gluing in the Lorentzian
excursion. Moreover, one need not continue all of the operator
insertions to the Lorentzian section. Operators left on the
Euclidean caps prepare different states in which the
expectation value is computed. 

In string theory, the Schwinger-Keldysh contour becomes a contour in the target space, over which the functional integral
that defines the worldsheet CFT is to be performed. Again, one
may proceed by starting from a worldsheet correlation function with the
Euclidean target and then continuing the vertex operator
insertions to the Lorentzian section. Of course,
it is not the location of the vertex
operator insertion on the worldsheet that one wishes to continue
in this case, but rather the quantum numbers 
that label how the operator transforms under the spacetime
symmetries. The Euclidean caps define the state in which the
string perturbation theory is constructed, and in AdS/CFT, the
string amplitudes compute expectation values of the boundary CFT
in the dual state associated to the caps.\footnote{More precisely,
  the string amplitudes compute contributions to the boundary
  CFT expectation values. Depending on whether or not the
  background is the dominant bulk saddle for a given boundary
  CFT observable, the corresponding string amplitudes will
  be the dominant contribution or a sub-dominant correction.} One may
also construct  
string theories with different initial and final states, e.g. by
fixing different operator insertions on the incoming and
outgoing Euclidean sections.

In fact, regarded as a deformation of the Euclidean integration
contour in target space, the Lorentzian excursion is
contractible. It does not alter the homology cycle of the
Euclidean functional integral, and therefore inserting the
excursion does not change the integral. One merely takes the
Euclidean answer and continues the labels to Lorentzian
time. 

Although we have phrased the above discussion in the sigma-model
approximation, which is convenient for
visualizing the target geometry as a complex contour for the
functional integral, it is not necessary to resort to a
Lagrangian description of the worldsheet CFT. One may begin from
an abstract definition of the CFT by its three-point function and
OPE, and continue the operator labels to define the Lorentzian
string theory. As we review now, string theory in $\mrm{AdS}_3$
in various states is constructed by continuation from the unitary
$\mrm{SL}(2,\bbC)_k/\mrm{SU}(2)$ coset CFT, and various orbifolds
thereof, without requiring any reference to a Lagrangian.

\subsection{States in $\mrm{AdS}_3$}

Let us make this discussion concrete for strings in
$\mrm{AdS}_3$ and related examples.
In Eqns. \ref{eqn:ads-metric} and \ref{eqn:eads-metric} we
defined $\mrm{EAdS}_3$ by the continuation $\xi
= it$ from $\mrm{AdS}_3$, with 
$\xi \in (-\infty,\infty)$.
The Lorentzian metric is time-independent, and so one may cut it
along any spatial slice and glue in a Euclidean cap to prepare a
state. Cutting and gluing the
two cylinders at $\xi = t = 0$ prepares the global
$\mrm{AdS}_3$ vacuum state.

\begin{figure}[t]
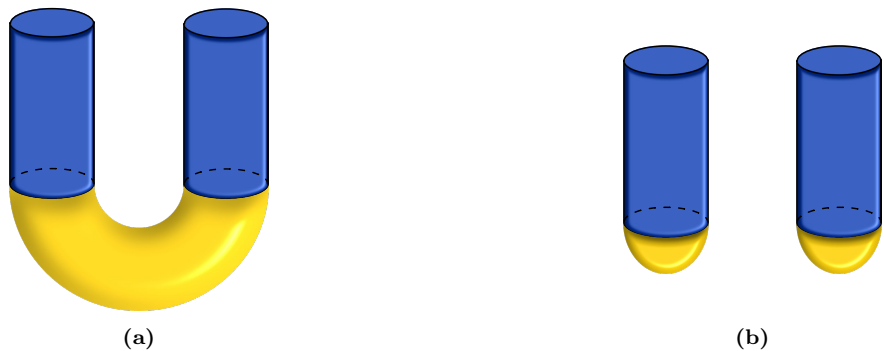

  \begin{subfigure}[t]{.43\textwidth}
    \centering
    \ig{0cm}{width=.5\linewidth}{ads-tfd}
    \caption{}
    \label{fig:ads-tfd}
  \end{subfigure}
  \hspace{1cm}
  \begin{subfigure}[t]{.43\textwidth}
    \centering
    \ig{.5cm}{width=.5\linewidth}{ads-vacuum}
    \caption{}
    \label{fig:ads-vacuum}
  \end{subfigure}
  \caption{\footnotesize \bd{TFD and Factorized Vacuum States in
      $\mrm{AdS_3}\cup \mrm{AdS_3}$}. Thermal $\mrm{AdS}_3$,
    obtained from Euclidean $\mrm{AdS}_3$ by compactifying the
    global time coordinate $\xi  \sim \xi  + \beta$,
    has the topology of a solid torus. Cutting the torus in half
    at $\xi  = 0$ and $\xi  = \beta/2$ and gluing a
    copy of Lorentzian $\mrm{AdS}_3$ at each prepares the TFD
    state in two disconnected copies of $\mrm{AdS}_3$ (left). By
    contrast, the factorized vacuum (right) is an unentangled
    state prepared by cutting $\mrm{EAdS_3}$ in half and gluing
    it to each copy of $\mrm{AdS}_3.$ In the figure, the
    $\mrm{EAdS_3}$ half-cylinder has been compactified to a half-ball.}
\end{figure}

Alternatively, one could define the Euclidean continuation with
compact $\xi \sim \xi + \beta$. The resulting manifold is
known as thermal $\mrm{AdS}_3$
($\mrm{TAdS}_3|_{\beta} = \mrm{EAdS_3}/\beta \bbZ$), and has the
topology of a solid torus whose non-contractible cycle is
parameterized by $\xi$. Cutting the torus at $\xi = 0$
and gluing in the Lorentzian cylinder prepares a thermal state
in $\mrm{AdS_3}$ at inverse temperature $\beta$. More generally,
one may slice the torus in half by
making cuts at both $\xi  = 0$ and
$\xi =\beta/2$, 
and then glue a copy of $\mrm{AdS_3}$ at each cut. The result is the TFD
state in two disconnected copies of $\mrm{AdS}_3$
(Fig. \ref{fig:ads-tfd}). Tracing over one slice returns the
thermal state on the other, corresponding to sewing one
of the cuts back up. Fig. \ref{fig:ads-vacuum} shows the same
two copies of $\mrm{AdS}_3$, now with each in its vacuum
state. In the figure we have compactified the half-cylinder to
the half-ball.

As reviewed in Sec. \ref{sec:ads}, $\mrm{AdS}_3$ is equivalent
to $\mrm{SL}(2,\reals)$, and
to describe a bosonic string in
$\mrm{AdS_3}$ it is natural to take the $\mrm{SL}(2,\reals)_k$
WZW model for the worldsheet CFT \cite{Maldacena:2000hw,Maldacena:2000kv,Maldacena:2001km,deBoer:1998gyt,Giveon:1998ns,Kutasov:1999xu}.
The $\mrm{SL}(2,\reals)_k$ WZW action is an 
inadequate definition of the worldsheet CFT, however, for the
reasons explained above. Instead one must choose a
state, and define the theory by the corresponding continuation
from a unitary CFT. The simplest choice is the vacuum
state, corresponding to continuation from $\mrm{EAdS}_3 = 
\mrm{SL}(2,\bbC)/\mrm{SU}(2).$
The $\mrm{SL}(2,\bbC)_k/\mrm{SU}(2)$ coset was studied in
\cite{Gawedzki:1991yu,Teschner:1997ft,Teschner:1999ug}. 
Its continuation to Lorentzian signature,
constructed in
\cite{Maldacena:2000hw,Maldacena:2000kv,Maldacena:2001km},
defines string perturbation
theory in $\mrm{AdS_3}$ in the spacetime vacuum state, as we
review now in some detail. We then describe string perturbation
theories with 
other choices of the spacetime state.

\subsubsection{Vacuum State}
\label{sec:vacuum}

Consider the theory of a string in $\mrm{AdS}_3$ in the
spacetime vacuum state.
As reviewed in Sec. \ref{sec:ads}, 
the isometry algebra of
$\mrm{AdS}_3=\mrm{SL}(2,\reals)$ is $\Lsl(2,\reals)_\mrm{L}
\oplus \Lsl(2,\reals)_\mrm{R} \simeq \Lso(2,2)$,
which is extended to an 
$\what\Lsl_k(2,\reals)_\mrm{L}\oplus
\what\Lsl_k(2,\reals)_\mrm{R}$
current algebra of the worldsheet CFT.
As usual in $\mrm{AdS/CFT}$, the isometry algebra of the bulk is
identified with the global conformal algebra 
of the dual CFT (BCFT) defined on the spacetime conformal
boundary. Namely, the BCFT global conformal
generators\footnote{We denote the worldsheet Virasoro generators
by $L_n$ and the BCFT Virasoro generators by $\sL_n$.}
$\{\sL_0,\sL_{\mp  
  1}\} \cup \{\bar{\sL}_0, \bar{\sL}_{\mp 1} \}$
satisfy the same 
algebra as the current zero-modes $\{ J^3_0,J^\pm_0\} \cup
\{\bar J^3_0, \bar J^\pm_0\}$ 
(Eqn. \ref{eqn:sl2-global}):
$[\sL_0,\sL_{\mp 1}] = \pm \sL_{\mp 1}$, 
$[\sL_{-1},\sL_1] = -2 \sL_0$, and
similarly for the anti-holomorphic sector.

Likewise, the 
$\mrm{SL}(2,\bbC)_k/\mrm{SU}(2)$ coset is equipped with an
$\what \Lsl_k(2,\bbC)$ current algebra, whose global sub-algebra
$\Lsl(2,\bbC) \simeq \Lso(1,3)$ is the isometry algebra of
$\mrm{EAdS_3}$ and the global conformal
algebra of the BCFT in Euclidean signature.\footnote{Prior to the quotient, one
has an $\what\Lsl_k(2,\bbC)_\mrm{L}\oplus
\what\Lsl_k(2,\bbC)_\mrm{R}$ current algebra corresponding to
the independent symmetries $g \to \Omega_\mrm{L} g
\Omega_\mrm{R}^{\dg}$ on the left and right. After the quotient,
one is left with a single copy of 
the symmetry to preserve Hermiticity, $g \to \Omega g \Omega^\dg$.} The
two current algebras share a common complexification, 
$\what\Lsl_k(2,\bbC)_\mrm{L}\oplus
\what\Lsl_k(2,\bbC)_\mrm{R}$. The complexified global
sub-algebra $\Lsl(2,\bbC)_\mrm{L}\oplus \Lsl(2,\bbC)_\mrm{R}$ is
the standard complexification of the global conformal algebra
from the perspective of the BCFT, wherein the complex coordinate
$x$ and its complex-conjugate $\bar x$ on the boundary sphere
are promoted to
independent complex coordinates, on which the
holomorphic and anti-holomorphic dual Virasoro generators
act independently.

Whereas string amplitudes in asymptotically flat space compute
$S$-matrix elements, string amplitudes in $\mrm{AdS}$ compute
correlation functions of local operators of the BCFT, or
expectation values thereof in Lorentzian signature.
The spectrum of the BCFT is organized in unitary representations of its
Virasoro algebra, 
\begin{align}
  [\sL_n,\sL_m] = (n-m)\sL_{n+m} +
  \frac{\textsf{c}}{12} (n^3-n)\delta_{n+m}, 
\end{align}
and similarly for $\bar{\sL}_n$, with
$[\sL_n,\bar{\sL}_m] = 
0$. The central charge is, at large $k$,
\cite{Brown:1986nw}
\begin{align}
 \textsf{c} =\frac{3}{2}\frac{l_\mrm{s}}{l_\mrm{p}} \sqrt{k}.
\end{align}

The representations are
built on Virasoro primary states labeled by their
spins $(J,\bar J)$, which are positive real numbers.
Focusing on the holomorphic factor, 
a Virasoro primary state $\ket{J,M=J}$ is annihilated by the
positive modes $\sL_{n>0}$ and is an eigenstate of
$\sL_0$ with 
eigenvalue $M=J$.
Being annihilated by $\sL_1$, 
it sits at the bottom of a
lowest-weight discrete-series representation $D_J^+ = \{
\ket{J,M}\cn M \in J +\bbN \}$ of the global conformal algebra
$\Lsl(2,\bbC)_\mrm{L}$, the descendent states being obtained
by action of $\sL_{-1}$,  
$\ket{J,M} \propto (\sL_{-1})^{M-J}\ket{J,J}$.

To each global representation $D_J^+$, one associates by the
state-operator map a primary operator $\cO_{J}(x)$ 
that transforms under $\Lsl(2,\bbC)_\mrm{L}$  according to
\begin{subequations}
  \label{eqn:global-conf}
\begin{align}
  &[\sL_{-1}, \cO_J(x)] =  \partial \cO_J(x)\\
  &[\sL_0, \cO_J(x)] =  (x \partial + J ) \cO_J(x)\\
  &[\sL_1, \cO_J(x)] =  (x^2 \partial + 2Jx ) \cO_J(x).
\end{align}
\end{subequations}

$\cO_J(x)$ prepares the lowest-weight state when
inserted at the origin, $\cO_{J}(0)\ket{0} = \ket{J,J}.$
By the $\mrm{AdS/CFT}$ dictionary,
there is a dual state of a string in $\mrm{AdS}_3$,
which likewise transforms as a lowest-weight state in $D_J^+$, now
with respect to the global $\Lsl(2,\bbC)_\mrm{L}$ sub-algebra of
the worldsheet current algebra. 
The descendent states $\ket{J,M} \in D_J^+$ are prepared in the
boundary by inserting derivatives of $\cO_J(x)$, which are dual to
excited string states in the bulk.

Similarly, when inserted at
the point-at-infinity, $\lim\limits_{x\to \infty}
x^{2J}\cO_J(x)$ prepares a highest-weight 
state $\ket{J,M=-J} \in D_J^-$. Thus, lowest-weight states may be interpreted as
in-states and highest-weight states as out-states.

The string amplitudes, which compute correlation functions of
such BCFT operators, are 
defined by continuation from
the $\mrm{SL}(2,\bbC)_k/\mrm{SU}(2)$ coset CFT.
The spectrum of the latter
is  organized in representations of its $\what\Lsl_k(2,\bbC)$
current algebra. 
These are standard current-algebra representations,
built upon a spin-$j$
representation of the global $\Lsl(2,\bbC)$ sub-algebra that is
primary with respect to the positive current modes.\footnote{As
  in Sec. \ref{sec:ads}, we
  denote the $\Lsl(2,\bbC)$ spin $j$ and $J^3_0$ eigenvalue $m$
  by lower-case letters for unflowed representations, and
  capital $J$, $M$ for spectral-flowed representations.}
The primaries appearing in the coset Hilbert space are the
continuous-series representations with $j \in \frac{1}{2} + i \reals_+$
\cite{Gawedzki:1991yu,Teschner:1997ft,Teschner:1999ug}. 
$j$ may be continued away from this line, however \cite{Teschner:1999ug,Maldacena:2001km}.
Indeed, these are not the representations of interest for string
theory in $\mrm{AdS}_3$---they correspond to the bosonic string
tachyon, and would map to dual representations with complex
conformal weights. 

As in the two-dimensional BCFT context recalled above,
the spin-$j$ representation of $\Lsl(2,\bbC)$ 
may be described in a function space basis, with
the generators implemented by differential operators acting on
functions of a complex coordinate $x$ \cite{Teschner:1997ft,Teschner:1999ug}:
\begin{align}
  \label{eqn:D-generators}
  &\cD^3 = x \pd{}{x} + j,\quad\quad
  \cD^+ = \pd{}{x},\quad\quad
  \cD^- = x^2 \pd{}{x} + 2 j x.
\end{align}

The primary operators of the worldsheet $\what \Lsl_k(2,\bbC)$
current algebra may then be written $\Phi_j(z, \bar
z; x, \bar x)$, where $(z,\bar z)$ are worldsheet coordinates
and $(x,\bar x)$ parameterize the representation.
Their OPEs with the currents are
\begin{align}
  \label{eqn:ope}
  J^a(z) \Phi_j(z', \bar z';x,\bar x) \sim 
  \frac{\cD^a \Phi_j(z', \bar z'; x, \bar x)}{z - z'}
  \quad\quad a = 3,\pm. 
\end{align}
They are worldsheet scalars of conformal weight
\begin{align}
  h_j= -\frac{j(j-1)}{k-2}.
\end{align}
$h_j$ is a real number for $j \in \frac{1}{2} + i \reals$, as
well as for $j \in \reals$. In the former case it is always
positive, while in the latter it is positive only in the window $0<j<1$.
Note also that $h_j$ is symmetric under $j \to 1 -j,$ a reflection
about $j = \frac{1}{2}$.

In the asymptotic region, the vertex operators behave for large
$k$ as \cite{Kutasov:1999xu,Teschner:1999ug} 
\begin{align}
  \label{eqn:phi-asymptotic}
  \Phi_j(z, \bar z; x, \bar x)
  \overright{\sigma \to \infty}
  e^{-2(1-j)\sigma}\delta^2(\gamma-x) + \frac{1}{\pi}(2j-1)e^{-2j\sigma}|\gamma - x|^{-4j}.
\end{align}
Accounting for the measure factor $\sqrt{g} = l_\mrm{AdS}^2 e^{2\sigma}$ 
from the metric Eqn. \ref{eqn:poincare-metric},
observe that the target wavefunction is
delta-function normalizable only for $\mrm{Re}(j) =
\frac{1}{2}$, consistent with the preceding statement that the
$\mrm{SL}(2,\bbC)_k/\mrm{SU}(2)$ Hilbert space contains
only the complex branch states. 

For $\mrm{Re}(j)<\frac{1}{2}$, the second exponential dominates,
and the operator is spread over the boundary sphere. 
For $\mrm{Re}(j)>\frac{1}{2}$, however, the first term
dominates, and the non-normalizability of the 
wavefunction is localized at $\gamma = x$, interpreted as
a source for an operator of the BCFT on the boundary $\mrm{S}^2$. 
Eqn. \ref{eqn:phi-asymptotic} is the standard asymptotic
expansion of a solution to the $\mrm{AdS}_3$ wave equation
for a scalar of mass $l_\mrm{AdS}^2m^2 = \Delta(\Delta -2)$,
with a delta-function
source at $(x, \bar x)$ for the dual operator of
dimension $\Delta = 2j$. The leading term inserts the source and
the sub-leading term is $e^{-2j\sigma}/(2\Delta -d)$ times the two-point
function of the dual operator \cite{Klebanov_1999}.

Thus, for $j>\frac{1}{2}$ one has a map from the worldsheet vertex operator
$\Phi_j(z, \bar z; x, \bar x)$ to a primary scalar BCFT operator
$\hat \Phi_j(x, \bar x)$
of real conformal weight $(j,j)$ inserted at 
$(x, \bar x)$, the transformation of the vertex operator under
the global sub-algebra (Eqn. \ref{eqn:ope}) matching the global
conformal transformation of the boundary operator
(Eqn. \ref{eqn:global-conf}).\footnote{More precisely, one has a
  map from $\Phi_j(z, 
  \bar z; x, \bar x)\otimes \cO_h(z,\bar z)$, where $\cO_h$ is a
primary of the internal CFT such that the combined vertex
operator is marginal. The construction of the full BCFT Virasoro
algebra was discussed in \cite{Giveon:1998ns,Kutasov:1999xu,deBoer:1998gyt}.} 
A string amplitude with many insertions
$\Phi_{j_i}(z_i,\bar z_i; x_i, \bar x_i)$ computes a BCFT
correlation function on $\mrm{S}^2$ with operator insertions at $(x_i,\bar
x_i)$ of dual conformal weights $(j_i,j_i)$.\footnote{In fact,
the string partition function is related to the BCFT generating
functional in an ensemble in which the BCFT central charge
fluctuates. To obtain a standard CFT with fixed central charge
one must perform a Legendre transform of the string partition
function, as explained in \cite{Kim_2015}.}
The vertex operators with $j > \frac{1}{2}$ are non-normalizable
from the perspective of the $\mrm{SL}(2,\bbC)_k/\mrm{SU}(2)$
coset, as appropriate for operators that insert
delta-function sources on the boundary of Euclidean $\mrm{AdS}$,
and are defined by analytic continuation in $j$
\cite{Teschner:1999ug,Maldacena:2001km}. 
String amplitudes of two, three, and four primaries
were studied in this way in \cite{Maldacena:2001km}. 

\begin{figure}[t]
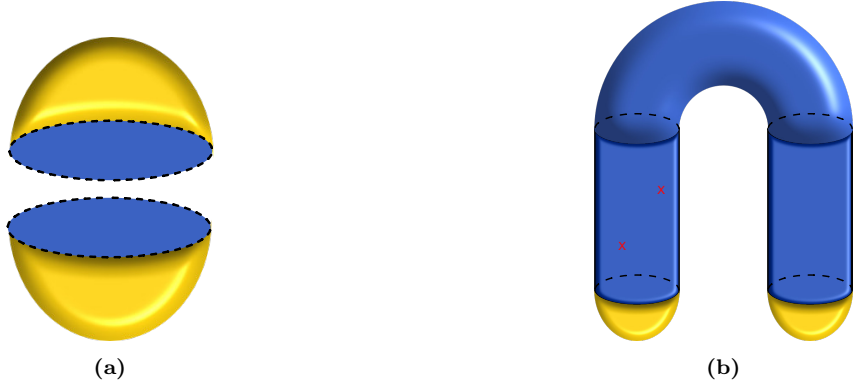

  \begin{subfigure}[t]{.43\textwidth}
    \centering
    \ig{0cm}{width=.4\linewidth}{half-ball}
    \caption{}
    \label{fig:half-ball}
  \end{subfigure}
  \hspace{1cm}
  \begin{subfigure}[t]{.43\textwidth}
    \centering
    \ig{0cm}{width=.5\linewidth}{vacuum-contour}
    \caption{}
    \label{fig:vacuum-contour}
  \end{subfigure}
  \caption{\footnotesize \bd{$\mrm{AdS}_3$ Vacuum
      Schwinger-Keldysh Contour}. String amplitudes for
    $\mrm{AdS}_3$ in the spacetime vacuum state compute vacuum
    expectation values of the dual CFT and are defined by
    continuation from $\mrm{SL}(2,\bbC)_k/\mrm{SU}(2).$ The
    Euclidean amplitudes compute BCFT correlation functions 
    on the $\mrm{S}^2$ conformal boundary of
    $\mrm{EAdS}_3$, with insertions at the points $(x,\bar x)$
    which labeled
    the worldsheet vertex operators. The spacetime is
    complexified and the insertions 
    continued to the Lorentzian section to obtain vacuum
    expectation values. Though the procedure does not rely on a
    Lagrangian, one may think of these amplitudes as defined by
    a worldsheet functional integral along the 
    Schwinger-Keldysh contour in target space shown.
    A Euclidean cap (in yellow) prepares the
    $\mrm{AdS}_3$ vacuum, which is glued to a Lorentzian
    excursion (in blue) that 
    flows forward and backward in time, and is then glued to
    the outgoing vacuum cap. String vertex operators insert dual
    operators on the conformal boundary, indicated by the red
    $\times$'s. Operators may also be left on the Euclidean
    caps to define string perturbation theory in $\mrm{AdS}_3$
    in different excited pure states.}
    \label{fig:vacuum-sk}
\end{figure}

From BCFT correlation functions on $\mrm{S}^2$,
the boundary insertion points $(x, \bar x)$
may be continued in  the usual way from the
Euclidean sphere to the Lorentzian cylinder in order to define
expectation values of the BCFT in the vacuum state
(Fig. \ref{fig:vacuum-sk}). The 
conformal transformation $(x=e^{\xi + i \theta}, \bar x =
e^{\xi - i \theta})$ maps the sphere to the Euclidean
cylinder, and continuing $\xi \to i t$ yields the
Lorentzian cylinder $(x = e^{i(t+\theta)}, \bar x = e^{i(t - \theta)}).$
In doing so, one has
continued $(x, \bar x)$ to independent complex coordinates,
and in turn the 
vertex operator $\Phi_j(z, \bar z; x, \bar x)$ is
continued to a representation of the complexified current algebra
$\what\Lsl_k(2,\bbC)_\mrm{L}\oplus
\what\Lsl_k(2,\bbC)_\mrm{R}$.
The original boundary $\mrm{S}^2$
and the Lorentzian cylinder correspond to two real sections of
the complexification. On the worldsheet, one interpolates
between vertex operators for insertions on the Euclidean sphere
or Lorentzian cylinder by restricting the complexified primary
$\Phi_j(z, \bar z; x, \bar x)$ to the appropriate section.

For example, the two-point amplitude of $\Phi_j(z, \bar z; x,
\bar x)$ is \cite{Maldacena:2001km}
\begin{align}
  \vev{\hat \Phi_j(x_1, \bar x_1) \hat\Phi_j(x_2,\bar x_2)}=
  \frac{1}{V_\mrm{conf}}\vev{\Phi_j(z = 0; x_1,\bar x_1)
  \Phi_j(z=0;x_2, \bar x_2)}   \propto |x_{12}|^{-4j},
\end{align}
as appropriate for a BCFT scalar of dimension $2j$. Making the
boundary conformal transformation $\Phi_j(z, \bar z; x, \bar x)
\to e^{-2j\xi} \Phi_j(z, \bar z; \xi,\theta)$, one obtains the
standard CFT two-point function for a scalar on the
cylinder:
\begin{align}
  \label{eqn:2-point-cylinder}
  \vev{\hat \Phi_j(\xi_1, \theta_1) \hat\Phi_j(\xi_2,\theta_2)}
  \propto\lp \cosh(\xi_{12}) -\cos(\theta_{12})\rp^{-2j}.
\end{align}
As expected for a Euclidean correlation function, this
expression is non-singular as long as the two insertions are not
coincident.

To obtain a Lorentzian expectation value, one slides each
insertion to the zero-time slice and then onto the Lorentzian
section: 
\begin{align}
  \label{eqn:2-point}
  \bra{0}\hat \Phi_j(t_1,\theta_1)\hat \Phi_j(t_2,\theta_2)\ket{0}
  \propto \lp \cos(t_{12}) -\cos(\theta_{12})\rp^{-2j}.
\end{align}
In doing so, one encounters singularities when one operator hits
the lightcone of the other, $t_{12} = \pm
\theta_{12}$. Following the 
usual $i\vep$ prescription to avoid
the singularity and obtain a time-ordered expectation value, one
replaces $t \to t (1- i \vep)$. We mostly suppress the $i\vep$'s
for notational simplicity.

The perturbative string states of Eqn. \ref{eqn:sl2r-spectrum}
may be identified with the modes of the boundary position basis
vertex operators. For example, the vertex operators
$\Phi_{jm\bar m}(z,\bar z)$ for the string states $\ket{j,m,\bar
  m}$ are formally given by the spacetime Fourier transform of
$\Phi_j(z,\bar z; t,\theta)$:
\begin{align}
  \label{eqn:vacuum-fourier}
  \Phi_{jm\bar m}(z,\bar z)
  \propto \int\limits_{-\infty}^\infty \diff t
  \int\limits_0^{2\pi} \diff \theta \,
  e^{-i (m+\bar m) t} e^{-i(m - \bar m)\theta}
  \Phi_j(z,\bar z; t,\theta),
\end{align}
where $m + \bar m$ is the spacetime energy and $m - \bar m$ is
the angular momentum, and similarly for the spectral-flowed
operators reviewed below. One may likewise compute string amplitudes
of such momentum-basis insertions \cite{Maldacena:2001km}.

When inserted at the origin of the Euclidean section, $\Phi_j(z,
\bar z; x, \bar x)$ 
prepares the lowest-weight 
state $\ket{j,j,j}$ in
the representation $ D_j^+\otimes D_j^+$ of the global
sub-algebra:
\begin{align}
  \label{eqn:string-lowest}
  \Phi_j(z=0;x=0) \ket{0} = \ket{j,j,j}.   
\end{align}
This is the bulk string state dual to the BCFT Virasoro primary
state of conformal weight $(j,j)$, the latter prepared by the
dual operator $\hat \Phi_j(x=0)$, both transforming as lowest-weight
states of $\Lsl(2,\bbC)_\mrm{L} \oplus
\Lsl(2,\bbC)_\mrm{R}$.\footnote{Again, once combined with an
  internal operator such that the Virasoro constraints are satisfied.}
The global descendent states  $\ket{j,m,\bar m} \in
D_j^+\otimes D_j^+$, related to
the lowest-weight state by the action of $J^+_0,\bar J^+_0$, are prepared
by inserting $x$ derivatives of $\Phi_j(z,\bar z; x, \bar x)$, and
are dual to the global conformal descendents of the BCFT Virasoro
primary. In the bulk effective field theory, $\ket{j,j,j}$ is
the lowest-energy state of a scalar field of mass\footnote{It is
hopefully clear from context when we use $m$ to refer to the
mass as opposed to the $J^3_0$ eigenvalue.}
$l_\mrm{AdS}^2m^2 = \Delta(\Delta-2)$ dual to the BCFT scalar primary
operator of dimension $\Delta = 2j$.

Of course, the Virasoro primaries of the dual CFT are not all
scalars. In addition to the current-algebra primaries $\Phi_j(z, \bar z;
x, \bar x)$ that prepare lowest-weight states $\ket{j,j,j} \in
\what D_j^+\otimes \what D_j^+$,
one has vertex operators denoted $\Phi_{jw}^{J\bar J}(z, \bar z;
x, \bar x)$  that prepare the spectral-flowed states of
Eqn. \ref{eqn:sl2r-spectrum} \cite{Maldacena:2001km}: 
\begin{align}
  \label{eqn:spectral-in}
  \Phi_{jw}^{J \bar J}(z = 0; x = 0)\ket{0}
  =\ket{j,m = J - kw/2,\bar m = \bar J -kw/2; w},\quad w> 0.
\end{align}
These operators are worldsheet Virasoro---but not
current-algebra---primaries of conformal 
weights as in Eqn. \ref{eqn:spectral-dimension}:
\begin{align}
  &h_{jw}^{J\bar J}= -\frac{j(j-1)}{k-2} - w J + \frac{1}{4} k
  w^2,\quad\quad
  \bar h_{jw}^{J\bar J}= -\frac{j(j-1)}{k-2} - w \bar J + \frac{1}{4} k w^2.
\end{align}
They carry the
additional labels\footnote{For unflowed
  operators one has $\Phi_j = \Phi_{j,w=0}^{J = \bar J = j}.$}
$(J,\bar J)$ specifying their spins with 
respect to the global $\Lsl(2,\bbC)_\mrm{L}\oplus
\Lsl(2,\bbC)_\mrm{R}$, as well as the spectral-flow label
$w$. 

Recall from Sec. \ref{sec:ads} that, for $w>0$, the spectral flow
$\ket{j,m,\bar m;w}$ of a current-algebra primary carries
$J^3_0$ weight $M = m + \frac{k}{2}w$ and transforms as a
lowest-weight state $\ket{J,M=J}\in D_J^+$ with respect to
$\Lsl(2,\bbC)_\mrm{L}$, and similarly for
$\Lsl(2,\bbC)_\mrm{R}$ with $\bar M = \bar J = \bar m +
\frac{k}{2} w$. Thus, each spectral-flowed primary 
$\ket{j,m,\bar m;w}$ with $w>0$ sits at the bottom of a discrete-series
representation\footnote{Note that, with respect to the global sub-algebra, one
obtains a lowest-weight discrete-series representation
regardless of whether $\ket{j,m,\bar m;w}$ came from a
spectral-flowed discrete-series $\what D_{j}^{+,w} \otimes \what D_j^{+,w}$
or continuous-series $\what C_{j,\alpha}^w\otimes \what C_{j,\alpha}^w$
representation of the current algebra, as
appropriate for a primary of the BCFT. In a continuous-series
representation, $J$ and $j$ (and $\bar J$ and $j$) are
unrelated, whereas in a lowest-weight discrete-series representation they are
related by $j=J - \frac{k}{2}w -\bbN$.} $D_J^+ \otimes D_{\bar J}^+$ of the global
sub-algebra, and is dual to a BCFT Virasoro primary state of spin $(J,
\bar J).$\footnote{Once combined with the
  internal CFT and subjected to the constraints. In particular,
  $J = m + \frac{k}{2} w$ is not 
  guaranteed to be
  positive on the spectrum of the $\mrm{SL}(2,\reals)_k$ WZW
  model. It is only after applying the Virasoro constraints
  that one obtains a physical state of positive $J, \bar J$,
  which maps to a primary of the BCFT.}
Associated to each such lowest-weight string state one has a vertex
operator $\Phi^{J\bar J}_{jw}(z, \bar z; x, \bar x)$ that
prepares it. The global descendent states
$(J_0^+)^{N}(\bar J_0^+)^{\bar N}\ket{j,m,\bar
  m;w}$ are as before prepared by insertions with
  derivatives.\footnote{Note that 
    $(J_0^+)^{N}(\bar J_0^+)^{\bar N}\ket{j,m,\bar
      m;w}=(J_{-w}^+[w])^{N}(\bar J_{-w}^+[w])^{\bar N}\ket{j,m,\bar m;w}$   
    transforms as a current-algebra descendent with respect to
    the unflowed generators, and should not be confused with
    the state $\ket{j,m+N,\bar m + \bar N;w}$. The
  latter is the lowest-weight state of its own discrete-series
  representation, and carries a different
  worldsheet conformal weight besides.}

For $w<0$, on the other hand, $\ket{j,m,\bar m;w}$ transforms as
a highest-weight state $\ket{J',-J'}\otimes \ket{\bar J',-\bar
  J'}\in D_{J'}^-\otimes D_{\bar J'}^-$ of spin $J' = -\lp m +
\frac{k}{2}w\rp$, $\bar J' = -\lp \bar m + \frac{k}{2} w \rp$. As
recalled earlier, a lowest-weight state is interpreted as
an in-state in the dual CFT, prepared by a spin $(J,\bar J)$ primary insertion at the origin,
whereas a highest-weight state is interpreted as an out-state,
prepared by an insertion at infinity. $\Phi_{jw}^{J\bar
  J}(z, \bar z; x, \bar x)$, which likewise transforms as a local
operator of spin $(J, \bar J)$ in $x$-space with respect to the global
sub-algebra, also prepares a lowest-weight in-state, and 
should therefore be labeled by $w>0$ as in
Eqn. \ref{eqn:spectral-in}. When inserted at infinity, it
prepares the highest-weight out-state $\ket{j,- \lp J -
  \frac{k}{2} w \rp,- \lp \bar J - \frac{k}{2} w \rp;-w}$ in
$D_{J}^-\otimes D_{\bar J}^-$.

The complete spectrum of the $\mrm{SL}(2,\reals)_k$ WZW model
given in Eqn. \ref{eqn:sl2r-spectrum}, when combined 
with a unitary internal CFT and subjected to the Virasoro
constraints, yields
the unitary spectrum of strings in $\mrm{AdS}_3$ in the
vacuum state \cite{Maldacena:2000hw}.
The upper bound on the real branch, $j <
\frac{k-1}{2}$, is required to obtain a unitary on-shell
spectrum, and ensures compatibility with the spectral-flow
isomorphism, which exchanges the upper and lower bounds under $j
\to \frac{k}{2} - j$. The spectrum also includes the unflowed
complex branch representations $\what C_{j=\frac{1}{2}(1+is),\alpha}$,
whose vertex operators $\Phi_j(z, \bar z; x, \bar 
x)$ do not map to well-defined local operators of the
BCFT. These are the bosonic string tachyons, whose spacetime
mass $l_\mrm{AdS}^2m^2 = -1-s^2$ falls below the tachyonic BF
bound $l_\mrm{AdS}^2m^2 < 
-1$ for a scalar in $\mrm{AdS}_3$ 
\cite{Breitenlohner:1982jf,Maldacena:2000hw}.\footnote{We point
  out that below $k= 3$, at which point it has been argued that the
  $\mrm{SL}(2,\reals)_k$ WZW model undergoes a
  phase transition \cite{Giveon:2005mi,Berkooz:2007fe}, the real branch spectrum $\frac{1}{2}
  < j < \frac{k-1}{2}$ falls within the BF window $-1 <
  l_\mrm{AdS}^2m^2 < 0$ in which 
  two normalizable fall-offs are admissible.}

String amplitudes with spectral-flowed vertex operators were also
computed in \cite{Maldacena:2001km}, and may similarly be
continued to BCFT expectation values by continuing the boundary
insertion points to the Lorentzian cylinder. 

For a Virasoro primary $\Phi_{jw}^{J\bar J}\otimes \cO_{h \bar h}$, with
$\cO_{h\bar h}$ a contribution from the internal CFT, 
the Virasoro constraint $L_0 - 1 = 0$ may be written
\begin{align}
  \label{eqn:J}
  J=    \frac{1}{4} k w +\frac{1}{w} \lp
  -\frac{j(j-1)}{k-2}+h-1\rp.
\end{align}
On the complex
branch, $j$ and $J$ are unrelated, and this equation gives a
continuous spectrum of spacetime conformal weights parameterized
by $\mrm{Im}(j).$\footnote{The constraint guarantees $J$ is
  positive, $-j(j-1) \geq \frac{1}{4}$ being bounded below on
  the complex branch and $h\geq 0$ being positive by unitarity
  of the internal CFT. Then Eqn. \ref{eqn:J} is positive with $k > 2$
  and $w>0$. Note that the constraint, $m+ \frac{k}{2} w=    \frac{1}{4} k w +\frac{1}{w} \lp
  -\frac{j(j-1)}{k-2}+h-1\rp$, is invariant under $w \to -w$ and
  $m \to -m$. Namely, if $\Phi_{jw}^{J\bar J}(z, \bar z; x, \bar
  x)\otimes \cO_{h,\bar h}(z, \bar z)$ prepares a physical lowest-weight
  in-state $\ket{j, J - \frac{k}{2}w, \bar J - \frac{k}{2} w;
    w}\otimes \ket{h, \bar h}$ when inserted at the origin,
  then the highest-weight out-state
  $\ket{j, -\lp J - \frac{k}{2}w \rp, -\lp \bar J - \frac{k}{2}w
  \rp; -w}\otimes \ket{h,\bar h}$
  obtained by inserting the operator at infinity is likewise physical.}
These are known as ``long string''
states. They are heavy in the semi-classical limit, $J$ being of
order $k$, and 
are believed to be a peculiarity of the pure NS background. 
For the real branch, on the other hand, $j$ and $J$ are
related by $j=J-\frac{k}{2} w- N$, with $N \in \bbN$. Solving the on-shell
condition for $J$ then yields a discrete spectrum of spacetime
conformal weights \cite{Maldacena:2000hw},
\begin{align}
  J = N + w + \frac{1}{2}+ \sqrt{\frac{1}{4}+(k-2) \lp h - 1 -N
  w-\frac{1}{2} w(w+1) \rp}.
\end{align}
These, by constrast, are known as ``short string'' states, and
are the more typical vertex operators.

Next we describe how string perturbation theory
in $\mrm{AdS}_3$ is defined
for other choices of the state. Note first of all that one need
not continue all the operator insertions in $x$-space to the
Lorentzian section. Leaving an insertion on the Euclidean cap
prepares the associated state on the cylinder from the
perspective of the BCFT, and defines string perturbation theory
in an excited pure state from the perspective of the worldsheet
theory.

\subsubsection{Thermal  State}
\label{sec:thermal}

Suppose now that one wishes to study string perturbation theory
in $\mrm{AdS_3}$ in a
thermal state. In the BCFT, thermal expectation values are
obtained by continuation from Euclidean correlation functions
on
$\mrm{T}^2$, the periodicity of the Euclidean time circle fixing the
inverse temperature of the state. One constructs the
Schwinger-Keldysh contour  
by cutting the torus at a single time-slice and gluing 
in the Lorentzian cylinder, or by cutting the torus
in half to prepare the TFD state in two copies of the
BCFT Hilbert space on a circle. In the bulk, this BCFT state is
dual to the Hartle-Hawking wavefunctional defined by the
gravitational functional integral with
thermal boundary conditions at infinity and ending on a spatial
slice bounded by 
two circles. Below the Hawking-Page temperature, this
wavefunctional is sharply peaked on two disconnected copies of
the $\mrm{AdS}_3$ metric, and above the Hawking-Page temperature
it is peaked on the zero-time slice of the two-sided,
asymptotically $\mrm{AdS}_3$ black hole\cite{Maldacena:2001kr}.

Thus, one may think of the bulk dual to the BCFT TFD
state as either the  bulk TFD state 
in two disconnected copies of $\mrm{AdS_3}$
or the Hartle-Hawking (HH) state in the $\mrm{AdS}_3$ black hole.
At the Hawking-Page temperature $T_\mrm{HP} = \frac{1}{2\pi}$ (in
$\mrm{AdS}$ units) there is a first-order phase 
transition between the two  
\cite{Maldacena:2001kr,Hawking:1982dh,Witten:1998zw}. At lower
temperatures, the dominant bulk Euclidean saddle is
$\mrm{TAdS}_3$, whose 
Euclidean time circle is non-contractible, while at higher
temperatures 
the bulk saddle is the Euclidean black hole, which has the same
solid-torus topology but whose Euclidean time direction is
now identified with the contractible cycle.

The worldsheet theory for a string in
$\mrm{TAdS_3}|_\beta$ is the $J^3_0+\bar J^3_0$ orbifold
$\beta \bbZ \bs 
\mrm{SL}(2,\bbC)_k/\mrm{SU}(2)$ that compactifies
the cylinder, $\xi \sim \xi + \beta$. 
String perturbation theory in a thermal state in $\mrm{AdS}_3$
is defined by continuation from this quotient. Its string
amplitudes compute the dominant contribution to BCFT $\mrm{T}^2$
correlation functions below the Hawking-Page temperature.
Above the Hawking-Page temperature, a thermal gas of strings in
$\mrm{AdS}_3$ collapses to a black hole, which becomes the
dominant contribution thereafter.\footnote{Until
  one reaches the Hagedorn 
  temperature, where the theory becomes unstable. Of course,
  the bosonic string theory is already unstable because of the
  tachyon. But as the temperature is increased, the thermal
  circle becomes small and the modes that wind it become
  lighter. At the Hagedorn temperature, the circle becomes so
  small that the lightest winding mode
  becomes tachyonic \cite{Atick:1988si,Berkooz:2007fe}.}
We discuss the black hole in the next sub-section.

The quotient
preserves two of the six isometries of $\mrm{EAdS}_3$,
corresponding to $\xi$ and $\theta$ translations. The
orbifold projection is
\begin{align}
  e^{i\beta(J^3_0+\bar J^3_0)} \Phi_j(z, \bar z; x, \bar x)
  e^{-i\beta (J^3_0 + \bar J^3_0)} = \Phi_j(z, \bar z; x, \bar x).
\end{align}
After the boundary conformal transformation $x= e^{\xi + i
  \theta}$ which sends
$\Phi_j(z, \bar z;x, \bar x)\to e^{-2j
  \xi}\Phi_j(z, \bar z; \xi,\theta)$, the
projection condition simply enforces periodicity in $\xi$:
\begin{align}
  \label{eqn:projection}
  \Phi_j(z, \bar z; \xi, \theta) = \Phi_j(z, \bar z;
  \xi+\beta, \theta).
\end{align}
Each unflowed 
primary $\Phi_j$ of $\mrm{SL}(2,\bbC)_k/\mrm{SU}(2)$ may be
projected to an operator in the untwisted sector of the orbifold by
summing over images,
\begin{align}
  \label{eqn:image-sum}
  \Phi_j(z, \bar z; \xi, \theta) \to
  \sum_{n \in \bbZ}  \Phi_j(z, \bar z; \xi + \beta 
  n,\theta). 
\end{align}
For spectral-flowed operators, one may likewise sum over images
of $\Phi_{jw}^{J\bar J}(z,
\bar z; x, \bar x) \to e^{-J(\xi+i \theta)} e^{-\bar
  J(\xi-i \theta)} 
\Phi_{jw}^{J\bar J}(z, \bar z; \xi, \theta)$ to obtain a
projection-invariant operator: 
\begin{align}
  \label{eqn:spectral-image-sum}
  \Phi_{jw}^{J\bar J}(z, \bar z; \xi, \theta)
  \to
  \sum_{n \in \bbZ}
    \Phi_{jw}^{J\bar J}(z, \bar z; \xi+\beta n,
  \theta). 
\end{align}

We emphasize that the projection acts on the Euclidean
$\mrm{SL}(2,\bbC)_k/\mrm{SU}(2)$ vertex operators and not on the
spectrum 
of Lorentzian string states $\ket{j,m,\bar m;w}\otimes
\ket{h,\bar h}$. The string states are particle-like excitations on 
top of the $\mrm{AdS}_3$ background---whether the background is
in the vacuum or thermal state does not affect the spectrum of
particles. 

The correlation
functions of the projected primaries
Eqns. \ref{eqn:image-sum}-\ref{eqn:spectral-image-sum} in the
orbifold are obtained 
by summing over images in the original
$\mrm{SL}(2,\bbC)_k/\mrm{SU}(2)$ correlators, and their
amplitudes produce the $\mrm{T}^2$ correlation functions of the
BCFT for $\beta > 2\pi$. Correspondingly, the BCFT
local operators and correlation functions on $\mrm{T}^2$ may
independently be obtained from the cylinder by summing over
images.

The continuation to Lorentzian expectation values in the thermal
state is as before. If one cuts the torus at $\xi =
0$ and continues the operators to the Lorentzian section, one
obtains string amplitudes in $\mrm{AdS}_3$ in the thermal state
at inverse temperature $\beta$. If one makes cuts at both
$\xi = 0$ and $\xi = \beta/2,$ one can continue the
operator labels to either $\xi = i t_\mrm{R}$ or
$\xi = \frac{\beta}{2} + i t_\mrm{L}$. The resulting
string amplitudes compute expectation values of the continued
insertions in two copies of $\mrm{AdS}_3$ in the TFD state. The
corresponding Schwinger-Keldysh contour is obtained by gluing
together two copies of Fig. \ref{fig:ads-tfd}.
One may also leave insertions on the $\mrm{T}^2$ Euclidean section to
define perturbation theory in a thermal or TFD state deformed by
sources.

For example, the two-point amplitude of $\Phi_j$ is obtained by
summing over images in
Eqn. \ref{eqn:2-point-cylinder},\footnote{Note that by simply
  replacing each operator by its sum over images,
  e.g. $\sum_{n,m\in \bbZ}\vev{\cO(\xi+n\beta) \cO(\xi' + m \beta)} = 
  \sum_{\substack{n-m\in \bbZ\\ n+m \in \bbZ}} \vev{\cO(\xi -
    \xi' +(n-m)\beta)\cO(0)}$, one obtains an extraneous
  divergent sum $\sum_{n+m}$, which should be discarded.} 
\begin{align}
  \vev{\hat \Phi_j(\xi_1, \theta_1) \hat\Phi_j(\xi_2,\theta_2)}\propto
  \sum_{n\in\bbZ}\lc \cosh(\xi_{12} + n \beta)
  -\cos(\theta_{12})\rc^{-2j} . 
\end{align}
By cutting the torus at $\xi = 0$ and continuing both insertions
to the Lorentzian cylinder one obtains the amplitude in a
thermal state, 
\cite{Keski_Vakkuri_1999,Birmingham_2003,Maldacena:2001kr} 
\begin{align}
  \mrm{tr}\lp e^{-\beta H} \hat \Phi_j(t_1, \theta_1)
  \hat\Phi_j(t_2,\theta_2)\rp \propto
  \sum_n\lc \cos(t_{12} + in \beta) -\cos(\theta_{12})\rc^{-2j},
\end{align}
again suppressing the $i\vep$'s.
Alternatively, one could cut the torus at both $\xi = 0$ and
$\xi = \beta/2$ preparing the TFD state, and e.g. continue one
operator to each side
\begin{align}
  \bra{\mrm{TFD}}
  (1_\mrm{L}\otimes&\hat \Phi_j(t_{\mrm{R},1}, \theta_1))
  (\hat\Phi_j(t_{\mrm{L},2},\theta_2)\otimes 1_\mrm{R})
  \ket{\mrm{TFD}}\\
  &\propto
  \sum_{n\in \bbZ}\lc \cos\lp t_{\mrm{R},1} -t_{\mrm{L},2} + i\lp n -
  1/2\rp  \beta\rp -\cos(\theta_{12})\rc^{-2j} .\nt
\end{align}
One may also consider twisted-sector operators of the orbifold,
but these are not expected to map to local operators of the
BCFT.

\subsubsection{Hartle-Hawking State}
\label{sec:black-hole}

Above the Hawking-Page temperature, the asymptotic $\mrm{AdS}_3$
black hole, known also as BTZ, is the
dominant contribution to thermal BCFT 
expectation values \cite{Banados:1992wn,Banados:1992gq,
  Maldacena:2001kr,Witten:1998zw}.  
BTZ is a particularly simple black hole because it is a quotient
of
$\mrm{AdS}_3$, performed with respect to the
$J^2_0+ \bar J^2_0$ isometry. The parameterization
Eqn. \ref{eqn:global-matrix} diagonalized the action of $J^3_0
\pm \bar J^3_0$. To describe the BTZ orbifold, it is more
natural to diagonalize $J^2_0 \pm \bar J^2_0$ via
\begin{align}
\label{eqn:hyp-g}
  g = e^{i (\til\Theta + \til T) \mrm{T}_2}e^{2i \til R
  \mrm{T}_1} e^{i (\til \Theta-\til T)\mrm{T}_2} \in 
  \mrm{SL}(2,\reals), 
\end{align}
on which the group metric evaluates to
\begin{align}
  \label{eqn:hyp-metric}
  \diff s^2_\text{AdS-Rindler} = l_\mrm{AdS}^2 \lp
  -\sinh^2(\til R)
  \diff \til T^2 + \diff  \til R^2 + \cosh^2(\til R) \diff
  \til\Theta^2\rp.    
\end{align}
We recognize in Eqn. \ref{eqn:hyp-metric} the
$\mrm{AdS}_3$-Rindler metric (Eqn. \ref{eqn:ads-rindler}), where
$\til T,\til \Theta \in
(-\infty, \infty)$ and $\til R \in (0,\infty)$. These coordinates
cover only a wedge 
of $\mrm{AdS}_3$, up to the coordinate horizon at $\til R = 0$ where
the coefficient of $\diff \til T^2$ vanishes. 

$\til T$ translations are implemented in Eqn. \ref{eqn:hyp-g}
by $g \to e^{i \delta \til T \mrm{T}_2}g e^{-i \delta \til T \mrm{T}_2}$,
and $\til \Theta$ translations
by $g \to e^{i \delta \til \Theta \mrm{T}_2}g e^{i\delta \til\Theta \mrm{T}_2}$.
These isometries are therefore generated in
$\mrm{SL}(2,\reals)_k$ 
by $J^2_0 - \bar J^2_0$ and $J^2_0 + \bar J^2_0$, respectively.
The (non-rotating) BTZ black hole of radius $R_\mrm{s}$ is
defined by the 
$J^2_0 + \bar J^2_0$  orbifold that compactifies
\begin{align}
  \til\Theta \sim 
  \til\Theta + 2\pi R_\mrm{s}/l_\mrm{AdS}.  
\end{align}
In $\mrm{SL}(2,\reals)$, this is the identification $g \sim h g
h$, with $h = e^{2\pi i R_\mrm{s} \mrm{T_2}/l_\mrm{AdS}}
= \mrm{Diag}(e^{\pi R_\mrm{s}/l_\mrm{AdS}},e^{-\pi 
    R_\mrm{s}/l_\mrm{AdS}})$. $h$ is a hyperbolic element of
$\mrm{SL}(2,\reals)$, meaning that $\mrm{tr}( h) = 2 \cosh(\pi
R_\mrm{s}/l_\mrm{AdS})> 2$. The quotient preserves the
translation isometries in $\til T$ and $\til \Theta$.

The more familiar form of the BTZ metric is obtained by the
coordinate transformation
\begin{align}
  \label{eqn:btz-coords}
  R = R_\mrm{s} \cosh(\til R),\quad
  T =\frac{l_\mrm{AdS}}{R_\mrm{s}} \til T,\quad
  \Theta =  \frac{l_\mrm{AdS}}{R_\mrm{s}} \til\Theta,
\end{align}
in terms of which Eqn. \ref{eqn:hyp-metric} becomes
\begin{align}
  \label{eqn:btz-metric}
  \diff s^2_\text{BTZ} =
  -( R^2 - R_\mrm{s}^2) \diff T^2 +
  \frac{l_\mrm{AdS}^2}{ R^2 - R_\mrm{s}^2} \diff   R^2 +
   R^2 \diff  \Theta^2,  
\end{align}
with the BTZ identification  $ \Theta \sim  \Theta +
2\pi$. These Schwarzschild-like coordinates cover the right
wedge $ R > R_\mrm{s}$ of the black hole, which may as usual
be extended to a two-sided geometry with left and right
asymptotic $\mrm{AdS}_3$ regions separated by the horizon
(Fig. \ref{fig:btz}). The black hole mass was given in
Eqn. \ref{eqn:btz-mass}.

The states of a string in BTZ are  given by the $J^2_0
+ \bar J^2_0$ orbifold of the $\mrm{SL}(2,\reals)_k$ spectrum
(Eqn. \ref{eqn:sl2r-spectrum}) with
the projection $e^{2\pi i
    R_\mrm{s}/l_\mrm{AdS}( J^2_0 + \bar J^2_0)} =1$, combined
with the twisted sectors that wind the compactified $\Theta$
cycle \cite{Natsuume:1996ij,Hemming:2001we,Hemming:2002kd}.
As in Eqn. \ref{eqn:hyp-g}, one therefore 
chooses a basis of $\mrm{SL}(2,\reals)_k$ that diagonalizes
$J^2_0,\bar J^2_0$, which
have continuous spectrum. The spectrum of the 
Hamiltonian $\frac{R_\mrm{s}}{l_\mrm{AdS}} (J^2_0 -  
\bar J^2_0)$ with respect to the Schwarzschild time $T$ is
likewise continuous, as  
is expected from the bulk effective field theory in the black hole
background in the $M_\mrm{p} \to \infty$  limit. 

Note that
in BTZ there is no simple relationship
between the perturbative string states and BCFT primary
operators, in contrast to the vacuum theory.
In the latter case, 
for example, the lowest-weight string state
$\ket{j,j,j} \in \what D_j^+ \otimes \what D_j^+$ is prepared on
the worldsheet by inserting $\Phi_j(z=0; x = 0)$
(Eqn. \ref{eqn:string-lowest}). It is dual to the BCFT Virasoro primary
state $\ket{j,j,j} \in D_j^+ \otimes D_j^+$, likewise
transforming in the lowest-weight discrete-series representation
of the boundary global conformal algebra, and itself prepared by
inserting the dual operator $\hat \Phi_j(x = 0)$ at the origin
of the boundary hemisphere. Both states
are eigenstates of the respective bulk and boundary
Hamiltonians, global $\mrm{AdS}_3$ time translations being
generated by $J^3_0+\bar J^3_0$, which maps to $\sL_0 +
\bar {\sL}_0$ in the dual. In BTZ, by contrast, a boundary local
operator insertion does not prepare an eigenstate of the bulk
Hamiltonian $J^2_0 - \bar J^2_0$, and one should not expect a simple relationship
between bulk string states and BCFT primaries.

The BTZ string theory is again defined by continuation from its
Euclidean counterpart.
Setting $\til T_\mrm{E} = i \til T$ in Eqn. \ref{eqn:hyp-metric}
defines the Euclidean BTZ black hole (EBTZ):
\begin{align}
  \label{eqn:ebtz}
  \diff s^2_\text{EBTZ} = l_\mrm{AdS}^2 \lp  \sinh^2(\til R)
  \diff \til T_\mrm{E}^2 + \diff \til R^2 + \cosh^2(\til R)
  \diff \til\Theta^2\rp.    
\end{align}
Near $\til R=0$, the metric $\til R^2 \diff \til T_\mrm{E}^2 + \diff
\til R^2 + \diff \til \Theta^2+\cdots$ describes the plane in polar
coordinates times a circle, and the angle
$\til T_\mrm{E}$ must be $2\pi$ periodic to obtain a smooth solution;
the near-horizon Lorentzian geometry is then
$\mrm{Rindler}\times \mrm{S}^1$.
Eqn. \ref{eqn:hyp-g} is likewise invariant under $\til T
\to \til T + 2\pi i$, as $e^{\pm 2\pi \mrm{T}_2} = -1_{2\times 2}$.
The
Euclidean Schwarzschild time $ T_\mrm{E}$ is
periodic in $\beta = 2\pi l_\mrm{AdS}/R_\mrm{s}$, which
is identified as the inverse Hawking temperature of the black
hole in $\mrm{AdS}$ units.

\begin{figure}[t]
  \centering
  \begin{subfigure}[t]{.46\textwidth}
    \centering
    \ig{0cm}{width=.6\textwidth}{black-hole}
    \caption{
      \footnotesize
      \bd{The BTZ Black Hole}. The BTZ black hole is the
      two-sided, asymptotically $\mrm{AdS}_3$ solution of
      three-dimensional gravity with negative cosmological
      constant. It is a quotient of $\mrm{AdS}_3$, obtained by
      compactifying $\til\Theta \sim \til\Theta + 2\pi
      R_\mrm{s}/l_\mrm{AdS}$ in the coordinates of
      Eqn. \ref{eqn:hyp-g}. The $\til\Theta$ circle is suppressed in
      the figure.
      The geometry is time-dependent, but $\mrm{Z}_2$ symmetric
      with respect to the dashed line. This line is the wormhole
      between the left and right causally disconnected
      regions. It has the topology of an annulus, with the two
      circle boundaries corresponding to slices of the
      asymptotic $\mrm{AdS}_3$ boundaries.}  
  \label{fig:btz}
\end{subfigure}
\hspace{1cm}
\begin{subfigure}[t]{.46\textwidth}
  \centering
  \ig{1cm}{width=.6\textwidth}{half-torus}
  \caption{\footnotesize\bd{The Hartle-Hawking Cap}. The Euclidean
    continuation of the BTZ black hole is a solid torus, whose
    contractible cycle is identified with the Euclidean
    time. Slicing the torus in half across this cycle produces a
    manifold in the shape of a halved bagel. Its annular
    boundary is identical to the zero-time slice of the
    Lorentzian black hole, indicated by the dashed line on the
    left. Gluing the Euclidean cap to this slice prepares the
    Hartle-Hawking state in the black hole background
    (Fig. \ref{fig:ads-hh}), whose 
    reduced density matrix in a single wedge is a thermal state
    of inverse temperature $2\pi l_\mrm{AdS}/R_\mrm{s}$. The
    purple boundary of the torus prepares the TFD state of the
    BCFT on the two circle boundaries of the annulus
    (Fig. \ref{fig:ads-hh-tfd}).}  
    \footnotesize
    \label{fig:hh}
  \end{subfigure}
  \caption{}
\end{figure}

EBTZ at inverse temperature $\beta$ is therefore a solid torus,
with contractible cycle 
$\til T_\mrm{E} \sim \til T_\mrm{E} + 2\pi$ and non-contractible cycle
$\til \Theta \sim \til \Theta + 4\pi^2/\beta$.
It is identical to the $\mrm{TAdS}_3|_{\til \beta}$ solid torus
at inverse temperature $\til\beta = 2\pi R_\mrm{s}/l_\mrm{AdS} =
4\pi^2/\beta.$ When referring to $\mrm{TAdS}_3$, however, it is
the non-contractible cycle that one identifies with the
Euclidean time as in the previous sub-section.
Whereas cutting the solid torus in half across its
non-contractible cycle prepared the TFD state in two
disconnected copies of $\mrm{AdS}_3$ (Fig. \ref{fig:ads-tfd}),
cutting across its contractible cycle produces
a Euclidean cap in the shape of a halved bagel, which
prepares the HH state in the connected black hole
(Figs. \ref{fig:hh} and \ref{fig:ads-hh})
\cite{Hartle:1976tp,Hartle:1983ai,Maldacena:2001kr}.

The zero-time slice of the two-sided black hole is a wormhole 
passing between the left and right asymptotic $\mrm{AdS}_3$
regions. Topologically, it is 
an annulus, with the two circle boundaries
corresponding to slices of the two $\mrm{AdS_3}$ boundary
cylinders.
The black hole is $\mrm{Z}_2$ symmetric with respect to this
zero-time slice, and
in preparing the HH state the annulus is
glued to the halved EBTZ torus, whose annular boundary is
identical.

Alternatively, the HH state may be thought of as a TFD state
with respect to the left and right 
wedges of the black hole, entangled by the Euclidean cap that
evolves between them in angular time (Fig. \ref{fig:vacuum}).
The reduced density matrix in a single wedge is a thermal state
at inverse temperature $\beta=2\pi l_\mrm{AdS}/R_\mrm{s}$. In
this sense, both 
slicings of 
the $\mrm{TAdS}_3$/EBTZ solid torus prepare TFD states, the
distinction being that in the first case the spacetime is
disconnected while in the second it is connected.

The worldsheet theory for a string in BTZ in the HH state is 
then similarly obtained by continuation from the
$\frac{4\pi^2}{\beta} \bbZ \bs 
\mrm{SL}(2,\bbC)_k/\mrm{SU}(2)$ orbifold, but with the
continuation now performed with respect to the contractible
cycle \cite{Hemming:2002kd,Hemming:2001we,Natsuume:1996ij}.
The equivalence of $\mrm{EBTZ}|_\beta$ and
$\mrm{TAdS}|_{4\pi^2/\beta}$ shows that the $J^3_0+\bar J^3_0$
and $J^2_0 + \bar J^2_0$ orbifolds are identical in the
Euclidean case, and we may proceed as in the previous
sub-section but with the temperature inverted.

Thus, the projection to the untwisted sector of the orbifold is 
\begin{align}
  e^{i\frac{4\pi^2}{\beta}(J^3_0+\bar J^3_0)} \Phi_j(z, \bar z; x, \bar x)
  e^{-i\frac{4\pi^2}{\beta} (J^3_0 + \bar J^3_0)} = \Phi_j(z,
  \bar z; x, \bar x). 
\end{align}
After the boundary conformal transformation $x =
e^{\frac{2\pi}{\beta} ( \Theta
+ i T_\mrm{E})}$, the projection demands periodicity in
$\Theta$,
\begin{align}
  \Phi_j\lp z, \bar z;  T_\mrm{E},\Theta +
  2\pi\rp
  =\Phi_j\lp z, \bar z;  T_\mrm{E}, \Theta \rp,
\end{align}
and likewise for $\Phi_{jw}^{J\bar J}(z, \bar z;  T_\mrm{E},
 \Theta).$ 
Their correlation functions in the orbifold are again
obtained by summing over images, and their string amplitudes
compute $\mrm{T}^2$ correlation functions of the BCFT for $\beta
< 2\pi$. For example,
\begin{align}
  \big\langle \hat \Phi_j( T_\mrm{E,1}, \Theta_1)&\hat
  \Phi_j(T_\mrm{E,2}, \Theta_2)\big\rangle\\
  &\propto
  \sum_{n \in \bbZ} \lc
  \cosh\lp \frac{2\pi}{\beta} \lp  \Theta_{12}+
  2\pi n 
  \rp\rp - \cos \lp \frac{2\pi}{\beta} T_\mrm{E,12}\rp
    \rc^{-2j} \nt.
\end{align}
Cutting the torus at $T_\mrm{E} = 0$ and gluing in the
Lorentzian cylinder prepares the thermal state. Continuing both
operators gives their thermal expectation value
\cite{Keski_Vakkuri_1999,Birmingham_2003,Maldacena:2001kr}
\begin{align}
  \mrm{tr} \big( e^{-\beta H}\hat \Phi_j( T_1, \Theta_1)&\hat
  \Phi_j( T_2, \Theta_2)\big)\\
  &\propto
  \sum_{n \in \bbZ} \lc
  \cosh\lp \frac{2\pi}{\beta} \lp  \Theta_{12}+
  2\pi n 
  \rp\rp - \cosh \lp \frac{2\pi}{\beta} T_{12}\rp
    \rc^{-2j} \nt.
\end{align}
Or, making cuts at both $T_\mrm{E} = 0$ and $\beta/2$ and
e.g. continuing an operator to each side,\footnote{We
  flip the sign of $ T_\mrm{L}$ here because the 
Schwarzschild time Killing vector points in opposite directions
on the left and right sides of the black hole.} 
\begin{align}
  \bra{\mrm{TFD}}
  (1_\mrm{L}&\otimes \hat \Phi_j( T_\mrm{R,1},
  \Theta_1))
  (\hat\Phi_j(T_\mrm{L,2}, \Theta_2)\otimes
    1_\mrm{R})
  \ket{\mrm{TFD}}\\ 
  &\propto
  \sum_{n \in \bbZ} \lc
  \cosh\lp \frac{2\pi}{\beta} \lp  \Theta_{12}+
  2\pi n 
    \rp\rp + \cosh \lp \frac{2\pi}{\beta}
    ( T_\mrm{R,1} +  T_\mrm{L,2})\rp
    \rc^{-2j} \nt.
\end{align}

As in Eqn. \ref{eqn:vacuum-fourier}, one may Fourier transform
the boundary position space basis of vertex operators to obtain
operators of definite BTZ energy and angular momentum. For example,
\begin{align}
  \label{eqn:btz-fourier}
  \Phi_{jK \bar K}(z, \bar z)
  \propto
  \int\limits_{-\infty}^\infty \diff T
  \int\limits_{0}^{2\pi} \diff \Theta\,
  e^{-i\frac{2\pi}{\beta}(K - \bar K) T} e^{-i
  \frac{2\pi}{\beta}(K + \bar K) \Theta}
  \sum_{n \in \bbZ}\,   
  \Phi_j(z,\bar z; T,\Theta+2\pi n),
\end{align}
where $\frac{2\pi}{\beta}(K\mp \bar K)$ are the energy and
angular momentum.
Examples of string amplitudes  in this basis are computed in
\cite{Hemming:2002kd}. Absent the sum over images $n$, one would
instead obtain a mode of $\mrm{AdS}_3$-Rindler.

\subsection{The Two-Dimensional Black Hole}
\label{sec:2d-bh}
Finally, we briefly discuss string perturbation theory in the
two-dimensional black hole in the HH state.
The two-dimensional Euclidean black hole (Eqn. \ref{eqn:cigar-bkgd})
followed from
$\mrm{SL}(2,\reals)_k$ (or $\mrm{SL}(2,\bbC)_k\ab/\mrm{SU}(2)$) by 
gauging the $J^3_0 + \bar J^3_0$ isometry that generates
translations along the length of the cylinder
(Eqn. \ref{eqn:cigar-action}).  The Lorentzian black hole is
then obtained by continuing in the compact coordinate
$\theta = i t$
(Eqn. \ref{eqn:lorentzian-metric}). Alternatively, recalling the
BTZ coordinates on 
$\mrm{SL}(2,\reals)$ describe an $\mrm{AdS}_3$-Rindler patch
(Eqns. \ref{eqn:hyp-g}-\ref{eqn:hyp-metric}), likewise related to
$\mrm{EAdS_3}$ by continuation in the compact cycle, one may arrive
directly at the two-dimensional Lorentzian black hole by gauging
the $J^2_0+\bar J^2_0$ symmetry of $\mrm{SL}(2,\reals)_k$ that
generates translations in $\Theta$.
Then whereas the two-dimensional Euclidean black hole spectrum
(Eqn. \ref{eqn:coset-spectrum}) followed from the
$\mrm{SL}(2,\reals)_k$ spectrum (Eqn. \ref{eqn:sl2r-spectrum})
by the coset construction that gauged $J^3_0 + \bar J^3_0$, the
Lorentzian spectrum is obtained by the coset construction with
respect to $J^2_0 + \bar J^2_0$. 

String perturbation theory in the two-dimensional black hole in
the HH state is again defined by continuation from the Euclidean
black hole. For example, continuing the
Euclidean vertex operators $\cO_{jn,w=0}$ with $j = \frac{1}{2}
(1 + i s)$ by sending $n \to i
E$ yields scattering states in the right
wedge.\footnote{Here $E$ is the energy measured in units of
  $1/\sqrt{\alpha' 
  k}$, and is conjugate to $t = -i \theta$. The proper time in
the large $r$ limit of the metric
(Eqn. \ref{eqn:lorentzian-metric}) is $-\alpha' 
k \diff t^2$.} 
From Eqn. \ref{eqn:asymptotic} one obtains
\begin{align}
  \label{eqn:bh-scatterers}
  \cO_{jE} \overright{r \to \infty}
  \lp e^{-2(1-j)r} + R(j,E) e^{-2jr} \rp
  e^{- i E t},
\end{align}
with wavefunction
\begin{align}
  \Psi_{jE} \underset {\propto} {\overright{r \to \infty}}
  \lp e^{is r} + R(j,E) e^{-is r} \rp 
  e^{- i E t},
\end{align}
describing an incoming particle in the right wedge that scatters
off the black hole 
horizon. One may likewise construct by continuation vertex
operators describing 
outgoing modes, as well as similar modes in the left wedge, and modes
behind the past and future horizons \cite{Dijkgraaf:1991ba}.

In the Euclidean theory, one found normalizable bound states at
poles of
the reflection coefficient. Now $R(j,E)$ is non-singular
on the real
branch, and one finds only these delta-function normalizable
scattering states on the complex branch.\footnote{As in Rindler,
  the modes are singular at the  
horizon, $\Psi_{jE} \propto r^{-i E}e^{-iEt} (1 +
\cO(r^2))$.}
These modes do not form a complete set, however. The
reason is tied to non-unitarity of the scattering
matrix---strings can fall behind the horizon. One can close the
OPE by including additional modes, including the operators with
$w \neq 0$, but at the cost of sacrificing mutual locality of
the vertex operators.

String amplitudes of such scattering operators in the HH state may be
obtained by continuation from the corresponding Euclidean
amplitudes. 
The Schwinger-Keldysh 
contour consists of the two Euclidean caps obtained by halving
the cigar (Fig. \ref{fig:half-cigar}), glued to the zero-time
slice of the Lorentzian black hole
(Fig. \ref{fig:black-hole}), which evolves forward and
backward in Lorentzian time.
The continuation is more technically challenging
than in $\mrm{AdS}_3$ where one simply continued the boundary
insertion point in $x$-basis worldsheet vertex operators. In
momentum basis, one instead computes the Euclidean amplitude as
a function of the discrete Matsubara frequencies $n$, and then
continues the result to continuous Lorentzian energies $E$
\cite{Dijkgraaf:1991ba}. The analogous objects in three
dimensions are the Fourier modes Eqns. \ref{eqn:vacuum-fourier},
\ref{eqn:btz-fourier}.

\section{Stringy $\mathrm{\mathbf{ER = EPR}}$}
\label{sec:erepr}

Having described the formulation of Lorentzian
string perturbation theory
in various states by continuation from unitary CFTs, we now
consider the string theories obtained by continuation from the
dual descriptions of the 
$\mrm{SL}(2,\bbC)_k/\mrm{SU}(2)$, $\bbZ \bs
\mrm{SL}(2,\bbC)_k/\mrm{SU}(2)$, and
$\mrm{SL}(2,\reals)_k/\mrm{U}(1)$ CFTs. Each of the three
Euclidean dualities shares the essential
feature that the Euclidean time circle is contractible in one
description and non-contractible in the other, with Euclidean
time winding conservation violated in the latter case by a
condensate of winding strings.
Upon continuation then, each gives a string duality realizing
$\mrm{ER = EPR}.$ On one side is string theory in a connected
spacetime with a horizon in its Hartle-Hawking state,
and on the other is  string theory in a disconnected
union of entangled spacetimes in the thermofield-double state. The
principal remaining  challenge in 
formulating these continuations is the Lorentzian interpretation
of the Euclidean time winding operators that play a critical
role in each of the examples. We will argue that these
insertions should be treated in angular quantization on the worldsheet---with a
corresponding deformation of the moduli space integration
contour where necessary---giving rise to a condensate of pairs of
entangled, 
folded strings  emanating from the strong-coupling region.

\subsection{Two-Dimensional Dilaton-Gravity}
\label{sec:2d-erepr}

We begin with the $\mrm{SL}(2,\reals)_k/\mrm{U}(1)$ CFT. As we
have reviewed in the previous sections, for large $k$ this CFT
admits a weakly-coupled description given by a string in the
cigar-shaped Euclidean black hole of two-dimensional
dilaton-gravity, with an asymptotically linear dilaton
(Eqn. \ref{eqn:cigar-S}).
Cutting the cigar across its contractible $\theta$ cycle and
continuing yields
the conventional description of
a string in the 2D Lorentzian black hole in the HH state
\cite{Dijkgraaf:1991ba}. String amplitudes computing the
$S$-matrix of particles scattering off the black hole horizon
may be obtained by continuation from the Euclidean amplitudes of
vertex operators $\cO_{jn,w=0}$ under $n \to i E$, where $n$ is
the discrete Matsubara frequency of the mode around the compact
Euclidean time circle and $E$ is the continuous Lorentzian
energy. This is the ER description of the
string theory.

We now turn to the EPR description, corresponding to the string
background obtained by continuation 
from the dual, sine-Liouville description of
$\mrm{SL}(2,\reals)_k/\mrm{U}(1)$ 
(Eqn. \ref{eqn:sl}). The sine-Liouville background consists of
the free $\text{linear-dilaton}\times \mrm{S}^1$ background
(Eqn. \ref{eqn:linear-dilaton}) plus the sine-Liouville
potential $4\pi \lambda (W_+ +W_-)$, where $W_\pm$
are the marginal $\text{linear-dilaton}\times
  \mrm{S}^1$ operators with winding $\pm 1$ 
(Eqn. \ref{eqn:wpm}). It is a strongly-coupled description of
the CFT at large $k$.

In the asymptotic region $r, \hat r \to \infty$ of the cigar and
sine-Liouville, the two backgrounds are identical,
with the coordinates related by
Eqn. \ref{eqn:canonical-coordinates}. In that limit, the
$\mrm{SL}(2,\reals)_k/\mrm{U}(1)$ Virasoro primaries $\cO_{jnw}$
behave as the superposition of $\text{linear-dilaton}\times
\mrm{S}^1$ primaries given in Eqn. \ref{eqn:asymptotic}. To
compute an $\mrm{SL}(2,\reals)_k/\mrm{U}(1)$ correlation
function of such operators in the sine-Liouville description,
one would insert the corresponding free-field
superpositions in the sine-Liouville functional
integral. 

Let us consider the
sine-Liouville potential as a large
deformation of the $\text{linear-dilaton} \times \mrm{S}^1$ background,
expanding the condensate in powers of the winding operators,
\begin{align}
  \label{eqn:sl-expansion}
  &\vev{e^{-\frac{\lambda}{2\alpha'} \int \diff^2 z\, (W_+ +
    W_-)} \cdots}_\mrm{LD\times S^1}\\[.3cm]
  &\hspace{.5cm}=
  \sum_{N =0}^\infty \frac{1}{(N!)^2}
  \lp\frac{\lambda}{2\alpha'} \rp^{2N}
  \vev{
  \lp \int \diff^2z\, W_+(z,\bar z) \rp^N
  \lp \int \diff^2z'\, W_-(z',\bar z') \rp^N
  \cdots
  }_\mrm{LD\times S^1},\nt
\end{align}
where the ellipses stand for additional operator insertions
$\prod_i e^{-2Q(1-j_i)\hat r} e^{i n_i \theta}$.
Note that 
the winding conservation law of the free background demands that an equal number of $W_+$ and
$W_-$ factors contribute in each term.\footnote{We restrict our attention
  here to non-winding $\mrm{S}^1$ primaries that, with $j =
  \frac{1}{2}(1+i s)$, continue to ordinary scattering
  states of the black hole.}
We wish to understand the Lorentzian string theory defined by
continuation from each term in this expansion, which we refer to
as the EPR microstate string backgrounds.\footnote{Note,
  however, that these continued backgrounds describe the
  thermally entangled microstates. We will not discuss string
  backgrounds for the pure EPR microstates, though it would be
  interesting to do so.}

Note that Eqn. \ref{eqn:sl-expansion} is not a perturbative expansion
around the $\text{linear-dilaton}\times \mrm{S}^1$ background, however.
As recalled in Sec. \ref{sec:sl}, because $\lambda$ may be
rescaled by field 
redefinitions, there is no sense in which it is a small
parameter. As a result, sine-Liouville correlation functions are
not in general analytic in $\lambda$, but rather scale with
$\lambda^\kappa$ as in Eqn. \ref{eqn:sl-zero}, where $\kappa$ is the
function of the primary momenta $\{j_i \}$ given in
Eqn. \ref{eqn:kappa}. In the special case that $\kappa \in
2\bbN$, however, one does obtain an analytic
function\footnote{Or, rather, the residue at the pole of the
  gamma function is an analytic function of $\lambda$.}
proportional to
$\lambda^{\kappa}\Gamma(-\kappa) \langle\lp \int W_+\rp^{\kappa/2}
\lp \int W_-\rp^{\kappa/2} \cdots \rangle_{\mrm{LD} \times
  \mrm{S}^1,\fs{0}}$.

Correspondingly, for compatible values of the momenta $\{j_i \}$,
one finds in Eqn. \ref{eqn:sl-expansion} a single term
consistent with the 
anomalous momentum conservation law of the linear dilaton
(Eqn. \ref{eqn:anomaly}), 
\begin{align}
  \label{eqn:sl-compatible}
  2Nb_\mrm{sL}+ Q\sum_i (1-j_i) = \frac{1}{2} Q \chi.
\end{align}
This is simply the condition that $\kappa = 2N$, reproducing
the preceding result. The zero-mode integral collapses to $\int
\diff \hat r_0$, which diverges with the volume of the target and is
reflected in the pole of the gamma function. The free-theory
correlators therefore compute the residues of sine-Liouville
correlation functions at these poles \cite{Fukuda_2001}.
In principle the correlation
functions for general momenta could be obtained by
``continuation'' from these residues computed from the free
theory, by determining the meromorphic function with the
corresponding pole structure, as in Liouville
\cite{Dorn:1994xn,Zamolodchikov:1995aa,Teschner:2001rv,Di_Francesco_1992}.

For generic values of $\{j_i\}$, including the
scattering states of interest for the black hole, the anomalous
conservation law need 
never be satisfied, and each $\text{linear-dilaton}\times
\mrm{S}^1$ correlation function on the right-hand-side of 
Eqn. \ref{eqn:sl-expansion} appears to vanish.  
Yet sine-Liouville admits no such anomalous conservation law, the
translation symmetry of the target linear-dilaton direction
being completely broken by the potential, and exact CFT correlation
functions of operators that violate Eqn. \ref{eqn:sl-compatible}
certainly need not vanish.

This puzzle is familiar from analogous manipulations in
Liouville CFT
\cite{Dorn:1994xn,Zamolodchikov:1995aa,Teschner:2001rv}.
The source of the trouble is the strong-coupling
region.
Namely, Eqn. \ref{eqn:sl-expansion} should be interpreted
as defining a perturbative
expansion of the sine-Liouville measure in powers of
$\lambda e^{-2b_\mrm{sL} \hat 
  r_0}$, where $\hat r_0$ is the zero-mode of $\hat r$.
Whereas $\lambda$ itself is not a small parameter, the
combination is
invariant under $\hat r_0 \to \hat r_0 + 
\delta$, $\lambda \to e^{2b_\mrm{sL}\delta}\lambda,$ and gives a
perturbative expansion about the weak-coupling region.

To proceed more carefully, one should introduce a regulator
that controls the strong-coupling region. In the
zero-mode integral Eqn. \ref{eqn:zero-mode-integral} that lead
to Eqn. \ref{eqn:sl-zero}, one could introduce a hard cut-off
$\hat r_\mrm{c}$ on the lower bound of the integral and attempt to
understand the limit as $\hat r_\mrm{c} \to -\infty$.
The regulator
will break the anomalous momentum conservation law of the free
background, eliminating the spurious constraint
Eqn. \ref{eqn:sl-compatible}. 
Alternatively,
rather than this hard cut-off step function $\Theta(\hat r_0 -
\hat r_\mrm{c})$, one could employ a soft cut-off by inserting
$\exp\lp -e^{-2b_\mrm{sL}(\hat r_0 - \hat r_\mrm{c})}\rp$ in the
integral, which
behaves similarly but varies smoothly. Then
Eqn. \ref{eqn:zero-mode-integral} becomes
\begin{align}
  \int\limits_{-\infty}^\infty \diff \hat r_0~
    e^{2b_\mrm{sL} \kappa  \hat r_0
  -\lp \frac{\lambda}{\alpha'} V_\mrm{sL}[\hat r', \hat \theta] 
  +e^{ 2b_\mrm{sL}\hat r_\mrm{c}}\rp
  e^{-2b_\mrm{sL} \hat r_0}}
  =&\frac{1}{2b_\mrm{sL}}
     \lp \frac{\lambda}{\alpha'} V_\mrm{sL}[\hat r', \hat
     \theta] +e^{ 2b_\mrm{sL}\hat r_\mrm{c}}\rp^\kappa
     \Gamma \lp -\kappa \rp.
\end{align}
Writing $\vep = e^{2b_\mrm{sL}\hat r_\mrm{c}}$, such that $\vep
\to 0$ when $\hat r_\mrm{c} \to -\infty$ meaning that the
regulator is removed, the regulated version of
Eqn. \ref{eqn:sl-zero} is 
\begin{align}
  &\vev{\prod_N  e^{-2Q(1-j_N) \hat r}\cS_N}_\mrm{sL,\vep}\\
  &\hspace{1cm}=
     \frac{1}{2b_\mrm{sL}}
     \Gamma \lp -\kappa \rp
    \vev{\lp \frac{\lambda}{\alpha'} V_\mrm{sL}[\hat r, \hat
     \theta]+ \vep\rp^\kappa \prod_N  e^{-2Q(1-j_N) \hat r}\cS_N}_\mrm{LD\times
     S^1,\fs{0}}.\nt
\end{align}
One may now attempt to expand the free-field correlator
in powers of $\frac{1}{\vep}$ by writing
\begin{align}
  \label{eqn:binomial}
  \lp  \frac{\lambda}{\alpha'} V_\mrm{sL}+ \vep\rp^\kappa
  =
  \vep^\kappa 
  \sum_{M = 0}^\infty {\kappa\choose M} \lp
  \frac{\lambda}{\alpha'}\frac{V_\mrm{sL}}{\vep}\rp^M ,
\end{align}
where ${\kappa\choose M}$ is the generalized binomial
coefficient.
In this way, one more properly obtains as in Eqn. \ref{eqn:sl-expansion}
an expansion for sine-Liouville correlation functions as a sum over
free-theory correlators with integer powers of the integrated potential
inserted.
This expansion may diverge, in
general. The situation is similar to conformal perturbation
theory, where one expands an exactly marginal deformation
$e^{-\lambda \int \cO}$,  obtaining a series with finite radius
of convergence. In that case the expansion is suppressed by
factors of $\frac{1}{M!}$, whereas the binomial coefficients in
Eqn. \ref{eqn:binomial} fall off less rapidly. Thus,
the above expansion may only be an asymptotic
series, which we speculate may be Borel resummable.
We have not attempted to verify this, however.

We will not
pursue further the explicit implementation of the regulator
here. Our goal is not to offer a new computational
framework for
obtaining string amplitudes in the black hole, but to give an
abstract understanding of the string backgrounds corresponding
to the thermal EPR microstates. 
To that end,
we now consider the Lorentzian continuation of the string
background defined by each term in Eqn. \ref{eqn:sl-expansion}.

Consider first the free $\text{linear-dilaton}\times \mrm{S}^1$
itself, i.e. $N = 0$.
This is the flat space solution of the dilaton-gravity equations
of motion (Eqn. \ref{eqn:2d-eom}).
The target cylinder has the topology of
an annulus, which, when halved and glued to its Lorentzian
continuation with respect to the $\mrm{S}^1$, prepares a
Schwinger-Keldysh contour for the disconnected union of two
copies of $\text{linear-dilaton} \times \mrm{time}$ in the TFD
state (Fig. \ref{fig:tfd}).

As in the Euclidean background, each Lorentzian spatial slice
extends from a weak-string-coupling limit at $\hat r \to \infty$
to a strong-coupling region $\hat r \to -\infty$. The two
asymptotic weak-coupling regions are identical to the left and
right asymptotic regions of the two-sided black hole. But
whereas the left and right regions of the black hole are
connected in the interior at the horizon, the two copies of
$\text{linear-dilaton}\times \mrm{time}$ are disconnected, with
strong-coupling boundaries in their interiors instead of
horizons.

Each remaining term in the expansion Eqn. \ref{eqn:sl-expansion}
inserts $N$ pairs of $W_+, W_-$ operators on top of the
$\text{linear-dilaton}\times \mrm{S}^1$ background. Thus, upon
continuation they
will introduce deformations of the TFD state in the
disconnected union of $\text{linear-dilaton}\times
\mrm{time}$. Because the $W_\pm$ are winding operators around
the Euclidean time circle, however, the Lorentzian
interpretation of these deformations is not immediately obvious.

Moreover, there is a problem of mutual locality that arises in
attempting to continue $n \to i E$ in the insertions represented
by the ellipses in Eqn. \ref{eqn:sl-expansion}.
Namely, the winding and momentum operators of
the compact boson obey an OPE
\begin{align}
  \label{eqn:momentum-winding-ope}
  e^{i n \theta(z,\bar z)} 
e^{\pm i k \til \theta(0)}
  \sim
  \lp \frac{z}{\bar z} \rp^{\pm n/2}
  e^{i n \theta(0) \pm i k \til \theta(0)} .
\end{align}
When the momentum operator circles the winding
insertion at the origin ($z \to e^{2\pi i}z, \bar z \to e^{-2\pi
i} \bar z$), the OPE coefficient transforms by a factor
$e^{\pm 2\pi i n}$. Then the operator algebra is well-defined
only for $n \in \bbZ$, which is of course the expected momentum
quantization of the compact boson. It follows that
by continuing the momentum 
labels $n \to i E$ in a $\text{linear-dilaton}\times \mrm{S}^1$
correlator with a given number of $W_\pm$ insertions, one will
in general obtain a multi-valued function on the worldsheet.
In order to define string perturbation theory in the EPR
microstates 
obtained by continuation from each term in the expanded background,
we
must establish how to compute a well-defined string amplitude
by integrating such an apparently multi-valued expression over
the moduli space.

We address these questions in the following two sub-sections. 
We first show that each pair of $W_+,W_-$ insertions introduces
on top of the disconnected $\text{linear-dilaton}\times \mrm{S}^1$ 
background a pair of strings in a TFD state in the sense of
angular quantization on the worldsheet
(Fig. \ref{fig:angular-tfd}). Each string is folded, 
with its ends in the strong-coupling region, and each folded
string is entangled with its pair, with one in the left and one
in the right copy of the spacetime (Fig. \ref{fig:tfd}). We then
argue that the multi-valued correlation functions obtained by
continuation from the  $\text{linear-dilaton}\times \mrm{S}^1$
should be integrated over a deformed contour in a
complexification of the string moduli space on which they are
single-valued in order to obtain 
amplitudes in the background of such entangled, folded
strings.

\subsubsection{Angular Quantization}
\label{sec:angular-quantization}

As recalled in the introduction, one sometimes obtains 
a target space picture in string theory by
adopting the ``static'' gauge after choosing a flat metric on
the worldsheet,
in order to fix the residual
conformal gauge redundancy that remains within the full
$\mrm{diffeomorphism\times Weyl}$ 
gauge redundancy of the functional integral. In static gauge, the
target Euclidean time coordinate is fixed to the worldsheet time
coordinate $\rho$ in the sense of radial quantization, where $z
= e^{\rho + i \phi}$.
This choice is unacceptable 
in the background of Euclidean time
winding operators, however.
For example, in the neighborhood of a winding  $\pm 1$ insertion
at the origin on 
the worldsheet, the target Euclidean time $\theta \sim \theta +
2\pi$ should obey
\begin{align}
  \label{eqn:winding}
  \theta \overright{\rho \to -\infty}   \pm \phi.
\end{align}
Rather than the static gauge $\theta = \rho$,
one may instead adopt an angular gauge condition
$\theta = \pm \phi$, in which the compact
coordinate $\phi$ is viewed as the worldsheet Euclidean time
direction. 

We are therefore motivated to treat the neighborhood of
Euclidean time winding insertions on the worldsheet in angular
rather than the usual radial quantization. 
In this section we elaborate on this quantization scheme.
In fact, the following is a pure CFT discussion, 
independent of the application to Lorentzian string theory that
we ultimately have in mind, and which we return to at the end of
the section.

\begin{figure}[t]
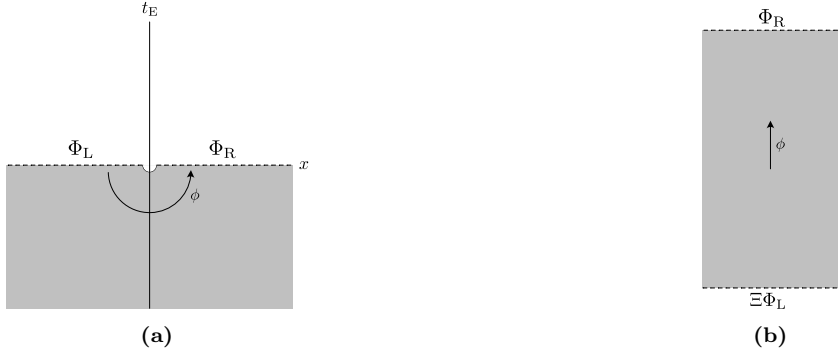

  \begin{subfigure}[t]{.43\textwidth}
    \centering
    \ig{0cm}{scale=1.5}{vacuum}
    \caption{}
    \label{fig:vacuum}
  \end{subfigure}
  \hspace{1cm}
  \begin{subfigure}[t]{.43\textwidth}
    \centering
    \ig{0cm}{scale=1.5}{vacuum-angle}
    \caption{}
    \label{fig:vacuum-angle}
  \end{subfigure}
  \caption{\footnotesize \bd{The Minkowski Vacuum and the
      Rindler TFD}. The Euclidean functional integral over the
    lower-half plane prepares the Minkowski vacuum state on the
    $t_\mrm{E} = 0$ slice (left). The same functional integral may be
    interpreted as a Euclidean transition amplitude between 
    states on the negative and positive $x$-axis, with the
    angular direction $\phi$ interpreted as Euclidean time
    (right). Then the Minkowski vacuum is equivalent to the TFD
    state in the angularly quantized Hilbert space $\cH_\mrm{L}
    \otimes \cH_\mrm{R}.$ In the same way, the HH state for the
    two-sided black hole may be interpreted as a TFD state with
    respect to the left and right wedges. The reduced density
    matrix in a single wedge is a thermal state of inverse
    Hawking temperature $\beta_\mrm{BH}.$}
\end{figure}

The basic idea of angular quantization is familiar\footnote{See e.g.
  \cite{Harlow:2014yka,Witten_2018} for useful reviews.}
from the functional integral 
derivation of the Unruh effect in Rindler spacetime
\cite{Unruh:1983ac}. Consider a field theory on $\reals^2$, with the
flat metric $\diff s^2 = \diff t_\mrm{E}^2 + \diff x^2$.
The functional integral over the lower-half plane
prepares the 
Minkowski vacuum state $\ket{\Omega}$ in the Hilbert space of
the theory on the $x$-axis (Fig. \ref{fig:vacuum}). It is a
wavefunctional, $\Psi_\Omega[\Phi(x)]= \brak{\Phi}{\Omega}$, that
takes field data 
$\Phi(x)$ on the fixed $t_\mrm{E}$ slice and returns the value
of the functional integral over the lower-half plane with that
boundary condition. 

Let $x = e^\rho \cos(\phi),$ $t_\mrm{E} = e^\rho \sin(\phi)$ define
cylinder coordinates, in terms of which $\diff s^2 = e^{2\rho}\lp
\diff \rho^2 +  \diff \phi^2\rp.$
The lower-half plane may be foliated by radial
lines at fixed $\phi$, rather than horizontal lines at fixed
$t_\mrm{E}$. Then the
same functional integral admits an alternative Hilbert space
interpretation as a transition amplitude between a state on the
negative $x$-axis and a state on the positive $x$-axis, with 
Euclidean time evolution given by rotation in $\phi$
(Fig. \ref{fig:vacuum-angle}):
\begin{align}
  \brak{\Phi}{\Omega} =  \bra{\Phi_\mrm{R}} e^{-\pi R} \Xi\ket{\Phi_\mrm{L}}.
\end{align}

We have divided the field data $\Phi(x)$ into
$\Phi_\mrm{L}(x)$ on the negative $x$-axis and $\Phi_\mrm{R}(x)$
on the positive $x$-axis. 
$R$ generates rotations in
$\phi$, and $\Xi$ is the CPT operator which maps the Hilbert
space on the left to the Hilbert space on the
right.

Inserting a complete set of eigenstates of $R$, the transition
amplitude may be written
\begin{align}
  \bra{\Phi_\mrm{R}} e^{-\pi R} \Xi\ket{\Phi_\mrm{L}}=
  \bra{\Phi_\mrm{L}}\otimes \bra{\Phi_\mrm{R}}
  \lp \sum_{i} e^{-\pi \omega_i}  \ket{i^*}\otimes \ket{i}\rp,
\end{align}
where $\ket{i^*} = \Xi^\dg \ket{i}$. Thus, the Minkowski vacuum
$\ket{\Omega}$ on the slice $t_\mrm{E} = 0$ is equivalent to
the TFD state in $\cH_\mrm{L} \otimes \cH_\mrm{R}$:
\begin{align}
  \label{eqn:mink-tfd}
  \ket{\Omega} = \sum_i e^{-\pi \omega_i} \ket{i^*} \otimes \ket{i}.
\end{align}

More precisely, the Hilbert space on the line does not factorize
into a product of left and right Hilbert spaces due to
divergences associated to degrees of
freedom in the neighborhood of the origin. The vacuum in the
full spacetime does define a state on the von Neumann operator
algebra in both Rindler wedges.

When $\reals^2$ is continued with respect to $t_\mrm{E} = i
t$, one obtains standard coordinates on Minkowski space,
with metric
$\diff s^2 = -\diff t^2 + \diff x^2$. Then expectation values in
the vacuum state may be computed by cutting the Euclidean
functional integral on $\reals^2$ at $t_\mrm{E} = 0$, gluing in
the Minkowski plane, and continuing operator insertions 
to the Lorentzian section.

Instead continuing with respect to $\phi = i \Lt$,
one obtains the right wedge of the Rindler decomposition of
Minkowski spacetime (Fig. \ref{fig:rindler}). In particular,
$x = e^\rho \cosh(\Lt)$ and $t = e^\rho \sinh(\Lt)$, and
therefore the coordinates $\rho$ and $\Lt$ cover the region $x
> |t|$, bounded by the 
Rindler horizon. The relation between the
Rindler coordinates on 
the right wedge and the full Minkowski spacetime is analogous to
the relation between Schwarzschild coordinates on the right
wedge of a black hole and the extended two-sided black hole.
One may similarly define
Rindler coordinates in the remaining wedges.
Lines of constant $\rho$
are hyperbolas $x^2 - t^2 = \mrm{const}.$, and lines of constant
$\Lt$ are straight lines through the origin, $t/x =
\mrm{const}.$ Translation in $\Lt$ is an isometry, timelike in
the right and left wedges (though with opposite orientations), and
spacelike in the top and bottom wedges.  
It corresponds to the boost isometry in the original Minkowski
coordinates.

Because $\phi$ is $2\pi$ periodic,
when a Euclidean functional integral on $\reals^2$ is cut at
$\phi = 0$ (i.e. the positive $x$-axis) and operator insertions are
continued to the right Rindler wedge $\phi \to i \Lt$, one
obtains an expectation value in a thermal state at inverse
temperature $2\pi$. This is the Unruh 
effect: the reduced density matrix of the Minkowski vacuum in
the right (or left) Rindler wedge is a thermal state. Indeed, 
from Eqn. \ref{eqn:mink-tfd} one 
obtains $\mrm{tr}_{\cH_\mrm{L}} \lp \ket{\Omega}\bra{\Omega} \rp=
e^{-2\pi R}.$ By slicing the Euclidean functional integral at
both $\phi = 0$ and $\phi = \pi$, one may continue operators to
either the left or right Rindler wedges, and so obtain
expectation values in $\cH_\mrm{L} \otimes \cH_\mrm{R}$ in the
TFD state.

The thermal state $e^{-2\pi R}$ and its TFD purification are simple
examples of states in an angularly quantized Hilbert space.
The angular quantization that we propose in what follows
generalizes this construction by allowing operator insertions at
one or both asymptotic endpoints of the spatial slices.

\begin{figure}[t]
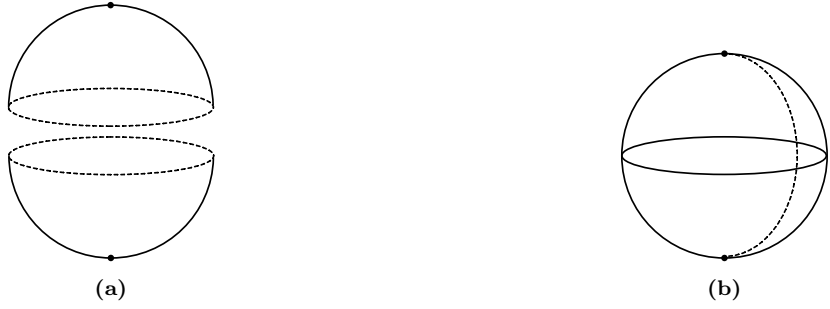

  \begin{subfigure}[t]{.43\textwidth}
    \centering
    \ig{0cm}{width=.4\textwidth}{radial-slicing}
    \caption{}
    \label{fig:radial-slicing}
  \end{subfigure}
  \hspace{1cm}
  \begin{subfigure}[t]{.43\textwidth}
    \centering
    \ig{0cm}{width=.4\textwidth}{angular-slicing}
    \caption{}
    \label{fig:angular-slicing}
  \end{subfigure}
  \caption{\footnotesize \bd{Radial and Angular Slicings}. The same
    Euclidean functional integral on a sphere may be assigned
    different Hilbert space interpretations by choosing
    different foliations. In radial quantization (left), one
    slices the sphere into circles centered at the poles. Each
    circle is a spatial slice, and Euclidean time flows along the
    radial direction transverse to the slices. Then the
    functional integral over the sphere computes the
    inner-product of the states prepared by the operator insertions at
    the poles in the Hilbert space e.g. on the equatorial
    circle. In angular quantization, the spatial slices are
    instead radial lines connecting the poles, and Euclidean
    time flows along the angular direction. Then the same
    two-point function on the sphere computes the thermal trace in the
    angularly quantized Hilbert space on the dashed line. Given
    a point on the sphere, its spatial slice in radial
    quantization is the orbit of the rotation generator, and its
    Euclidean time evolution is the orbit of the dilation
    generator. In angular quantization, the roles of the two
    symmetries are exchanged.}
\end{figure}

CFTs are usually treated in radial quantization. That is, when
one refers to the Hilbert space of a 2D CFT, one typically means
the Hilbert space $\cH(\mrm{S}^1)$ on a circle, corresponding to a  slice of the
cylinder $\mrm{S}^1 \times \bbR$ with Lorentzian time running along
its length. One has an isomorphism
between this Hilbert space and the space of local operators at a
point. 
Each local operator is mapped to a state on the circle by the
Euclidean functional integral over a hemisphere glued to the
circle slice, with the operator inserted at the pole. And each
state may be mapped to its corresponding local operator by a conformal
transformation that shrinks the circle to a point.

Thus, a Euclidean CFT correlation function on $\mrm{S}^2$, such as the two-point
function illustrated in Fig. \ref{fig:radial-slicing}, may be
cut, e.g. on the equator, and interpreted as an inner-product in
the radially quantized Hilbert space on that circle:
\begin{align}
  \vev{\cO(0)\cO'(\infty)}_{\mrm{S}^2} = \brak{\cO'}{\cO}_{\cH(\mrm{S}^1)}.
\end{align}
The Euclidean functional
integral over the southern hemisphere prepares the state
associated to the operator insertion at the south pole, and the
integral over the northern hemisphere prepares the state
associated to the operator at the north pole. By sewing up the
cut on the dashed circle, the functional integral on the sphere computes
the inner-product of these
two states. One may also insert the Lorentzian cylinder at the
cut, and by continuing additional operator insertions to the Lorentzian
section one may compute their expectation value between the two states.

In that continuation, Euclidean time evolution is defined by the dilation 
symmetry of the CFT, which scales the local complex coordinate
$z \to \lambda z$
centered at the operator insertion at the pole (and therefore translates
the cylinder time coordinate $\log |z| \to \log |z| + \log
\lambda$). The slices of the sphere at fixed Euclidean time are 
circles centered at the poles.
In this foliation,
given a point on the sphere, its spatial slice is the orbit of
the rotation generator around the poles, and its Euclidean time
evolution flows along the orbit of the dilation generator.

This foliation is not unique, however. Let us instead slice
the sphere in radial lines connecting the poles
as in Fig. \ref{fig:angular-slicing}, and define Euclidean time
evolution by rotation. In this slicing, the roles of the dilation
and rotation generators are exchanged: the spatial slice on
which a point lies is its
orbit under dilation, and Euclidean time runs along
the orbit of rotation. 

Then the Euclidean
functional integral over the sphere with the radial cut
prepares a thermal state $e^{-2\pi R}\in \cH_{\cO\cO'}(\reals)$ in the
Hilbert space  of angular quantization on the
dashed line, where the subscripts label the operator insertions
at the poles.
The same two-point function, obtained by sewing up the 
cut, is therefore assigned a
different interpretation in this quantization scheme---it is a
thermal trace in $\cH_{\cO\cO'}(\reals)$:
\begin{align}
  \brak{\cO'}{\cO}_{\cH(\mrm{S}^1)} = \mrm{tr}_{\cH_{\cO\cO'}(\reals)} \lp
  e^{-2\pi R} \rp.
\end{align}

One may continue additional operator insertions with
respect to $\phi \to i \Lt$, corresponding to gluing in a
Rindler wedge at the cut, and so obtain expectation values of
operators in Rindler spacetime in this thermal state.
And one may cut the sphere in half to define the TFD
state in two copies of $\cH_{\cO\cO'}(\reals)$
(Fig. \ref{fig:angular-tfd}). By continuing 
additional operator insertions via $\phi \to i \Lt_\mrm{R}$ or
$\phi \to \pi + i \Lt_\mrm{L}$, one obtains an expectation value in
the TFD state in the left and right Rindler wedges.

\begin{figure}[t]
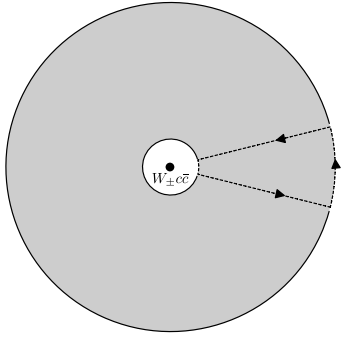

  \centering
    \ig{0cm}{scale=1.3}{brst-contour}
    \caption{\footnotesize \bd{BRST Invariance}. The physical
      vertex operator $W_\pm c \bar c$ is BRST invariant and
      prepares a BRST invariant state in the sense of radial
      quantization. The thermal state of angular quantization
      prepared with an asymptotic insertion of $W_\pm c \bar c$
      is likewise BRST invariant. With the neighborhood of the
      insertion excised from the worldsheet and replaced by the
      appropriate asymptotic conditions on the resulting
      boundary circle, the BRST current vanishes on the circle in
      the limit that it shrinks away. Because the line integral
      of the current around the dashed loop shown vanishes by
      current conservation, it follows that the BRST charges measured
      on the two radial slices are identical. }
    \label{fig:brst-contour}
  \end{figure}

One way to regulate the UV divergences associated to the
left/right split
of the Hilbert space here is to end the Rindler wedges at a
boundary with appropriate boundary conditions. In the Euclidean
functional integral this corresponds to excising a small
neighborhood 
of the insertion point of the operator being treated in angular
quantization (Fig. \ref{fig:brst-contour})
\cite{Ohmori:2014eia}. In the limit that the 
regulator is removed, the 
boundary conditions become the asymptotic conditions associated
to the inserted operator. Of course, the regulator breaks
conformal symmetry, which is only restored in the limit.

Let us now apply this machinery to the
$\text{linear-dilaton}\times \mrm{S}^1$ background with
insertions of $W_\pm$ operators in order to understand the
deformations they introduce on top of the $\reals^{1,1}\cup
\reals^{1,1}$ string theory upon continuation. 
The Hilbert space $\cH_{\cO \cO'}(\reals)$ is defined by the
asymptotic conditions imposed by the insertions $\cO,\cO'$
at
the ends of the line. For a $W_\pm$ insertion, these asymptotic
conditions  follow from Eqn. \ref{eqn:ld-circle-green} with
$\alpha = b_\mrm{sL}$,\footnote{When the operator insertion
  coincides with a curvature 
singularity, the asymptotic condition is modified by a
contribution from the linear-dilaton background charge. But the
correction is sub-leading in the large $k$ limit because $Q \sim
\frac{1}{\sqrt{k}}$.}
\begin{align}
  \label{eqn:ld-asymp-cond}
  \hat r  \overright{\rho \to -\infty} \sqrt{\alpha' (k-2)} \rho,
\end{align}
together with the winding condition
Eqn. \ref{eqn:winding}.
Note in particular that the neighborhood of the insertion point
is mapped to the strong-coupling region $\hat r \to -\infty$ in
spacetime.\footnote{More precisely, the asymptotic condition
  fixes the derivative $\partial_\rho \hat r$ in the
  neighborhood of the operator insertion, and
  Eqn. \ref{eqn:ld-asymp-cond} may be shifted by a
  constant. This constant shift may be an imaginary
  number, since in general the saddles of the functional
  integral may be complex. The asymptotic condition nevertheless
  implies $\mrm{Re}(\hat r)\to - \infty$, meaning that
  the string is sent to the strong-coupling region.}

Consider the leading term $N = 1$ on top of the
$\text{linear-dilaton}\times \mrm{S}^1$ string background in
Eqn. \ref{eqn:sl-expansion}, with a pair of winding operators
$W_+, W_-$ inserted on $\mrm{CP}^1$. Let $W_+(0)$ be fixed at
the origin and $W_-(\infty)$ at the point-at-infinity using the
global conformal redundancy of the string worldsheet
(Fig. \ref{fig:angular-tfd}). 
In the neighborhood of each of the
two insertions, the string is mapped to the strong-coupling
region as it wraps the Euclidean spacetime cylinder 
with unit winding. In between, the 
worldsheet extends along the cylinder toward finite string
coupling before folding back on itself
(Figs. \ref{fig:annulus} and \ref{fig:folded-string}).

When the Euclidean spacetime annulus is halved to prepare the TFD state
on the two zero-time slices $t_\mrm{R} = 0$ at $\theta = 0$ and
$t_\mrm{L} = 0$ at $\theta = \pi$, one finds a pair of
entangled folded strings, one on each spatial slice,
emanating from the strong-coupling region
(Fig. \ref{fig:tfd}).
The respective folded strings are the
images of the worldsheet spatial slices---in the sense of angular
quantization---at $\phi = 0$ and $\phi = \pi$.
The pre-image of
the halved embedded worldsheet that connects the two folded
strings across the target Euclidean cap is the halved worldsheet
shown in Fig. \ref{fig:angular-tfd}, bounded by $\phi = 0$ and $\pi$.
Thus, the pair of folded
strings are prepared in the worldsheet TFD state in two copies
of the angularly quantized Hilbert space, $\cH_{+-}(\reals)\otimes
\cH_{+-}(\reals)$.  

When the angular gauge condition $\theta = \phi$ is continued
with respect to the target and angularly quantized worldsheet
(i.e. $\theta = it_\mrm{R}$ and $\phi = i\Lt_\mrm{R}$, and
$\theta = \pi + i t_\mrm{L}$ and $\phi = \pi + i \Lt_\mrm{L}$),
one may think of the folded strings as evolving along their
respective Lorentzian spacetimes, with both ends continuing to
asymptote to the strong-coupling boundaries.

The asymptotic conditions can be implemented by a boundary
condition on an excised neighborhood of the operator insertion,
in the limit that the boundary shrinks away
(Fig. \ref{fig:brst-contour}). For example, the 
$\text{linear-dilaton}\times \mrm{S}^1$ with $W_+$ and $W_-$
insertions at the origin and point-at-infinity is described by
the $L \to \infty$ limit of the action\footnote{Here we choose
  the cylinder metric, which is 
  responsible for the background-charge shifts of the boundary
  terms by $-Q/2$. These terms are unimportant in the large $k$
  limit, however. One should also add an $L$-dependent
  counterterm to render the on-shell action finite.}
\begin{align}
  \label{eqn:ld-wpm-action}
  S =& \frac{1}{4\pi \alpha'}
  \int\limits_{-L}^L \diff \rho \int\limits_0^{2\pi} \diff \phi
  \lc 
  (\partial_\rho \hat r)^2 + (\partial_\phi \hat r)^2
  +  (\partial_\rho \hat \theta)^2 + (\partial_\phi \hat
  \theta)^2
  \rc\\
  &+2 \lp b_\mrm{sL} - \frac{Q}{2} \rp 
    \int\limits_0^{2\pi} \frac{\diff \phi}{2\pi} 
    \lp \hat r|_{L} + \hat r|_{-L}  \rp
  +\int\limits_0^{2\pi} \frac{\diff \phi}{2\pi}
       \lc
       \sigma_+ (\partial_\phi \hat \theta|_{L} - R)
       + \sigma_- (\partial_\phi \hat \theta|_{-L}-R)
       \rc.\nt
\end{align}
Then the boundary equations of motion,
\begin{subequations}
\begin{align}
  &\partial_\rho \hat r |_{\rho = \pm L}
    = \mp 2\alpha'  \lp b_\mrm{sL} - \frac{Q}{2} \rp \\
  & \partial_\phi \hat \theta|_{\rho = \pm L} = R,
\end{align}
\end{subequations}
imply the asymptotic conditions in the $L \to \infty$
limit. $\sigma_\pm$ are Lagrange multipliers that implement the
winding condition around $\mrm{S}^1$. 

Similarly, for $N>1$ each pair of $W_+,W_-$ insertions may be thought of
as introducing an additional pair of  folded strings to the
background in the TFD state of angular
quantization (Fig. \ref{fig:local-tfd}). Then the exponentiated
sine-Liouville potential amounts in the Lorentzian continuation
to a condensate  of folded strings emanating from
strong coupling on top of the disconnected union of two copies of
$\text{linear-dilaton}\times \mrm{time}$.

Angular quantization should be understood as applying to the
neighborhood of each pair of winding operators, and a
genus-zero string 
diagram with $2N$ such insertions is analogous to
a diagram with 
$2N$ loops of open strings (i.e. $2N$ holes). The associated
process is the amplitude of ordinary closed string scattering
states to interact with $N$ pairs of folded strings in a state
defined by 
the associated Euclidean functional 
integral. This also includes interactions by string exchange
between the folded strings.

The folded string worldsheets we describe here are somewhat formal in the
sense that there is no such saddle of the
$\text{linear-dilaton}\times \mrm{S}^1$ functional
integral---e.g. the two-point function of $W_+$ and $W_-$ alone
vanishes by the anomalous conservation law.
To obtain interesting saddles one must of course insert
additional operators, which will connect to the folded string
asymptotic conditions at strong coupling and fill in the
solution in the interior.
As discussed earlier, in general one
must also regulate the linear-dilaton to suppress its strong-coupling
region.

Alternatively, one may consider the continued sine-Liouville
background itself without expanding the condensate. Here it is
the sine-Liouville potential that regulates the strong-coupling
region. In this 
context, one may interpret the physics of the condensate by
considering the effect of introducing
an additional pair of $\cW_\pm$ insertions on top of it. Then the
asymptotic conditions associated
to these insertions must be modified
from Eqns. \ref{eqn:winding} and \ref{eqn:ld-asymp-cond}
because those map the
string out of the weak-coupling region where the free-field
description is no longer valid.\footnote{In the cigar
  description, the asymptotic conditions for a $\cW_\pm$
  insertion tell the string to wrap the tip of the cigar with
  winding $\pm 1$  (Eqn. \ref{eqn:tip-asymptotic}). We do not
  know the full asymptotic conditions in the sine-Liouville
  description, however.} 
However, the free-field
asymptotic conditions do show that a
string initially 
found at large $\hat r$ will head toward strong coupling at the
rate $\sqrt{\alpha'(k-2)}$ and with winding $\pm 1$ as one
begins to approach the insertion point on the worldsheet. Then
in the weak-coupling regions one will again find a pair of
folded strings, with their ends headed toward strong coupling.
The state of the pair of strings is again the TFD state of
angular quantization on the worldsheet, formally defined by the Euclidean
functional integral on a hemisphere with insertions
of $\cW_+$ and $\cW_-$ on its boundary.
In other words, one has added
another pair of folded strings on top of the
full EPR background, whose ends dissolve into the condensate at
strong coupling.

To properly define the state of the folded strings, one should
include the contribution from the $bc$ gauge-fixing ghost
system. Given a weight $(1,1)$ Virasoro primary operator of the
matter CFT such as $W_\pm$, one forms a physical
operator $W_\pm c \bar c$ of the full string background
by including a factor $c \bar c$ from the
ghosts. In radial quantization, this operator prepares the BRST
invariant state
$\ket{W_\pm c \bar c} = \ket{W_\pm}\otimes \ket{\downarrow
  \downarrow} \in \cH(S^1)$ of the $\text{linear-dilaton}\times
\mrm{S}^1 
\times bc \times \bar b \bar c$ CFT, where $\ket{\downarrow}$ is
the $bc$ vacuum state of ghost number $-1/2$ prepared by
$c(0)$.

In angular quantization, with $W_+c\bar c$ and $W_- c \bar c$
inserted at the poles of $\mrm{CP}^1$, one obtains the TFD state
in two copies of $\cH_{+-}(\reals)\otimes
\cH_{cc}(\reals)\otimes \cH_{\bar c \bar c}(\reals)$, where
$\cH_{cc}(\reals)$ is the angularly quantized Hilbert space
of the $bc$ CFT with $c$ insertions at either end, and likewise
for $\cH_{\bar c \bar c}(\reals)$. To define the latter Hilbert
spaces, it is convenient to recall that the $bc\times \bar b
\bar c$ CFT is itself equivalent to a linear-dilaton of complex
background charge $Q_{X} = 3i / \sqrt{2\alpha'}$
\cite{Friedan:1985ge}.
On the one hand, one has the ghost CFT of central charge $c_{bc}
= -26$, with 
holomorphic stress tensor 
\begin{align}
  T_{bc}(z) = -(\partial b) c - 2 b \partial c,
\end{align}
and with OPEs
\begin{align}
  b(z) c(0) \sim \frac{1}{z},\quad\quad
  b(z) b(0),~ c(z)c(0) \sim 0,
\end{align}
and the analogous anti-holomorphic formulas. The dual description
is a linear dilaton $X(z,\bar z) = X(z) + \bar
X(\bar z)$ with holomorphic stress tensor 
\begin{align}
  T_X(z) = -\frac{1}{\alpha'} (\partial X)^2 - Q_X \partial^2 X
\end{align}
and OPE
\begin{align}
  X(z) X(0) \sim - \frac{\alpha'}{2} \log z.
\end{align}
The fields are related by the identifications
\begin{align}
  \label{eqn:bosonization}
  b(z) = e^{i \sqrt{\frac{2}{\alpha'}}X(z)},\quad\quad
  c(z) = e^{-i \sqrt{\frac{2}{\alpha'}}X(z)},
\end{align}
and similarly for the anti-holomorphic fields.
In particular, the linear-dilaton central charge $c_X = 1 + 6
\alpha' Q_X^2$ reproduces $-26$.
$b$ and $c$ transform with conformal weights $(2,0)$ and
$(-1,0)$ with respect to $T_{bc}$, and $e^{\pm i
  \sqrt{\frac{2}{\alpha'}}X}$ carry the same weights with respect to
$T_X$.

To ensure fractional powers of $b$ and $c$ do not appear, $X$
must be compact with periodicity $2\pi \sqrt{\alpha'/2}$
\cite{Verlinde:1986kw}. There 
is no contradiction with the presence of the linear dilaton,
however,  because the background charge $Q_X$ is complex; though
the dilaton action $-\frac{Q_X}{4\pi} \int \diff^2 \sigma
\sqrt{h} \, \cR X$ is not invariant under the shift, its variation
is a multiple of $2\pi i$, and therefore the functional integral
measure is well-defined.

The $bc$ equations of motion imply the classical conservation of the
ghost-number current $J(z) = - bc$, with respect to which $c$
carries ghost number $+1$ and $b$ carries ghost number
$-1$. Then the identifications Eqn. \ref{eqn:bosonization},
taking care to account for the implied normal ordering,
yield $J = -i \sqrt{\frac{2}{\alpha'}} \partial X$. $X$
therefore corresponds to the bosonization of the
ghost-number current, the ghost-number symmetry of the $bc$
description mapping to the translation symmetry in $X$ of the
linear-dilaton description.

The ghost-number symmetry is
anomalous due to the mismatch in the number of $c$ and $b$
zero-modes on a worldsheet of genus $g$, which, by the
Riemann-Roch theorem, is 
$\mrm{dim}\{ c_0\} - \mrm{dim}\{b_0\} = 3 - 3g$. Then the functional
integral measure $\mrm{D}c\, \mrm{D}b \supset \prod \{\diff c_0\} \{\diff
b_0\}$ carries charge $-(3-3g)$ under the symmetry, and 
the charges of any insertions must sum to $3-3g  = \frac{3}{2}
\chi$ to obtain a non-zero correlation function, where $\chi = 2
- 2g$ is the Euler characteristic of the worldsheet. Correspondingly,
the target translation symmetry of $X$ in the linear-dilaton
background is anomalous (Eqn. \ref{eqn:anomaly}).
With $\alpha_b = -i/\sqrt{2\alpha'}$ for
a $b$ insertion and $\alpha_c = +i/\sqrt{2\alpha'}$ for a $c$
insertion, the
linear-dilaton anomalous conservation law likewise ensures that
a correlator with $N_c$ insertions of $e^{-i
  \sqrt{\frac{2}{\alpha'}}X}$ and $N_b$ insertions of $e^{i
  \sqrt{\frac{2}{\alpha'}} X}$ may be non-zero only if $N_c -
N_b = \frac{3}{2} \chi$.

The anomaly also implies that the ghost-number current is not a
Virasoro primary,  and therefore under the conformal
map $z = e^w$ between the plane and cylinder, the transformed
current is given not by $\til J(w) = z J(z)$ but by $\til J(w) =
z J(z) - \frac{3}{2}$. Then the ghost-number charges of a state
on the cylinder and a local operator on the plane differ by
$-3/2$, accounting for the discrepancy between the ghost-number
1 operator $c$ and the ghost-number $-1/2$ vacuum state
$\ket{\downarrow}$ it prepares. The corresponding anomaly in the
linear-dilaton current is likewise responsible for the shifted
momentum of states proportional to the background charge.

The equivalence of the $bc\times \bar b \bar c$ CFT and the
linear-dilaton $X$ makes the construction of the TFD state in two
copies of the
Hilbert space $\cH_{cc}(\reals) \otimes \cH_{\bar c \bar
  c}(\reals)$ of angular quantization analogous to the TFD in
two copies of $\cH_{+-}(\reals)$. The
asymptotic condition  Eqn.  \ref{eqn:ld-asymp-cond} for a
$W_\pm$ insertion at the end of 
the line is replaced by
\begin{align}
  \label{eqn:ghost-asymp}
  X \overright{\rho \to - \infty}  i \sqrt{2\alpha'} \rho,
\end{align}
or $X \to -i\sqrt{\frac{\alpha'}{2}} \rho$ when combined with the
background-charge contribution in the far past
on the cylinder. Along with the $\text{linear-dilaton}\times
\mrm{S}^1$ action Eqn. \ref{eqn:ld-wpm-action}, one has the
bosonized ghost action
\begin{align}
  \label{eqn:bosonized-action}
  S_X =& \frac{1}{4\pi \alpha'}
  \int\limits_{-L}^L \diff \rho \int\limits_0^{2\pi} \diff \phi
  \lc 
   (\partial_\rho X)^2 + (\partial_\phi X)^2
  \rc
  +2 \lp \alpha_c - \frac{Q_X}{2} \rp 
    \int\limits_0^{2\pi} \frac{\diff \phi}{2\pi} 
    \lp X|_{L} + X|_{-L}  \rp.
\end{align}

The combination $W_\pm c \bar c$ is BRST invariant in the
operator sense, and prepares a BRST invariant state on the
circle in the sense of radial quantization. The thermal (and TFD)
state  in the Hilbert space of angular quantization is likewise
 BRST invariant. Indeed, consider the integral of
$j_\mrm{BRST}$ around the dashed loop shown in
Fig. \ref{fig:brst-contour}, where $j_\mrm{BRST} = e^{-i
  \sqrt{\frac{2}{\alpha'}} X} \lp T_{\hat r} + T_{\hat
  \theta}+T_X\rp$ \cite{DAdda:1987tkh}. The grey 
annulus represents the 
worldsheet with the neighborhood of 
the origin and point-at-infinity excised and replaced by the
asymptotic conditions on their boundaries. The
integral around the closed loop vanishes by conservation of the
current, and therefore the difference in the charges evaluated
along the two radial line segments equals the
integrals of the current over the two circular arcs. The latter
vanish in the limit $L \to \infty$ because the conformal weight
of the excised insertion was zero. That is, when the
$\text{linear-dilaton}\times \mrm{S}^1$ stress tensor $T_{\hat
  r} + T_{\hat \theta}$ (Eqn. \ref{eqn:ld-stress}) is evaluated
on the asymptotic conditions Eqns. \ref{eqn:winding},
\ref{eqn:ld-asymp-cond} one obtains
\begin{align}
  T_{\hat r} + T_{\hat \theta} \overright{z \to 0}
  \frac{1}{z^2},   
\end{align}
whereas on the ghost asymptotic condition
Eqn. \ref{eqn:ghost-asymp} one finds
\begin{align}
  T_X \overright{z \to 0} -\frac{1}{z^2}.  
\end{align}
Together therefore, $j_\mrm{BRST} 
\to 0$. It follows that the BRST charge in the sense of angular
quantization commutes with the Hamiltonian (i.e. the rotation
generator), and therefore the thermal state of angular
quantization associated to $W_\pm c \bar c$ is BRST
invariant.\footnote{In principle, a trace defined with the BRST
  invariant density matrix we construct here should be
  computable by summing over BRST invariant states in the angularly
  quantized Hilbert space.
  We leave it for future work to determine this
  BRST cohomology \cite{daniel-nick}.
  In the classical limit,
  \cite{Maldacena:2005hi} found there were no solutions with vanishing
  stress energy.
  However, the $T = 0$ condition in the linear-dilaton theory
  is not itself a classically conformally invariant equation,
  necessitating a fully quantum treatment.}

\subsubsection{Mutual Locality and the String Moduli Contour}
\label{sec:iepsilon}

We now address the question of mutual locality that arises in
attempting to continue the labels of vertex operator insertions
in a given term in the expansion Eqn. \ref{eqn:sl-expansion}
from their Euclidean Matsubara frequencies $n$ to Lorentzian
energies $E$. Namely, the OPE
Eqn. \ref{eqn:momentum-winding-ope} of a compact boson primary
operator $e^{i n \theta(z, \bar z)}$ of momentum $n$ with a
unit-winding operator $e^{\pm i k \til \theta(z, \bar z)}$ is 
single-valued on the $z$-plane only for integral $n$.
It follows that a correlation function where these insertions
approach one another will behave as
$(z/\bar z)^{\pm n/2} = e^{\pm i n \phi}$, whose continuation 
$e^{\mp E  \phi}$ is a multi-valued function of $\phi$.
Thus, a
correlation function with a given number of winding
operators $W_\pm$ and scattering operators $\cO_{jn,w=0}|_{\hat
  r \to \infty}$ will
generically yield
a multi-valued function of $z$ upon continuation.
This multi-valuedness appears to be an obstacle to
defining Lorentzian string 
perturbation theory  for the EPR thermal microstates obtained
by continuation from each
term of Eqn. \ref{eqn:sl-expansion}. 

On the one hand, one could evaluate a Lorentzian string
amplitude for a given $N$ by first computing the Euclidean
amplitude with $n \in \bbZ$---i.e. integrating the 
Euclidean correlation function over the moduli space---and only then
continuing $n \to i E$ in the final answer. 
In principle, this is a
satisfactory definition of string perturbation theory
in each of the EPR microstates. However, we
would like to give a fully Lorentzian prescription for
computing EPR string amplitudes directly from the apparently
multi-valued correlation functions of winding operators and
scattering operators.

The multi-valuedness of the continued
$\text{linear-dilaton}\times \mrm{S}^1$ correlation functions is not
in and of itself problematic, as these are not the objects of
interest in string perturbation theory. Rather, the worldsheet CFT
correlation functions serve to produce a measure on the
string moduli space, which, when integrated, yields the desired
string amplitude. The relevant conditions on the measure are
then that it should be single-valued along the integration
contour and such 
that the integral converges.

We will argue that the continued correlation functions should be
integrated along a deformed contour in a complexification of the
moduli space for which these conditions are satisfied.

The necessity of complexification and deformation of the
integration contour is in fact encountered already
in ordinary string
backgrounds, though it is not always described in that language
\cite{Witten:2013pra}. The deformation is required to avoid
divergences that appear at points in the moduli space
corresponding to the 
collision of two operator insertions on the worldsheet.
For example,
the Virasoro-Shapiro amplitude for tree-level $2 \to 2$ tachyon
scattering 
in flat spacetime is typically expressed as
an integral over the 
complex plane with three marked points, corresponding to summing
over the insertion point $(z, \bar z)$ of one operator, with
the other three being 
fixed at arbitrary points using the conformal
redundancy. The integral over the $z$-plane
diverges whenever the momenta are above the threshold set by the
tachyon mass, due to singularities when the integrated
operator approaches the fixed insertions
\cite{Polchinski:1998rq}. In this example, it so happens that
the amplitude may be defined by 
evaluating the integral over $z$ in an unphysical region where
it does converge
and then continuing the answer to the physical region. But more
generally, the naive integral 
over the moduli space may never converge, and so a more
systematic approach is necessary.

As explained in \cite{Witten:2013pra}, to obtain a finite string
amplitude one complexifies the moduli space, corresponding in
the $2 \to 2$ example to
treating the coordinates $(z, \bar z)$ of the integrated
operator as independent complex coordinates $(z, \til z)$.
The original moduli space is the section $\til z = \bar z$ on which
$\til z$ coincides with the complex conjugate of $z$.
An appropriate integration contour for the
sum over moduli may then be taken to be
the usual cycle
$\til z = \bar z$ almost everywhere, but
with a small disk neighborhood of each of the fixed insertions
where the original integral diverged replaced
by the Lorentzian cylinder of radial quantization. In general one
must also ascend to a cover of the complexified moduli space on
which the integrand is single-valued.

In the neighborhood of the origin, for example, one
continues $(z = e^{\rho + i \phi}, \bar z = e^{\rho - i \phi})$
to independent complex coordinates, and 
replaces the original contour (on which $\rho$ and $\phi$ were real)
by continuing $\rho = \rho_0 + i \Lt$ in a small neighborhood
$|z|< e^{\rho_0}$ of the insertion. $\Lt$ is the worldsheet
Lorentzian time in the sense of radial quantization, and so the
deformation effectively glues the Lorentzian cylinder at $\Lt =
0$ to the $z$-plane with an excised disk along the boundary
circle $|z| = e^{\rho_0}$. The 
singular neighborhood of the collision point at the origin is
thereby removed from the integration contour, and one instead
integrates along $\Lt \in (-\infty,0].$ $\phi$ remains real on
this contour.

In a string background with Euclidean time winding operator
insertions, we conjecture that Lorentzian string amplitudes
should be defined with an analogous angular rather than radial
contour deformation.
That is, in the neighborhood of a point on the moduli space where 
a momentum operator and a winding operator collide, the original
contour is replaced not
with the Lorentzian cylinder of radial quantization, but
with a Rindler wedge (or wedges) of angular quantization.
In this case it is $\rho$ that remains real while
$\phi = i \Lt$ is continued.\footnote{It would also be interesting to
  understand the appropriate contour deformation when two
  Euclidean time winding operators collide.} 
The contour is sketched in Fig. \ref{fig:contour-deformation},
referring to the portion of the moduli integral over the
insertion point of an ordinary scattering operator. Whereas the
radial deformation prevents two scattering insertions from
colliding, the angular deformation prevents a scattering
insertion from looping around a winding insertion.

With this deformation, the measure
obtained from a $\text{linear-dilaton}\times
\mrm{S}^1$ correlation function after continuing $n \to i E$, which was
multi-valued on the 
original moduli space $\til z = \bar z$, becomes single-valued
on the new contour in the complexification. 
In particular, the continuation of the problematic OPE
coefficient $e^{\mp E \phi}$ that followed from 
Eqn. \ref{eqn:momentum-winding-ope} 
is now a single-valued and oscillatory 
function $e^{\mp i E \Lt}$ of $\Lt$.

For example, consider the two-point function in the $N = 1$
background of
$e^{-2Q(1-j)\hat r}e^{i n \theta}\cO_h$ and $e^{-2Q(1-j)\hat
  r}e^{-i n \theta}\cO_h$, where $\cO_h$ is an internal primary of
weight $h$. Using the conformal gauge redundancy,
fix the winding operators at the
origin and point-at-infinity, and fix one of the momentum
operators at $z = 1$.  The insertion point of the remaining
momentum operator is to be integrated. To avoid the need for a
regulator, let us for simplicity pick $j = \frac{k-1}{2}$, such that
Eqn. \ref{eqn:sl-compatible} is satisfied.\footnote{Albeit this
  is not a
scattering operator, for which $j$ would lie on the complex
branch. When
we discuss the $\mrm{AdS}_3$ duality in the next section,
however, it will be the real branch $j$'s that are of physical
interest. }
Continuing $n \to i 
E$, the internal weight $h$ is chosen to satisfy the on-shell
condition,
\begin{align}
  -\frac{j(j-1)}{k-2} - \frac{E^2}{4k}  +h =1.
\end{align}

The moduli space contour over which the location of the unfixed momentum
operator is to be integrated is shown in
Fig. \ref{fig:contour-deformation}.

\subsection{Asymptotic $\mrm{AdS}_3$ Gravity}
\label{sec:3d-erepr}

Next we consider the examples of $\mrm{ER=EPR}$ for asymptotic
$\mrm{AdS}_3$ gravity obtained by
continuation from the dual descriptions of the
$\mrm{SL}(2,\bbC)_k/\mrm{SU}(2)$ and $\bbZ \bs
\mrm{SL}(2,\bbC)_k/\mrm{SU}(2)$ CFTs. These examples are in
several respects simpler than in two-dimensional dilaton-gravity.

For large $k$, the weakly-coupled description of 
$\mrm{SL}(2,\bbC)_k/\mrm{SU}(2)$ is of a string
in a solid cylinder geometry (Eqn. \ref{eqn:eads-metric})
supported by a $B$-field (Eqn. \ref{eqn:eads-action-cov}). 
A Virasoro primary operator $\Phi_{jw}^{J\bar J}(z, \bar z;
\xi_0, \theta_0)$ is labeled by a point $(\xi_0, \theta_0)$
on the spacetime conformal boundary, where a delta-function
source for the
dual BCFT primary operator of conformal weight $(J, \bar J)$ is
inserted. A Euclidean string amplitude computes a BCFT correlation
function of such dual operator insertions.

In the weakly-coupled description of  $\bbZ \bs
\mrm{SL}(2,\bbC)_k/\mrm{SU}(2)$, the length $\xi 
\sim \xi + \til \beta$ is compactified to form a solid torus
geometry. Untwisted vertex operators of the orbifold may be
obtained from the primaries $\Phi_{jw}^{J\bar J}(z, \bar z;
\xi_0,\theta_0)$ by 
summing over images in $\xi_0$ to enforce periodicity, and the
string amplitudes of such operators compute BCFT correlation
functions on the spacetime boundary $\mrm{T}^2$.

In the dual description of $\mrm{SL}(2,\bbC)_k/\mrm{SU}(2)$
(Eqn. \ref{eqn:3d-sl-cov}), 
which is strongly coupled for large $k$, the $\mrm{EAdS}_3$
solid cylinder is replaced by the first-order cylinder system
$\cF(\bbC/\bbZ)$ 
and an infinite linear-dilaton direction
(Eqn. \ref{eqn:first-order-asymp}),
deformed by the three-dimensional sine-Liouville potential
$ 4\pi \lambda (W_++ W_-)$, where $W_\pm$
wind the now non-contractible $\theta$ cycle of the $W = \xi +
i \theta \in \bbC/\bbZ$ cylinder
(Eqn. \ref{eqn:3d-sl-comps}).
After writing the $\mrm{EAdS}_3$ action in first-order form
(Eqn. \ref{eqn:first-order}), the 
two backgrounds coincide in the weak-coupling region, where
the vertex operators
$\Phi_{jw}^{J\bar J}(z, \bar z; \xi_0,\theta_0)$ 
asymptote to a superposition of $\text{linear-dilaton}\times
\cF(\bbC/\bbZ)$ operators. For an
unflowed operator, for example, transforming
Eqn. \ref{eqn:phi-asymptotic} to cylinder coordinates one finds
for $j > 1/2$,\footnote{Here $\xi,\theta$
are sigma-model fields of the first-order system and
$\xi_0,\theta_0$ are labels for the boundary insertion
point.} 
\begin{align}
  \label{eqn:first-order-cyl-limit}
  \Phi_j(z, \bar z; \xi_0,
\theta_0) \to e^{-2Q(1-j) \hat r}\delta(\xi - \xi_0)\sum_{n \in
  \bbZ} \delta( \theta-
  \theta_0 - 2\pi n) + \text{sub-leading}.
\end{align}
Inserting these $\text{linear-dilaton}\times
\cF(\bbC/\bbZ)$ operators in the sine-Liouville functional
integral, one may in principle compute
$\mrm{SL}(2,\bbC)_k/\mrm{SU}(2)$ correlation functions in the 
dual description.
The dual description of $\bbZ \bs 
\mrm{SL}(2,\bbC)_k/\mrm{SU}(2)$ is the same, but with the
first-order cylinder $\cF(\bbC/\bbZ)$ replaced by the torus
$\cF(\bbC/(\bbZ \times \bbZ))$, corresponding to compactifying
$\xi = \mrm{Re}(W)$.

When the $\mrm{EAdS}_3$ cylinder is cut and continued with respect to
the Euclidean time coordinate $\xi$ along its length, the
Schwinger-Keldysh contour prepares the spacetime vacuum state in
$\mrm{AdS}_3$. Then continuing the
operator labels $\xi_0  \to i t_0$ in the Euclidean string
amplitudes yields BCFT expectation values of local operator
insertions on the Lorentzian boundary cylinder in the dual  vacuum
state of radial quantization.
When $\xi \sim \xi + \til\beta$ is compactified, continuing with
respect to $\xi$
defines string theory
two copies of $\mrm{AdS}_3$ in
the bulk TFD state of inverse temperature $\til \beta$. 

Performing the same continuation in the sine-Liouville
background yields a dual description of string perturbation
theory in $\mrm{AdS}_3$ in the vacuum or thermal state. This is
the better description of these theories when $k-2$ is a small
number. 

Our interest in the context of $\mrm{ER = EPR}$,
however, is the continuation with respect to the compact
Euclidean time coordinate $\theta$---contractible in the
original description and non-contractible in the dual, where the
condensate is responsible for breaking the winding symmetry.

When the original description of
$\mrm{SL}(2,\bbC)_k/\mrm{SU}(2)$ is
continued with respect to $\theta \to i T$, the
Schwinger-Keldysh contour prepares  the TFD state in a connected
pair
of $\mrm{AdS}_3$-Rindler wedges (Eqn. \ref{eqn:ads-rindler}), and
the continued string amplitudes compute expectation
values of the BCFT in its TFD state in two copies of the
angularly quantized Hilbert space on a line.
After the $\xi \sim \xi + \til \beta$ orbifold, 
the same continuation of $\bbZ \bs \mrm{SL}(2,\bbC)_k/\mrm{SU}(2)$
prepares the HH state in
the two-sided BTZ black hole of inverse Hawking temperature
$\beta = 4\pi^2/\til \beta$.\footnote{The black hole coordinates
  from Sec. \ref{sec:black-hole} are obtained by rescaling
$\Theta =
\frac{\beta}{2\pi}\xi$, which is the non-contractible cycle of
periodicity $2\pi$, and $T_\mrm{E} = \frac{\beta}{2\pi}\theta$,
which is the contractible cycle of periodicity $\beta$.} In both
cases, one obtains the 
theory of a string in a connected, two-sided geometry with a
horizon. These are the ER descriptions of string theory in
$\mrm{AdS}_3$-Rindler and BTZ.

The $\theta$ continuation of the dual backgrounds defines the
EPR description of each of these two string theories. As in
Eqn. \ref{eqn:sl-expansion}, we treat the sine-Liouville
potential as a large deformation of the free theory, now being the
$\text{linear-dilaton}\times \cF(\bbC/\bbZ)$, or
$\cF(\bbC/(\bbZ\times \bbZ))$, background.  At leading order,
continuing $\theta = \mrm{Im}(W)$ yields the 
Schwinger-Keldysh contour for the TFD state in two copies of
$\text{linear-dilaton}\times \cF(\reals^{1,1})$ or $\cF(\reals
\times \mrm{S}^1)$ (Fig. \ref{fig:tfd}).

The remaining terms in the expansion insert pairs of $W_+,W_-$
operators on top of the free-field Euclidean background, and in turn
introduce deformations of the spacetime TFD state upon
continuation. As in 
the two-dimensional case, we interpret each set as inserting a pair of
folded strings emanating from the strong-coupling region in the
TFD state of angular quantization on the worldsheet. The
asymptotic conditions defining a $W_\pm$ insertion are as in
Eqn. \ref{eqn:ld-asymp-cond} in the linear-dilaton direction,
and, from Eqn. \ref{eqn:winding-ope}, 
\begin{subequations}
\begin{align}
  &W(z) \overright{z \to 0} \pm \log(z)\\
  &\bar W ( \bar z) \overright{\bar z \to 0} \pm \log(\bar z).
\end{align}
\end{subequations}
Or, in terms of $\xi = \frac{1}{2} (W + \bar W)$ and $\theta =
\frac{1}{2i} (W - \bar W)$,
\begin{subequations}
  \begin{align}
    \label{eqn:xi-asymptotic}
      &\xi \overright{\rho \to - \infty} \pm \rho\\
  &\theta \overright{\rho \to -\infty} \pm \phi.
\end{align}
\end{subequations}

\begin{figure}[t]
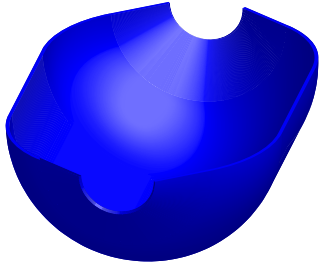

  \centering
  \ig{0cm}{scale=1.4}{3d-folded-string}
  \caption{\footnotesize\bd{Folded Strings in $\mrm{EAdS}_3$}. The
    spacetime image of the 
    halved worldsheet from Fig. \ref{fig:angular-tfd} with a
    pair of $W_\pm$ insertions is shown. The inner semicircles
    correspond to the strong-coupling region $\hat r \to -
    \infty $ of the linear-dilaton, the $\theta$ direction winds
    around these circles, and the $\xi$ direction extends along
    the depth of the figure. The left and right boundaries are
    folded strings extending from the strong-coupling region
    partway toward finite $\hat r$ before falling back to
    strong coupling, while $\xi$ ranges from minus infinity to
    infinity. The folded strings live on the zero-time slices
    $\theta =t_\mrm{R} = 0, \theta = it_\mrm{L}+ \pi = \pi$ of
    the Lorentzian continuation.} 
  \label{fig:3d-folded-string}
\end{figure}

Thus, with $W_+(0)$ and $W_-(\infty)$ fixed on $\mrm{CP}^1$, the
worldsheet is mapped to the strong-coupling region with unit
winding in the neighborhood of each insertion, while $\xi \to \mp
\infty$. In between, the string formally extends partway toward
finite $\hat r$ before folding back toward strong-coupling,
while $\xi$ ranges over the real line. When the target is halved
to prepare the spacetime TFD state, one finds on the left and right
zero-time slices a folded string emanating from the
strong-coupling region and extended along the $\xi$ direction
(Fig. \ref{fig:3d-folded-string}).

The same asymptotic conditions
describe the sine-Liouville potential in the Euclidean black
hole. Although Eqn. \ref{eqn:xi-asymptotic} may at first appear
to be in 
tension with the compactification of $\xi$, recall
that in these variables the 
current with respect to which the
orbifold is performed is not simply $\xi$'s conjugate variable
$\chi$, but $\chi + \partial 
\xi$, and the second term introduces an additional twist. As a
result it is not obvious at a glance that the asymptotic
condition respects the orbifold projection, but 
one is guaranteed as much because
the operators $W_\pm$ are invariant.

Unlike the two-dimensional $\text{linear-dilaton}\times \mrm{S}^1$
background, where the winding and continued scattering operator
insertions were mutually non-local
(Eqn. \ref{eqn:momentum-winding-ope}) and 
required the Rindler deformation of the moduli contour
discussed in the previous section, the 
boundary position basis vertex operators admitted by the $\mrm{AdS}_3$
asymptotics do not suffer from this issue of mutual locality with the
winding operators that appear in the expanded
background. Indeed, the Euclidean boundary position basis vertex 
operators are manifestly periodic in Euclidean time, and the
Lorentzian vertex operators, being defined by continuation,
likewise respect the imaginary time periodicity. 
Explicitly, the relevant OPE to check is between\footnote{The 
  additional 
  factors of Eqns. \ref{eqn:3d-sl-comps} and
  \ref{eqn:first-order-cyl-limit} are not relevant to the mutual
  locality question.} 
the winding operators $e^{\mp \sqrt{\frac{k}{\alpha'}}
  \int^{(z,\bar z)}
  \diff z' \hat \chi(z') + \diff \bar {z}'\hat {\bar \chi}(\bar
  z')}$ of the first-order system and the 
delta-function operators
\begin{align}
  \sum_{n \in \bbZ} &\delta^2\lp \hat W(z) - \hat W_0- 2\pi i
  \sqrt{\alpha' k}  n\rp\\
  &\propto
  \sum_{n \in \bbZ}  \int \diff p\, e^{\frac{i}{2} \lp p - i
    \frac{n}{\sqrt{\alpha' k}}
  \rp \lp \hat W(z)- \hat W_0\rp}
  e^{\frac{i}{2} \lp p + i \frac{n}{\sqrt{\alpha' k}} \rp  \lp \hat{\bar
  W}(\bar z) - \hat{\bar W}_0\rp}, \nt
\end{align}
which are conveniently expressed via their inverse Fourier
transform. Using
Eqn. \ref{eqn:w-chi-ope}, one finds the OPE
\begin{align}
  &e^{\frac{i}{2} \lp p - i\frac{n}{\sqrt{\alpha' k}}\rp
  \lp \hat W(z)- \hat W_0\rp}
  e^{\mp \sqrt{\frac{k}{\alpha'}} \int^0 \diff z' \hat \chi(z')}\\
  &\hspace{1cm}=
    z^{\pm \frac{i}{2}\sqrt{\alpha'k} \lp p - i\frac{n}{\sqrt{\alpha'
  k}}\rp}
  e^{\frac{i}{2} \lp p - i\frac{n}{\sqrt{\alpha' k}}\rp
  \lp \hat W(0)- \hat W_0\rp
    \mp \sqrt{\frac{k}{\alpha'}} \int^0 \diff z' \hat \chi(z')}\nt
    (1 + \cO(z)),
\end{align}
and similarly for the anti-holomorphic factors. Together, the
$z,\bar z$ dependence of the pre-factors is
\begin{align}
  z^{\pm \frac{i}{2}\sqrt{\alpha'k} \lp p - i\frac{n}{\sqrt{\alpha'
  k}}\rp}
  \bar z^{\pm \frac{i}{2}\sqrt{\alpha'k} \lp p + i\frac{n}{\sqrt{\alpha'
  k}}\rp}
  =
  |z|^{\pm i\sqrt{\alpha'k} p}
  \lp \frac{z}{\bar z} \rp^{\pm n/2},
\end{align}
which is single-valued and oscillatory, regardless of whether
$\hat W_0, \hat {\bar W}_0$ sit on the Euclidean or Lorentzian section.
Then the only $i\vep$ prescription necessary to define
Lorentzian string amplitudes with vertex operator insertions in
the boundary position basis is the usual prescription that
replaces the target Lorentzian time insertion point $T \to
T(1-i\vep)$, as expected from the perspective of the dual CFT,
and required to identify the branch of the
continued of Euclidean string amplitudes 
appropriate for time-ordered expectation values. 

Of course, if one wishes to compute string amplitudes in the
momentum basis rather than the boundary position basis
(Eqns. \ref{eqn:vacuum-fourier}, \ref{eqn:btz-fourier}), an analogous mutual locality issue
as in the two-dimensional case will arise, and the Rindler
deformation of the moduli integration contour will again be
necessary.

Unlike the two-dimensional black hole, 
in $\mrm{AdS}_3$ the short-string vertex operators of interest
include the real branch of $j$, and the unregulated expansion
Eqn. \ref{eqn:sl-expansion} is of greater practical utility. For
the two-point function of $\Phi_j(z, \bar z; \xi_0, \theta_0)$,
for example, the solution of the compatibility condition
Eqn. \ref{eqn:sl-compatible} at genus zero is
\begin{align}
  j_N = \frac{1}{2} \lp (k-2)N+1 \rp.
\end{align}
In e.g. the $N = 1$ background, one finds a compatible correlator
for $j = \frac{k-1}{2}$.\footnote{Curiously, this coincides with
  the upper bound of the discrete-series 
spectrum.} Then one could compute
\begin{align}
  \label{eqn:ads-correlator}
  &\vev{\Phi_j(z_0, \bar z_0; \xi_0,\theta_0) \Phi_j(z_1, \bar
  z_1; \xi_1, \theta_1)}_{\mrm{SL}(2,\bbC)_k/\mrm{SU}(2)}\\
  &~~=
  \lp \frac{\lambda}{2\alpha'} \rp^{2}
  \vev{ \cV_{j,\xi_0,\theta_0}(z_0, \bar z_0)
    \cV_{j,\xi_1,\theta_1}(z_1, \bar z_1)
  \int \diff^2 z_+ \diff^2z_-~ W_+(z_+, \bar z_+) 
  W_-(z_-,\bar z_-)}_{\mrm{LD \times \cF(\bbC/\bbZ)}},\nt
\end{align}
where $\cV_{j,\xi_i,\theta_i}$ is given by the right-hand-side
of Eqn. \ref{eqn:first-order-cyl-limit}. To extract the 
correlator for general $j$ one would, as in Liouville, determine
the meromorphic function whose poles coincide with the
free-theory correlators of $j_N$. 

To compute the string
amplitude obtained from Eqn. \ref{eqn:ads-correlator}, one would
fix $z_+=0$, $z_- = \infty$, and $z_1 = 1$, 
and integrate over $z_0$. No contour deformation is necessary in
this basis. One must also tensor the operators with an internal
CFT scalar primary $\cO_h$ of conformal weight $h$, chosen such that
the on-shell condition is satisfied,
\begin{align}
  -\frac{j(j-1)}{k-2}  +h =1.
\end{align}

\subsection{Infinitesimal Lorentzian Dualities}
\label{sec:inf-lorentzian}

Lastly, we briefly consider the Lorentzian continuation of the
infinitesimal dualities described in
Sec. \ref{sec:inf-fzz}. Each infinitesimal duality identifies
two equivalent descriptions of the effect of a conformal
perturbation by the sine-Liouville
operator $\cO_\mrm{sL} = \cW_+ + \cW_-$
(Eqn. \ref{eqn:sl-op})
in the ER description of the respective CFTs.
In one description, the deformation introduces a condensate of
strings that wrap the Euclidean horizon $r = 0$, and  in the other
the constant mode of the dilaton is shifted. The black hole mass
is in turn shifted under the deformation, in the case of
$\mrm{SL}(2,\reals)_k/\mrm{U}(1)$ and 
$\bbZ \bs \mrm{SL}(2,\bbC)_k/\mrm{SU}(2)$.

The conformal perturbation deforms the Euclidean background that
defines the Lorentzian string theory upon
continuation. Expanding the perturbation as in
Eqn. \ref{eqn:sl-perturbation}, a
superposition of $\cW_\pm$ insertions is introduced on the
worldsheet. Note 
that, in contrast to Eqn. \ref{eqn:sl-expansion}, we are now
treating the perturbation as an expansion around the exact CFT
background as opposed to the free-field background. In
particular, the winding number need not be
conserved. 

In the cigar, the asymptotic conditions
Eqn. \ref{eqn:tip-asymptotic} for 
$\cW_\pm$ map the neighborhood of the insertion point to the tip
as pictured in Fig. \ref{fig:cigar-wrapping}. On the $\theta =
0$ fixed-time slice, one finds a string with one end at
the tip of the cigar, and likewise on the $\theta = \pi$ slice.
When the cigar is
cut in half across its $\theta$ cycle to prepare the HH state of
the black hole, the $\cW_\pm$ insertion therefore adds a pair of
entangled strings in the TFD state of
$\cH_{\pm,1}(\reals)\otimes \cH_{\pm,1}(\reals)$, one in the left wedge
and one in the right, each with 
one end pinned at the horizon bifurcation point. Or, with a pair
of $\cW_+,\cW_-$ insertions as in Fig. \ref{fig:angular-tfd},
one finds on the spacetime zero-time slice a pair of folded
strings emanating from the horizon in the TFD state of
$\cH_{+-}(\reals) \otimes \cH_{+-}(\reals)$, as 
sketched in Fig. \ref{fig:horizon-string}. These folded strings
should be compared 
with those emanating from the strong-coupling
region of the EPR background (Fig. \ref{fig:tfd}). One may think
of the deformation as adding to the condensate of strings that
make up the black hole and thus shifting the mass. The dual
dilaton-shifting description of course shifts the mass as well.

In $\mrm{SL}(2,\bbC)_k/\mrm{SU}(2)$ or $\bbZ \bs
\mrm{SL}(2,\bbC)_k/\mrm{SU}(2)$ the picture is similar. The 
$\cW_\pm $ insertions again add pairs of strings to the
$\mrm{AdS}_3$-Rindler or $\mrm{BTZ}$  backgrounds with one or
both ends on the 
horizon bifurcation locus, the strings now extending in the
$\xi$ direction as well due to the additional asymptotic
condition $\xi \to \pm \frac{1}{2} e^{2\rho} + \mrm{const}.$
(Eqn. \ref{eqn:uplifted-solution}). The
conformal perturbation by $\cO_\mrm{sL}$ introduces a condensate
of such strings.

\section*{Acknowledgements} We would like to thank Nick Agia, Ofer
Aharony, Bruno
Balthazar, Cindy Keeler, Zohar Komargodski, David Kutasov, Juan
Maldacena, Shiraz Minwalla, Victor Rodriguez, Eva Silverstein, Edward Witten, and Xi Yin for 
stimulating and helpful discussions. This work was supported in
part in its inception by NSFCAREER grant PHY-1352084, and then by DOE grant DE-SC0021528.

\bibliographystyle{utphys}

\singlespacing
\bibliography{bib_fzz}
\nocite{*}

\end{document}